\documentclass[prl,aps,psfig,preprintnumbers,amsmath,amssymb,showpacs]{revtex4}
\usepackage{color,graphicx}\usepackage{epsfig}

\begin{document}

\title{Statistical description of 
complex nuclear phases in supernovae and proto-neutron stars}

\author{Ad. R. Raduta$^{1}$}
\author{F. Gulminelli$^{2}$ }

\affiliation{$^{1}$~NIPNE, Bucharest-Magurele, POB-MG6, Romania,\\
$^{2}$~LPC (IN2P3-CNRS/Ensicaen et Universit\'e), 
F-14076 Caen c\'edex, France}

\begin{abstract}
We develop a phenomenological statistical model for dilute star matter
at finite temperature, in which free nucleons
are treated within a mean-field approximation and nuclei are considered to
form a loosely interacting cluster gas. Its domain of applicability,
that is baryonic densities ranging from about $\rho>10^8$ g $\cdot$ cm$^{-3}$  
to normal nuclear density, temperatures between 1 and 20 MeV 
and proton fractions between 0.5 and 0, makes it suitable for the 
description of baryonic matter produced in
supernovae explosions and proto-neutron stars. 
The first finding is that, contrary to the common belief,
the crust-core transition is not first order, and for all subsaturation
densities matter can be viewed as a continuous fluid mixture between  
free nucleons and massive nuclei. As a consequence, the equations of 
state and the associated observables do not present any discontinuity
over the whole thermodynamic range. 
We further investigate the nuclear matter composition over a wide range of 
densities and temperatures.
At high density and temperature our model accounts for a much larger mass fraction bound in
medium nuclei with respect to traditional approaches as Lattimer-Swesty,
with sizeable consequences on the thermodynamic quantities. 
The equations of state agree well with the presently used EOS 
only at low temperatures and in the homogeneous matter phase, while 
important  differences are present in the crust-core transition region.
The correlation among the composition of baryonic matter and neutrino
opacity is finally discussed, and we show that the two problems can be 
effectively decoupled.
\end{abstract}

\today

\maketitle

\section{I. Introduction}

Nuclear matter is not only a theoretical idealization providing benchmark studies of the
effective nuclear interaction, but it is also believed to constitute the major baryonic
component of massive objects in the universe, as exploding supernovae cores and neutron stars.
The structure and properties of these astrophysical objects at baryonic densities exceeding normal
nuclear matter density is still highly speculative
\cite{knorren,takatsuka,steiner,blaschke,panda,baldo,nakazato,yasutake}. 
Conversely for subsaturation densities it is well established that matter is 
mainly composed of neutrons, protons, electrons, positrons and photons in thermal 
and typically also chemical equilibrium \cite{prakash_science,haensel_book}. 
Depending on the thermodynamic condition, neutrinos 
and anti-neutrinos can also participate to the equilibrium. 

Such composite matter is subject to the contrasting couplings of the 
electromagnetic and the strong interaction.
Because of the electron screening, the two couplings act on comparable 
length scales giving rise to the phenomenon of frustration
\cite{horowitz,watanabe_prl,newton,sebille}, 
well-known in condensed matter physics \cite{frustration}.
Because of this, a specific phase diagram, different from the one of nuclear matter and 
including inhomogeneous components, is expected in stellar matter \cite{ising_star}.  

Many theoretical studies exist at zero temperature. In a cold neutron star, 
going from the dilute crust to the dense core, a transition is known 
to occur from a solid phase constituted of finite nuclei 
on a Wigner lattice immersed in a background of delocalized electrons and neutrons, 
through intermediate inhomogeneous phases composed of non-spherical nuclei (pasta phases), 
to a liquid phase composed of uniform neutrons, protons and electrons
\cite{Lattimer85,cold_NS,pethick,douchin,watanabe,maruyama}.

At finite temperature the matter structure and properties are not as well settled.
The most popular phenomenological approaches are the Lattimer-Swesty \cite{LS91} (LS) 
and the Shen \cite{shen} equation of state, recently updated in Ref. \cite{shen_horowitz}.
In these standard treatments currently used in most supernovae codes, the
dilute stellar matter at finite temperature is  described 
in the baryonic sector as a statistical equilibrium 
between protons, neutrons, alphas and a single heavy nucleus. 
The transition to homogeneous matter in the neutron star core is 
supposed to be first order in these modelizations and obtained through a 
Maxwell construction in the total density at fixed proton fraction.

It is clear that such single nucleus approximation (SNA) is highly schematic 
and improvements are possible.
Concerning integrated quantities as thermodynamic functions and equations of state, such 
variables may be largely insensitive to the detailed matter composition \cite{burrows}, 
though we show in this paper that this is not always the case. However it is also
known that the composition at relatively high density close to saturation, 
together with the pressure and symmetry energy, governs the electron capture rate, which in turn 
determines the proton fraction at bounce and the size of the homologous core, a key quantity 
to fix the strength of the shock-wave and the output of the 
supernovae explosion
\cite{bethe79,zeldovich,horowitz_nu,pinedo2006,janka,sonoda}.
Moreover the composition may also affect the nucleosynthesis of heavy elements, which is still
poorly understood \cite{Hix2003,Pinedo2004,cowan,qian}, 
as well as the neutrino scattering through the core after bounce \cite{margueron,caballero}, and the
cooling rate of neutron stars \cite{page,yakovlev}. 
For these reasons, in the recent years, many efforts have been done to improve 
over the simplistic representation of stellar matter given by the SNA approach.   

The different modelizations which consider a possible distribution of all 
different nuclear species inside dilute stellar matter are known under the 
generic name of Nuclear Statistical Equilibrium (NSE) \cite{NSE,mishustin,blinnikov,souza}.
The basic idea behind these models is the Fisher conjecture that strong 
interactions in dilute matter may be entirely exhausted by clusterization
\cite{fisher}. 
In these approaches stellar matter in the baryonic sector is then viewed as a 
non-interacting ideal gas of all possible nuclear species in thermal equilibrium.
The result is that thermodynamic quantities like entropies and pressure appear
very similar to the ones calculated with standard approaches, 
while noticeable differences are seen in the matter composition.
In particular an important contribution of light and intermediate mass
fragments is seen at high temperature, 
which is neglected in standard SNA approaches.

The strongest limitation of NSE-based approaches is that they completely
neglect in-medium effects, which are known to be very important in nuclear
matter. Since the only nuclear interactions are given by the cluster self-energies,
the homogeneous matter composing the neutron star core cannot be
modelized, nor it can be the phenomenon of 
neutron drip in the inner crust, well described by mean-field models \cite{negele}. 
As a consequence, these models cannot be applied at densities close to saturation 
and the crust-core transition cannot be described.  
 
To overcome this problem, different microscopic 
\cite{samaddar,schwenk,heckel,typel} as well as 
phenomenological \cite{hempel2010} approaches have been developed
in the very recent past. 
In this paper we would like to introduce a phenomenological model that treats the nuclei 
component within an improved NSE, while it describes the unbound protons and
neutrons in the finite temperature Hartree-Fock approximation.

The plan of the paper is as follows.
The first part of the paper is devoted to the description of the model. 
The clusterized component, the homogeneous component, 
the properties of the mixture and the lepton sector are described in 
successive sections together with their thermodynamic properties.
A particular attention is devoted to the modelization of the crust-core
transition. We show that the inclusion of excluded volume is sufficient 
to describe the transition from the clusterized crust to the homogeneous core,
and that this transition is continuous.  
Different generic as well as specific arguments are given against 
the possibility of a first-order transition.
The second part of the paper gives some results relevant for the star matter phenomenology. 
The first section shows observables following constant chemical potential
paths, in order to connect the observables with the properties of 
the phase diagram. Then the behavior of the different quantities for 
constant proton fractions is displayed in order to compare with more 
standard treatments of supernova matter.
Finally the last section addresses the problem of neutrino trapping and the 
interplay between the matter opacity to neutrinos and matter composition.
Conclusions and outlooks conclude the paper. 

\section{II. The model}

The model aims to describe the thermal and chemical properties of
nuclear matter present in supernovae and (proto)-neutron stars at
densities ranging from the normal nuclear density 
$\rho_0$ to $\approx 10^{-6} \rho_0$, temperatures between 0 and 20 MeV
and proton concentration between 0.5 and 0.
In this regime, the star matter typically consists of a mixture
of nucleons, light and heavy nuclear clusters,
neutrinos (if we consider the thermodynamic stage where neutrinos are trapped), 
photons and a charge neutralizing background of electrons and positrons.

As there is no interaction among electrons, neutrinos, photons and nuclear
matter, the different systems may be treated separately and their contributions
to the global thermodynamic potential and equations of state added-up.

In the grancanonical ensemble this reads,
\begin{equation}
G(\beta,\beta \mu_n,\beta \mu_p,\beta \mu_e,V)=
G^{(bar)}(\beta,\beta \mu_n,\beta \mu_p,V)
+G^{(lep)}(\beta,\beta \mu_e,V)
+G^{(\gamma)}(\beta,V),
\label{independent}
\end{equation}
where the grancanonical potential,  
\begin{eqnarray}
G(\beta,\beta \mu_n,\beta \mu_p,\beta \mu_e,V)&=&
\ln {\cal Z}_{gc}(\beta,\beta \mu_n,\beta \mu_p,\beta \mu_e,V)
\nonumber\\
&=& \bar S\left[\beta,-\beta \mu_n,-\beta \mu_p\right],
\end{eqnarray} 
is the Legendre transformation of the
entropy $S$ with respect to the fixed intensive variables. In the previous equations
$V$ is an arbitrary macroscopic volume and ${\cal Z}_{gc}$ 
is the grancanonical partition sum.

The observables conjugated to the ones fixed by the reservoir and geometry
can be immediately calculated as partial derivatives of $G$.
Thus, the total energy density is
\begin{eqnarray}
e&=&-\frac{1}{V}\left(\frac{\partial G}{\partial \beta}\right)|_{\beta\mu_n,\beta\mu_p,V}
\nonumber
\\
&=&e^{(bar)}+e^{(lep)}+e^{(\gamma)},
\label{eq:TotalEnergy}
\end{eqnarray}
the different particle densities are
\begin{equation}
\rho_{i}=\frac{1}{V}\left(\frac{\partial G}{\partial \left( \beta
  \mu_{i}\right)}\right)|_{\beta,\beta\mu_{j},V},
\end{equation}
where $i={n,p,e}$,
and finally the total pressure is
\begin{equation}
p =\frac{G}{\beta V}=p^{(bar)}+p^{(lep)}+p^{(\gamma)}.
\label{eq:TotalP}
\end{equation}
\subsection{A.The baryon sector}

The light and heavy nuclei are assumed to form a gas of loosely-interacting
clusters which coexist in the Wigner-Seitz cell with a homogeneous
background of delocalized nucleons.
To avoid exceeding the normal nuclear density and naturally allow for
homogeneous-unhomogeneous matter transition, nuclei and nucleons are forbidden
to occupy the same volume.
In the following we start describing the modelization of these two components 
separately, and we turn successively to the properties of the mixture obtained
when the two are supposed to be simultaneously present in the Wigner-Seitz cell.


\subsubsection{1. The homogeneous nuclear matter component}

Mean-field models constitute a natural choice for approaching interacting particle
systems. By introducing a mean-field potential, the physical problem is
reduced to the simplified version of a system of non-interacting particles. 
Effective nucleon-nucleon interactions allow one to express the system average 
energy as a simple single-particle density functional, and to 
cast the nuclear matter statistics in a way which is formally very similar 
to an ideal Fermi gas\cite{vautherin}.

The mean field energy density of an infinite homogeneous system
$e^{(HM)}=<\hat H>_0/V$ is a functional of the particle
densities $\rho _{q}$ and kinetic densities $\tau _{q}$ 
for neutrons ($q=n$) or protons ($q=p$). 
At finite temperature, the mean field approximation consists in expressing 
the grancanonical partition function of the
interacting particle system as the sum of the
grancanonical partition function of the corresponding 
independent-particle system associated to the mean-field single-particle energies, 
with the temperature weighted difference between the average
single-particle energy ($<\hat W>_0=-\partial_{\beta} \ln {\cal Z}_0$) 
and the mean-field energy
\begin{equation}
\ln {\cal Z}^{(HM)} \approx \ln {\cal Z}_0^{(HM)}+\beta \left( \left< \hat W\right>_0
-\left<\hat H  \right>_0\right).
\end{equation} 

The one body partition sum is defined as:

\begin{equation}
{\cal Z}_{0}^{(HM)}=Tr[e^{-\beta (\hat{W}_{0}-\mu _{n}\hat{N}_{n}-\mu _{p}
\hat{N}_{p})}]={\cal Z}_{0}^{n}{\cal Z}_{0}^{p},
\end{equation}
and can be expressed 
as a function of the neutron and proton kinetic energy density : 

\begin{equation}
\frac{\ln {\cal Z}_{0}^{q}}{V}=2 \int_{0}^{\infty }
\ln(1+e^{-\beta (\frac{p^{2}}{2m_{q}^{*}}-\mu _{q}^{\prime })})
\frac{4\pi p^{2}}{h^{3}} dp
=\frac{\hbar^2}{3m_{q}^{*}}\beta \tau 
_{q},
\end{equation}

with

\begin{eqnarray}
\rho _{q} &=&2\int_{0}^{\infty } n_{q}(p)  
\frac{4\pi p^{2}}{h^{3}}  dp,\label{EQ:density}\\
\tau _{q} &=&2\int_{0}^{\infty } \frac{p^{2}}{\hbar ^{2}} 
n_{q}(p) \frac{4\pi p^{2}}{h^{3}} dp. \label{EQ:tau}
\end{eqnarray}

In these equations the effective chemical potential $\mu _{q}^{\prime }$ includes the self energies 
according to $\mu _{q}^{\prime }=\mu _{q}-\partial _{\rho _{q}}e^{(HM)}$,
${m_{n,p}^{*}}=\left ( \partial e^{HM}/\partial \tau_{n,p}\right )^{-1}/2$ 
are the neutron (proton) effective mass,  and 
the factor $2$ comes from the spin degeneracy. 

Equation (\ref{EQ:density}) establishes a self-consistent relation between the
density of q-particles $\rho _{q} $ and 
their chemical potential $\mu _{q}$. Introducing
the single particles energies 
$\epsilon _{q}^{i}=\frac{p_{i}^{2}}{2m_{q}^{*}}+\partial _{\rho _{q}}e^{(HM)}$,
the above densities 
can be written as regular Fermi integrals by shifting the chemical potential
according to $\mu _{q}^{\prime }=\mu _{q}-\partial _{\rho _{q}}e^{HM}.$ 
The Fermi-Dirac
distribution indeed reads:
 
\begin{equation}
n_{q}(p)=\frac{1}{1+\exp(\beta (p^{2}/2m_{q}^{*}-\mu _{q}^{\prime }))}.
\label{EQ:distribution}
\end{equation}

Eqs. (\ref{EQ:distribution}) and (\ref{EQ:density}) 
define a self-consistent problem since $m_{q}^{*}$ depends 
on the densities. 
For each couple $(\mu _{n}^{\prime },\mu _{p}^{\prime })$
a unique solution $(\rho _{n},\rho _{p})$ is found by iteratively
solving the self-consistency between $\rho _{n,p}$ and $m_{n,p}^{*}$. 
Then Eq. (\ref{EQ:tau}) is used to calculate $\tau_{n,p}$. 

At the thermodynamic limit the system volume $V$ diverges  
together with the particle numbers $\langle \hat{N}_{n} \rangle$, 
$\langle \hat{N}_{p} \rangle$, and the thermodynamics is completely defined
as a function of the two particle densities $(\rho_n,\rho_p)$, or equivalently
the two chemical potentials $(\mu _{n},\mu _{p})$.

With Skyrme based interactions, the energy density of homogeneous, spin saturated matter with no
Coulomb effects is written as:  

\begin{eqnarray}
e^{(HM)}&=& \frac{\hbar ^{2}}{2m}(\tau _{n}+\tau _{p}) \nonumber \\
&+&t_{0}(x_{0}+2)(\rho _{n}+\rho _{p})^{2}/4
-t_{0}(2x_{0}+1)(\rho _{n}^{2}+\rho_{p}^{2})/4 \nonumber \\
&+& t_{3}(x_{3}+2)(\rho _{n}+\rho_{p})^{\sigma +2}/24
-t_{3}(2x_{3}+1)(\rho _{n}+\rho_{p})^{\sigma }(\rho _{n}^{2}+\rho _{p}^{2})/24 \nonumber \\
&+& \left(t_{1}(x_{1}+2)+t_{2}(x_{2}+2)\right)(\rho _{n}+\rho _{p})(\tau _{n}+\tau_{p})/8
+\left(t_{2}(2x_{2}+1)-t_{1}(2x_{1}+1)\right) (\rho _{n}\tau _{n}+\ \rho _{p}\tau_{p})/8 ,
\end{eqnarray}

where $t_0,t_1,t_2,t_3, x_0,x_1,x_2,x_3,\sigma$ are Skyrme parameters.

Several Skyrme potentials have been developed over the years for describing the
properties of both infinite nuclear matter and atomic nuclei, and address the
associated thermodynamics. 
It was thus in particular shown that nuclear matter manifests liquid-gas like first-order
phase transitions up to a critical temperature
\cite{ducoin_paper1,barranco,shlomo,meyer,rios}.

In order to make direct quantitative comparisons between our model and the one
of Lattimer and Swesty \cite{LS91}, through this paper
nucleon-nucleon interactions are accounted for according to 
the SKM* parameterization \cite{skmstar} if not explicitely mentioned otherwise. 
Table \ref{table:skmstarparam} summarizes the force parameters while the main
properties of nuclear matter are summarized in Table \ref{table:skmstarprop}. 

\begin{table}
\begin{center}
\caption{SKM* force parameters \cite{skmstar}.\label{table:skmstarparam}}
\begin{tabular}{c|c}
\hline
\hline
Parameter & Value   \\
\hline
$t_0$ (MeV fm$^3$) & -2645 \\
$t_1$ (MeV fm$^5$) & 410 \\
$t_2$ (MeV fm$^5$) & -135 \\
$t_3$ (MeV fm$^{3+3\sigma}$) & 15595 \\
$x_0$ & 0.09 \\
$x_1$ & 0. \\
$x_2$ & 0. \\
$x_3$ & 0. \\
$\sigma$ & 1/6 \\
\hline
\end{tabular}
\end{center}
\end{table}

\begin{table}
\begin{center}
\caption{Infinite nuclear matter and surface properties of SKM* force 
\cite{skmstar}.\label{table:skmstarprop}}
\begin{tabular}{c|c}
\hline
\hline
Observable & Value   \\
\hline
$E/A$ (MeV) & -15.78 \\
$K_{\infty}$ (MeV) & 216.7 \\
$m^*/m$ & 0.79 \\
$C_{sym}$ (MeV) & 30.03 \\
$\rho_0$ (fm$^{-3}$) & 0.1603 \\
$a_s$ (MeV) & 17.51 \\
$k_s$ & 3.74 \\
\hline
\end{tabular}
\end{center}
\end{table}

The thermodynamic properties of the system are best studied introducing the constrained entropy:

\begin{equation}
s_c^{(HM)}=s^{(HM)}-\beta e^{HM},
\end{equation}
where the entropy density is nothing but the Legendre transform of the mean field partition sum:
\begin{equation}
s^{(HM)}=\ln {\cal Z}^{(HM)}/V+\beta \left ( e^{(HM)} -\mu_n\rho_n -\mu_p\rho_p \right ).
\end{equation}
 
Instabilities are then recognized as local convexities of the constrained
entropy at a given temperature $T=1/\beta$ in the two-dimensional 
density  plane $(\rho_n,\rho_p)$. In the presence of an instability, the
system entropy can be maximized by phase mixing, which corresponds 
to a linear interpolation in the density plane. This geometrical 
construction corresponds to the well-known Gibbs equality between all
intensive parameters of two coexisting phases in equilibrium:

\begin{eqnarray}
\beta^{(1)}&=&\beta^{(2)}; \nonumber \\
\mu_q^{(1)}&=&\mu_q^{(2)},~~~ q=n,p;\nonumber \\
P^{(1)}&=&P^{(2)}.
\end{eqnarray}

\begin{figure}
\begin{center}
\includegraphics[angle=0, width=0.48\columnwidth]{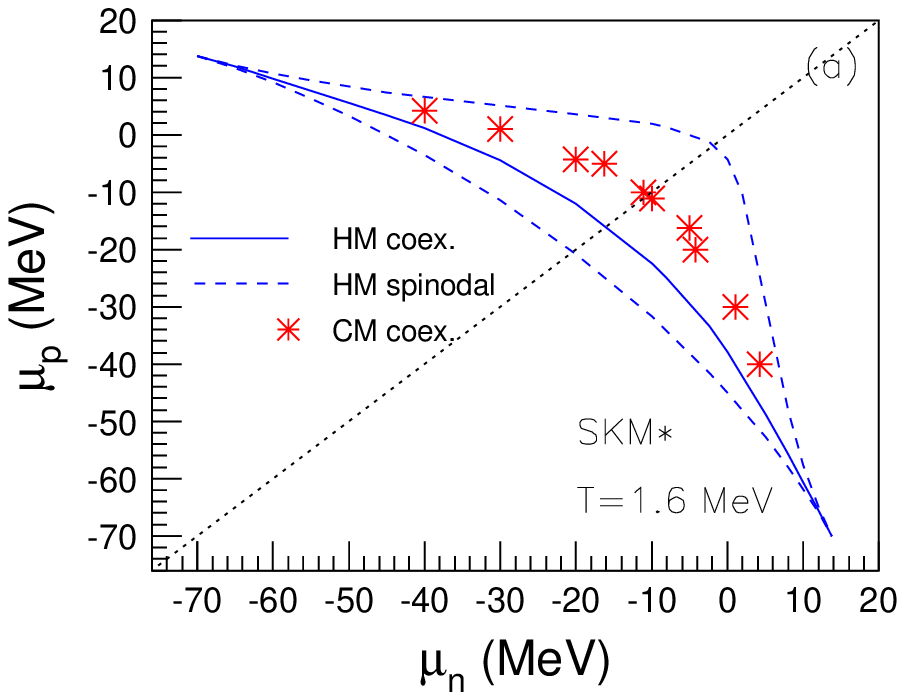}
\includegraphics[angle=0, width=0.48\columnwidth]{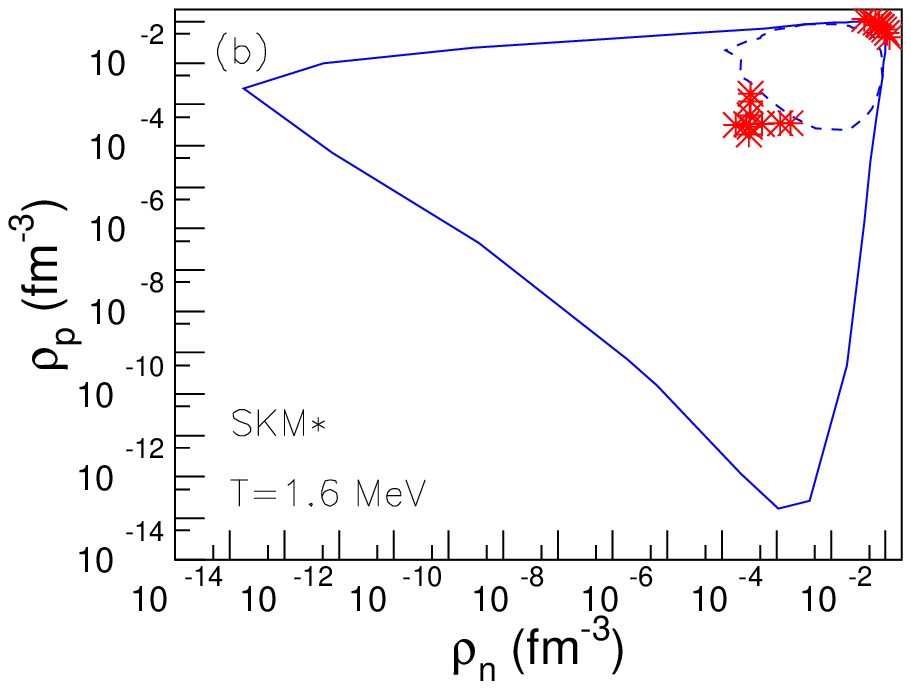}
\end{center}
\caption{(Color online) 
$\mu_p-\mu_n$ (left) and $\rho_p-\rho_n$ (right) representations of the
phase diagrams of the charge-neutral homogeneous infinite nuclear matter 
described by the SKM* interaction (lines) and 
net-charge neutralized clusterized nuclear matter (stars)
corresponding to $T$=1.6 MeV. 
In the case of the uniform system, the dashed lines mark the borders of the
spinodal zone, while the solid line marks the coexistence region.
These results are independent of the chosen confining volume $V=2.9 \cdot 10^4$ fm$^3$
except in the low density region of the phase
diagram of the clusterized matter system in $\rho_p-\rho_n$ coordinates, 
which is affected by finite size effects (see text).
}
\label{fig:phd_t=1.6}
\end{figure}

\begin{figure}
\begin{center}
\includegraphics[angle=0, width=0.48\columnwidth]{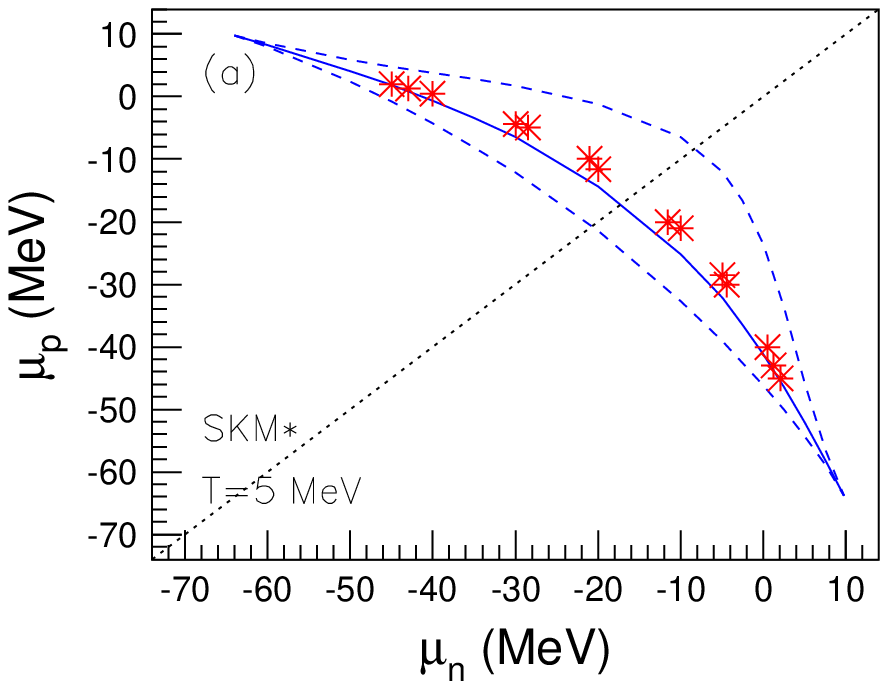}
\includegraphics[angle=0, width=0.48\columnwidth]{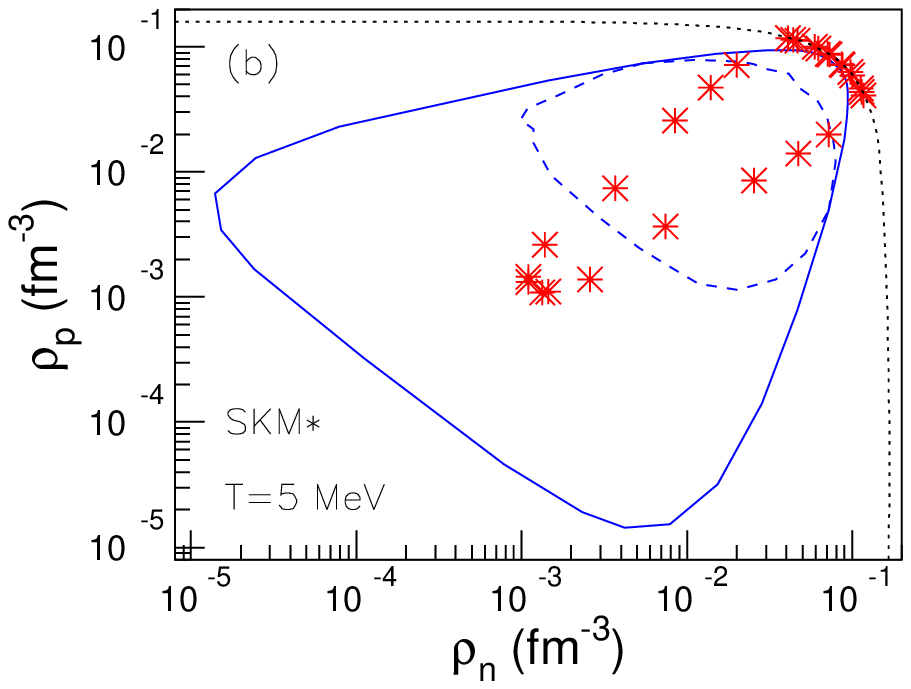}
\end{center}
\caption{(Color online)
The same as in Fig. \ref{fig:phd_t=1.6} for $T$=5 MeV.
For symbol and line codes, see  Fig. \ref{fig:phd_t=1.6}.
}
\label{fig:phd_t=5}
\end{figure}

\begin{figure}
\begin{center}
\includegraphics[angle=0, width=0.48\columnwidth]{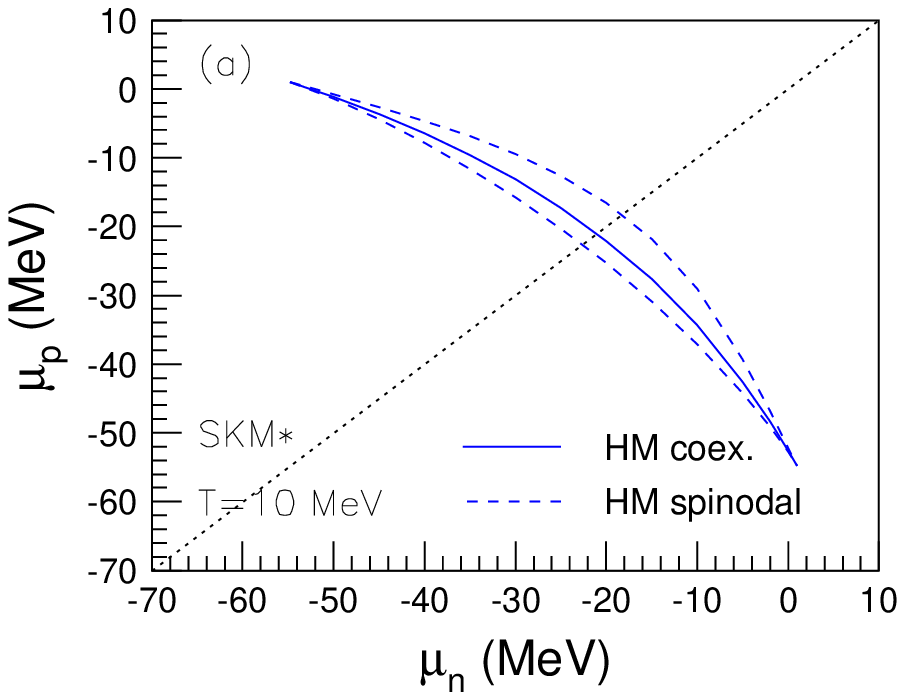}
\includegraphics[angle=0, width=0.48\columnwidth]{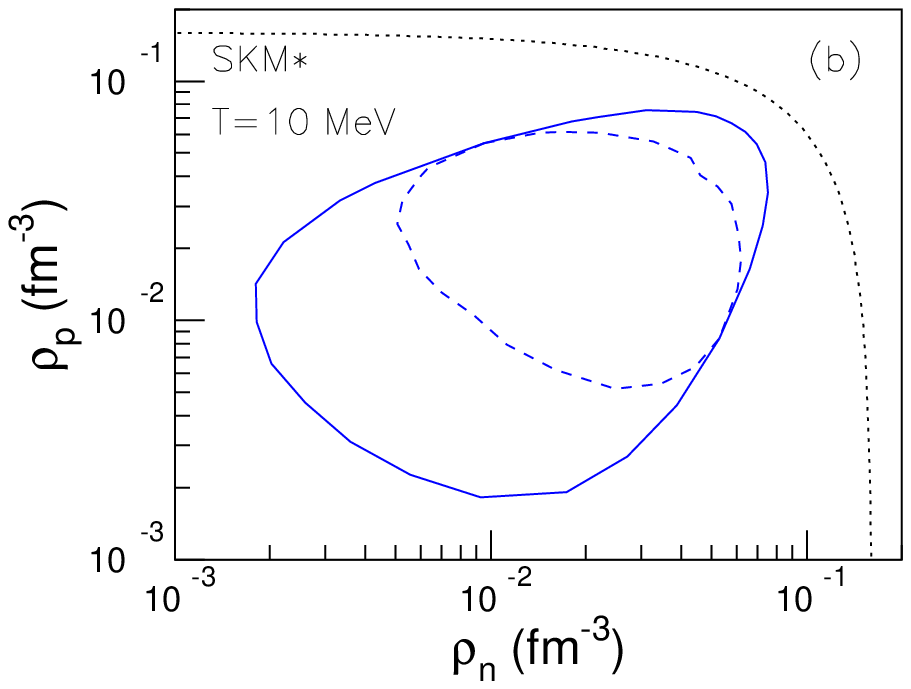}
\end{center}
\caption{(Color online)
The same as in Fig. \ref{fig:phd_t=1.6} for $T$=10 MeV.
}
\label{fig:phd_t=10}
\end{figure}

Phase diagrams of SKM* nuclear matter are illustrated in 
Figs. \ref{fig:phd_t=1.6},  \ref{fig:phd_t=5} and  \ref{fig:phd_t=10} 
for three values of temperature $T=$1.6, 5 and, respectively, 10 MeV 
in $\mu_n-\mu_p$ (left panels) and $\rho_n-\rho_p$ (right panels) representations.
As it is customary in nuclear matter calculations, the system is 
charge neutral in the sense that the electromagnetic 
interaction has been artificially switched off. 
In the star matter case the proton electric charge has to be taken into account, and 
compensated by the electron charge to give a net zero charge.
The resulting Coulomb interaction energy will be discussed in the next chapter. 
Full lines correspond to the phase coexistence region and dashed lines mark
the borders of the instability region, the so-called spinodal surface. 

As one may notice, a large portion of the density and chemical potential plane
is characterized by the instability even at temperatures as high as 10 MeV. 
Inside the phase coexistence in the density representation, 
the mean field solution is unphysical
at thermal equilibrium if matter is uniquely composed of homogeneously distributed 
(uncharged) protons and neutrons. In the case of stellar matter, the entropy 
convexity properties have to be examined only after considering all the different
constituents, which can very drastically change the properties of the phase 
diagram. As we will show in the next sections, the introduction of electrons 
has the effect of making the Gibbs construction unphysical, while  
the introduction of clusterized matter has the effect of stabilizing 
the unstable mean field solution.
 
\subsubsection{2. The clusterized nuclear matter component}

As we have discussed in the previous section, in a wide region of temperature and density,
uncharged uniform nuclear matter is unstable with respect to density fluctuations.
If such fluctuations occur on a macroscopic scale, a first-order liquid-gas phase transition
follows.
Inside the spinodal surface however, finite wavelength fluctuations may also occur,
leading to spontaneous cluster formation. Spinodal decomposition and nucleation are indeed
known to constitute the dynamical mechanisms leading to phase separation in macroscopic
uncharged systems. As we will show in the following, in the presence of the repulsive
long range Coulomb interaction among protons, which is screened by the leptons only on 
macroscopic scales, phase separation is quenched in stellar matter and the instability
is cured by the formation of nuclear clusters, that is clusters in the femtometer scale.  

One possibility to account for the thermal and phase properties of a 
clusterized matter sub-system is to use a statistical model with cluster
degrees of freedom.
Several such models have been so far proposed for the study of
condensation close to the critical point \cite{fisher},
nuclear multifragmentation \cite{smm,mmmc,koonin,mmm} and 
compact stars \cite{mishustin,blinnikov,souza}. They basically consist in the estimation
of the number of microscopic states compatible with the thermodynamic
macroscopic constraints.

Considering that the center of mass of the clusters can be treated as classical 
degrees of freedom, the grancanonical partition function of the clustered system
reads:

\begin{eqnarray}
{\cal Z}^{(cl)}_{\beta,\beta\mu_n,\beta\mu_p}
&=& \sum_C W_C = \sum_C \frac{1}{N_C!} \int
\frac{d^3r_1...d^3r_{N_C}d^3p_1...d^3p_{N_C}}{h^{3N_C}}
\exp \left[ -\beta \left( E_C-\mu_n N_C -\mu_p Z_C\right)  \right]
\nonumber
\\
&=&\sum_C \frac{1}{N_C!} 
\exp\left(-\beta E_{int}(C)\right) \cdot \nonumber
\\
& \cdot& \prod_{i=1}^{N_C}
\left[ V_i^{free}(C)
\left( \frac{mA_iT}{2\pi \hbar^2} \right)^{3/2} \rho(A_i,Z_i,\epsilon_i)
\exp\left( \beta B_i -\beta \epsilon_i +\beta \mu_n N_i +\beta \mu_p Z_i \right)  
\right].
\label{eq:Zgc_exact}
\end{eqnarray}

Here $C$ denotes a generic cluster configuration
$C=\{ (A_1,Z_1,\epsilon_1),...,(A_{N_C},Z_{N_C},\epsilon_{N_C})\}$
defined by the mass number $A_i$, the proton number $Z_i$
and the excitation energy $\epsilon_i$ of each constituent nuclear state indexed by $i$,
whose binding energy is $B_i$ and level density is $\rho(A_i,Z_i,\epsilon_i)$; 
$E_{int}(C)$ represents the inter-cluster configuration dependent interaction energy,
and $V_i^{free}(C)<V$ is the volume accessible to cluster $i$ which is not excluded by the 
presence of the other clusters. Because of the short range of the nuclear force, 
this excluded volume correction can be viewed as a first-order approximation to 
the nuclear part of the interaction energy in the spirit of the Van der Waals gas.
More sophisticated in-medium corrections can in principle be also added as modifications
of the cluster self-energies $B_i$ \cite{typel}. 
Because of this, we will consider that the interaction energy only contains
the Coulomb electrostatic energy of the configuration.

To obtain the so-called NSE approach, one has to neglect all 
inter-particle interactions except the cluster self-energies,
leading to :

\begin{equation}
{\cal Z}^{(cl)}_{\beta,\beta\mu_n,\beta\mu_p}
\approx {\cal Z}^{(cl)}_{0}= \sum_C \frac{V^{N_C}}{N_C!} 
 \prod_{i=1}^{N_C}
\left[ 
\left( \frac{mA_iT}{2\pi \hbar^2} \right)^{3/2} \rho(A_i,Z_i,\epsilon_i)
\exp\left( \beta B_i -\beta \epsilon_i +\beta \mu_n N_i +\beta \mu_p Z_i \right)  
\right].
\label{eq:Zgc_NSE}
\end{equation}

The advantage of this simplification is that 
a completely factorized expression for the partition sum
can be obtained, leading to analytical expressions for all thermodynamic quantities.
Indeed any configuration $(C)$ can be ordered as 

\begin{equation}
(C)=\{
  \underbrace{T_1,T_1,\dots,T_1}_{n_1(C)},\underbrace{T_2,T_2,\dots,T_2}_{n_2(C)}, 
\dots \},
\end{equation}
where $n_i(C)=n_{A_i,Z_i,\epsilon_i}(C)$ gives the multiplicity of 
clusters of type $T_i=\{A_i,Z_i,\epsilon_i\}$.
Equation (\ref{eq:Zgc_NSE}) becomes:

\begin{equation}
{\cal Z}^{(cl)}_{0} =  \prod_{i=1}^{\infty} \sum_{n_{i}=0}^{\infty}\frac{1}{n_{i}!}
\left [z_{\beta,\beta\mu_n,\beta\mu_p}\left ({A_i,Z_i,\epsilon_i}\right) \right ]^{n_{i}},
\label{fisher}
\end{equation}

with

\begin{equation}
z_{\beta,\beta\mu_n,\beta\mu_p}\left ({A_i,Z_i,\epsilon_i}\right)
=\left( \frac{mA_iT}{2\pi \hbar^2} \right)^{3/2} \rho(A_i,Z_i,\epsilon_i)
\exp\left( \beta B_i -\beta \epsilon_i +\beta \mu_n N_i +\beta \mu_p Z_i \right) . 
\end{equation}

Eq. (\ref{fisher}) is not yet an analytically tractable expression, because of the 
infinite number of possible excited states for each nuclear isotope $(A_i,Z_i)$. 
The internal degrees of freedom can be however integrated over giving

\begin{equation}
 {\cal Z}^{(NSE)}_{\beta,\beta\mu_n,\beta\mu_p} =  
\prod_{A,Z=1}^{\infty} \sum_{n_{A,Z}=0}^{\infty}\frac{1}{n_{A,Z}!}
\left [z_{\beta,\beta\mu_n,\beta\mu_p}\left ({A,Z}\right) \right ]^{n_{A,Z}},
\label{eq:Zcluster}
\end{equation}
where $z_{\beta,\beta\mu_n,\beta\mu_p}\left ({A,Z}\right)$ 
is the standard grancanonical partition sum 
for a single nucleus of mass $A$ and charge $Z$:

\begin{eqnarray}
z_{\beta,\beta\mu_n,\beta\mu_p}\left ({A,Z}\right)
&=&\left( \frac{mAT}{2\pi \hbar^2} \right)^{3/2} \int d\epsilon \rho(A_i,Z_i,\epsilon)
\exp\left( \beta B_i -\beta \epsilon +\beta \mu_n N +\beta \mu_p Z \right)  \nonumber \\
&=& \left( \frac{mAT}{2\pi \hbar^2} \right)^{3/2} g_\beta(A,Z)  
\exp\left( \beta B_i +\beta \mu_n N +\beta \mu_p Z \right),  
\end{eqnarray}
and $g_\beta(A,Z)$ is a temperature dependent degeneracy.

While elegant and analytically tractable, Eq. (\ref{eq:Zcluster}) represents
a correct estimation of Eq. (\ref{eq:Zgc_exact}) if and only if the 
interactions are fully exhausted
by clusterization.

For this reason, in this work we will consider only Eq. (\ref{eq:Zgc_exact}),
which we will estimate numerically with a precise, efficient and well-tested 
Metropolis-Monte-Carlo technique \cite{mmm},  and calculate
relevant thermodynamic observables other than the ones fixed by the exterior
as ensemble averages.
For the configuration-defined quantities, $X_C$=$A_{tot}(C)$, $Z_{tot}(C)$, $E_{tot}(C)$:
\begin{eqnarray}
A_{tot}(C)&=&\sum_{i=1}^{N_C} A_i \nonumber \\
Z_{tot}(C)&=&\sum_{i=1}^{N_C} Z_i \nonumber \\
E_{tot}(C)&=&\sum_{i=1}^{N_C} \left(-B_i+\epsilon_i \right),  \nonumber \\
\end{eqnarray}
the corresponding ensemble averaged density reads:
\begin{equation}
<x>=\frac{1}{V}\frac{\sum_C X_C W_C}{\sum_C W_C},
\end{equation}
and stability is systematically checked against an increase of the considered volume $V$.
 
For each sampled configuration $(C)$, we evaluate the Coulomb interaction energy 
in the Wigner-Seitz approximation \cite{Lattimer85},
\begin{equation}
V_{Coulomb}(C)=\sum_{i=1}^{N_C} \frac35 c(\rho) \frac{e^2 Z_i^2}{r_0
  A_i^{1/3}},
\label{eq:WS}
\end{equation}
with
\begin{equation}
c(\rho)=1-\frac32 \left(  \frac{\rho_e}{\rho_{0p}} \right)^{1/3}+
\frac12 \left(\frac{\rho_e}{\rho_{0p}} \right),
\end{equation}
accounting for the screening effect of electrons.
$\rho_{0p}=Z/A \rho_0$ denotes the proton density inside the nuclei and
$\rho_e$ is the electron density. 
Net charge neutrality of the system imposes the electron density $\rho_e$ 
to be equal to the proton density $\rho_p$.

The total pressure results:
\begin{eqnarray}
\beta p&=&\left( \frac{\partial G^{(cl)}}{\partial V}
\right)|_{\beta,\beta\mu_n,\beta\mu_p}
\nonumber
\\
&=&\frac{N_C}{V}-\frac35 \frac{\beta}{2V} \left(\frac{1}{n^{1/3}}-\frac{1}{n}
\right)
\sum_i \frac{e^2 Z_i^2}{r_0 A_i^{1/3}}
\nonumber
\\
&=&\beta \left(p^{(cl)}+p^{(lattice)} \right).
\end{eqnarray}
Here $p^{(cl)}$ originates from clusters translational motion and
$p^{(lattice)}$ is the so-called lattice Coulomb pressure.

Concerning the excluded volume, the result of the $N_C$ integrals over the coordinate space
corrected for the volume excluded by each cluster, $V_i$, can be re-casted as a global 
configuration-dependent factor $\chi(C) V^{N_C}$ as, 
\begin{equation}
V(V-V_1)(V-V_1-V_2)...(V-V_1-V_2-...-V_{N_C-1})=\chi V^{N_C}.
\end{equation}

For the clusters we assume the following properties:
\begin{itemize}

\item clusters binding energies are described by a liquid-drop
  parameterization, 
 \begin{equation}
   B(A,Z)=
 \left (a_v A -a_s \left( 1-T f(T)\right) A^{2/3}\right)\left ( 1- a_I(A-2Z)^2/A^2\right ) 
\label{eq:wi_star}
\end{equation}
with  $a_v$=15.4941 MeV, $a_s$=17.9439 MeV, $a_I=1.7826$
  \cite{ld_mass}
  where the surface term is additionally made to vanish at the 
  critical temperature $T_C$=12 MeV \cite{clustermatter}; 
  the Coulomb energy contribution is accounted for 
  in $V_{Coulomb}(C)$ as discussed above;
  structure effects thoroughly considered by other authors
  \cite{hempel2010} are disregarded because of the
  relatively high temperatures we are focusing on, but they can be straight-forwardly 
  included;

\item in order to avoid multiple counting of the free nucleons, 
  already considered in the uniform background, the clusters
  should have $Z \geq 2$; this arbitrary limit may be replaced by $A>1$, such
  as to allow also for $^2$H and $^3$H, but this choice is not expected to be important
  for the presently considered quantities;

\item no limit is imposed for the largest cluster which, in principle, may
  reach the total system size; 
  given the fact that the fragment mass partition
  enters in Eq. (\ref{eq:Zgc_exact}) not only in the translational energy
  factor and binding energy, but also in the calculations of excluded volume 
  and Coulomb energy, we expect it to play a dramatic role in the limit of low
  temperatures and densities close to $\rho_0$;

\item to allow creation of exotic species, there is no restriction for the
  neutron/proton composition; a cluster is allowed to exist as soon as 
  $B(A,Z)>0$;

\item cluster internal excitation energy $\epsilon$ is upper limited by the
  cluster binding energy and the corresponding level density is of Fermi-gas
  type with cut-off correction \cite{iljinov},
\begin{equation}
\rho(\epsilon)=\frac{\sqrt{\pi}}{12 a^{1/4}\epsilon^{5/4}}
~ \exp(2 \sqrt{a \epsilon}) ~ \exp(-\epsilon/\tau),
\label{eq:nuclrho}
\end{equation}
with  $a=0.114 A+0.098 A^{2/3}$ MeV$^{-1}$ 
and $\tau$=9 MeV.
\end{itemize}

The phase diagram of clusterized stellar matter 
can be inferred from the bimodal behavior of the total particle number 
in the grancanonical ensemble \cite{clustermatter}. This model presents 
a first-order liquid-gas phase transition similar to the case of uniform nuclear matter,
but contrary to the latter it is never unstable.
The two phase diagrams are compared for $T$=1.6 and 5 MeV in
Figs.  \ref{fig:phd_t=1.6} and, respectively, \ref{fig:phd_t=5}. 
Fig.  \ref{fig:phd_t=10} presents exclusively the phase diagram of 
neutral uniform matter at $T$=10 MeV, as this temperature is supra-critical
for the clusterized matter. 

We can see that the thermodynamics of a clusterized system is very
different from the one of neutral nuclear matter in the mean field approximation,
even though the employed effective interaction parameters are typically tuned on 
the same experimentally accessible observables of cold nuclei close to saturation
density, which are implemented in the cluster energy functional. 
As a consequence, the two phase diagrams are only compatible at low temperatures
and at densities close to saturation. 

The extension of the coexistence region in density and temperature 
is noticeably reduced for clusterized matter. 
This effect is partially not physical. 
Indeed the numerical procedure adopted for estimating
Eq. (\ref{eq:Zgc_exact}) implies that clusterized matter results 
depend on the system size for the lowest densities:
the lowest accessible density is determined by the chosen finite volume 
and the minimum allowed cluster size ($A_{min}$)
eventually multiplied by the minimum
allowed number of clusters ($N_{min}$). 
Clusterized matter phase diagrams considered in this section have been obtained
working within a cell of volume $V=2.9 \cdot 10^4$ fm$^3$.
If $N_{min}=1$ and $A_{min}=3$, the lowest accessible density
$\rho_{min} = N_{min} A_{min}/V$ is of the order $1.1 \cdot 10^{-4}$ fm$^{-3}$, 
which implies that no point below this limit is present.
Actually this value is even higher as for numerical efficiency purposes the
minimum allowed cluster number is larger than 1.  

By consequence, the low density limits
of the coexistence zone have to be considered as a numerical artifact.
To have trustable predictions at lower densities, the simulation volume will 
have to be increased in future calculations.
Conversely the shrinking of the coexistence zone with increasing temperature and asymmetry 
is a physical effect.

For a given baryonic density and temperature, we can also see  
that coexisting phases of clusterized matter are much more isospin symmetric
with respect to the ones of the uniform matter, even if the energetics of the two systems
corresponds to compatible values for the symmetry energy. This can be physically 
understood from the fact that higher densities are locally explored 
if a system is clusterized,
and the symmetry energy is an increasing function of density.

Concerning the reduction of the limiting temperature in the clusterized
system, this is understood as a cumulated effect of Coulomb 
  and cluster center of mass translational degrees of freedom,
which are not accounted for in the homogeneous nuclear
matter calculation \cite{clustermatter}. 

These general observations already show the importance of properly accounting
for the cluster properties of matter, even if one is only interested in global 
integrated thermodynamic quantities as equations of state.

From the $\mu_p-\mu_n$ representation it comes out that the coexistence line
of the clusterized matter is embedded in the spinodal region of the uniform
matter, and shifted towards higher values of $\mu$ respect to the coexistence
of nuclear matter, that is it corresponds to the low density part of the 
spinodal region of homogeneous matter. 
This fact, together of the absence of instability
of the clusterized system, is already an indication that the homogeneous matter
instability is effectively cured by accounting for the possibility of cluster
formation, as we will develop in the next section.
In the end, we mention that the phase diagram of clusterized matter might
  slightly change while modifying cluster properties.

\subsubsection{3. The mixture properties}\label{sec_mixture}

In the previous section we have examined the thermodynamics of purely 
uniform and purely clusterized nuclear systems.
In the physical situation of supernovae and proto-neutron star matter
it is clear that these two components have to be simultaneously present,
and the question arises of how describing their mutual equilibrium conditions.
In many past works 
\cite{Lattimer85,pethick,douchin,hempel2010,ohnishi}, 
the evolution of clusterized matter
towards homogeneous matter at high density was described 
as a first-order phase transition, governed by Gibbs or Maxwell equilibrium
rules. The simultaneous presence in the Wigner-Seitz cell of localized clusters
and homogeneously distributed free protons and neutrons in the nucleonic drip 
region is then also depicted as a manifestation of the same first order transition,
though corrected by surface and Coulomb effects arising from the finite size of the cell.

However, as we discuss now, the proper way of treating density inhomogeneities occurring
at a microscopic (femtometer scale) level is the concept of a gas mixture, if the coexisting
components (here: nucleons and fragments) are non-interacting. If interaction effects 
arising from the excluded volume are included, we show that such a mixture naturally
produces a continuous (second order) transition to uniform matter at densities close 
to saturation. 

Restricting for the moment to the baryonic sector, the thermodynamics of a system with two
conserved charges $\rho_n$ and $\rho_p$ is a two-dimensional problem. The other charges 
do not affect the thermodynamics at the baryonic level because they are not 
coupled to the baryons (see Eq. (\ref{independent})). 
The net electron density is coupled to the proton density through
the electromagnetic interaction, but this does not imply an additional degree of freedom
as charge neutrality imposes $\rho_p=\rho_e$.
This two-dimensional problem can be easily taggled if we switch from the grancanonical 
ensemble, where all the intensive parameters are constrained and 
the densities can freely vary, 
to a statistical ensemble where one single extensive observable (i.e. a single density) 
is free to fluctuate \cite{ducoin_paper1}.
In our two-dimensional problem, two equivalent possibilities exist, namely
the proton-canonical, neutron-grancanonical ensemble where the thermodynamic
potential is Legendre transformed to its conjugated extensive variable $\rho_p$,
\begin{eqnarray}
s_{pc}\left (\beta,\beta\mu_n,\rho_p\right )&=&
\frac{1}{V} \bar S \left [ \beta,-\beta\mu_n \right ] \nonumber \\
&=&\frac 1V \ln {\cal Z}_{gc}(\beta,\beta\mu_n,\beta\mu_p) - \beta \mu_p\rho_p ,
\end{eqnarray}
and the neutron-canonical, proton-grancanonical ensemble described by the thermodynamic
potential,
\begin{eqnarray}
s_{nc}\left (\beta,\rho_n,\beta\mu_p\right )&=&
\frac{1}{V} \bar S \left [ \beta,-\beta\mu_p \right ] \nonumber \\
&=&\frac 1V \ln {\cal Z}_{gc}(\beta,\beta\mu_n,\beta\mu_p) - \beta \mu_n\rho_n.
\end{eqnarray}


\begin{figure}
\begin{center}
\includegraphics[angle=0, width=0.85\columnwidth]{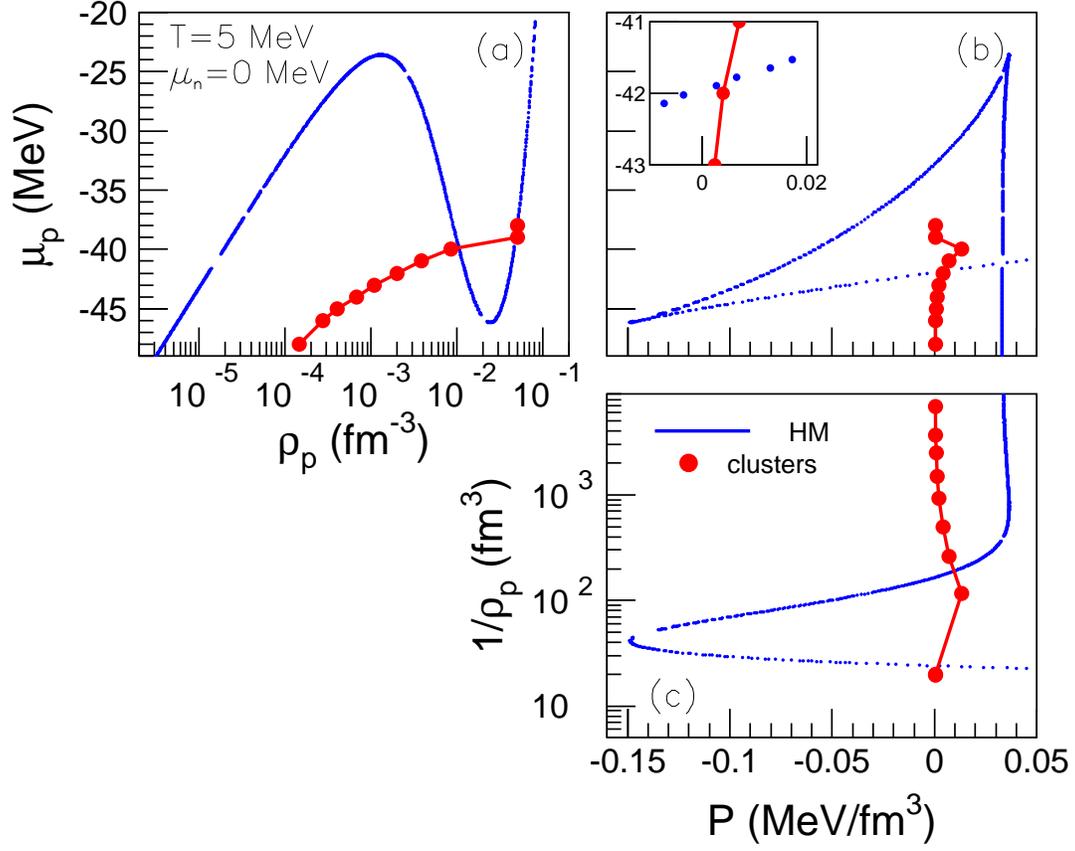}
\end{center}
\caption{(Color online)
Properties of the uniform (solid lines) and net-charge neutralized clusterized 
(solid circles connected by lines)
matter at $T$=5 MeV and $\mu_n$=0 MeV.
Homogeneous matter is described by SKM*.
}\label{fig:phases_t=5}
\end{figure}

\begin{figure}
\begin{center}
\includegraphics[angle=0, width=0.85\columnwidth]{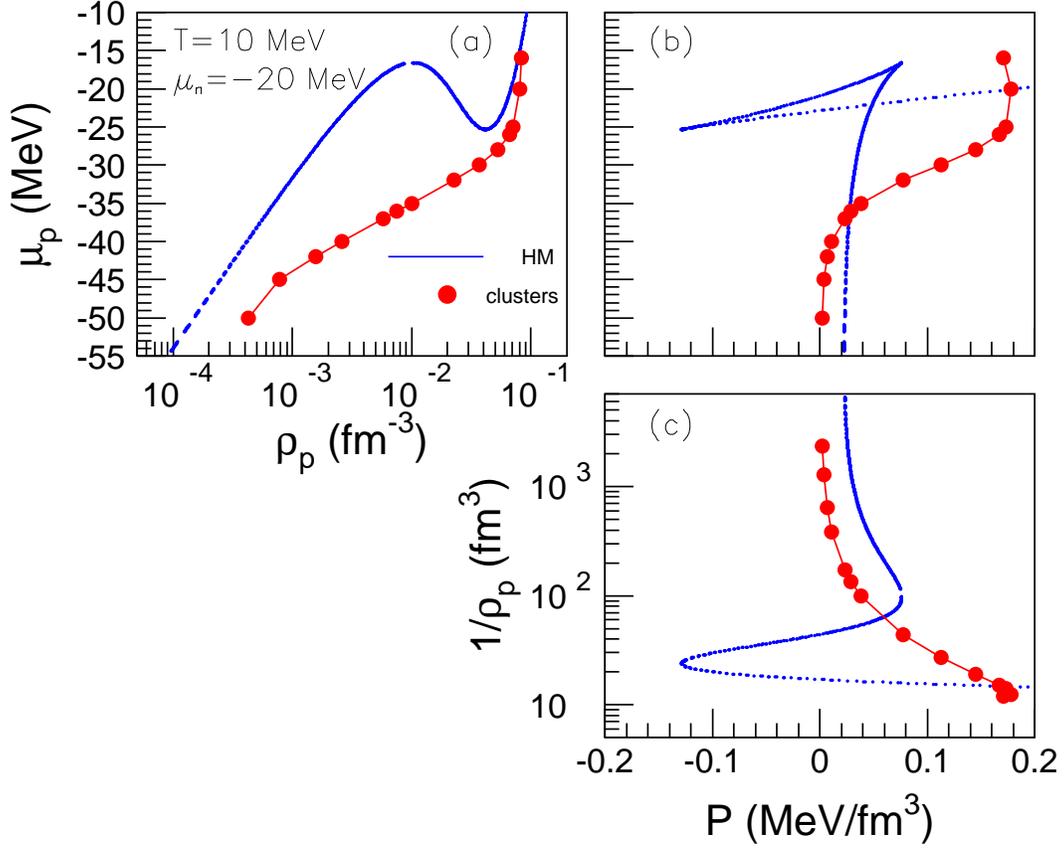}
\end{center}
\caption{(Color online)
The same as in Fig. \ref{fig:phases_t=5} for $T$=10 MeV and
$\mu_n=-20$ MeV.
}\label{fig:phases_t=10}
\end{figure}

We now come to illustrate the way of mixing clusters and homogeneous matter
and the differences between such mixture and a possible phase coexistence
which would be associated to a first-order phase transition.
Fig. \ref{fig:phases_t=5} shows the equations of state 
$\mu_p(\rho_p)$ and $P(1/\rho_p)$ for the proton-canonical ensemble, as 
well as the functional relation between the two intensive parameters $\mu_p$ and $P$  
for the uniform (solid lines)
and clusterized (solid circles connected by lines) matter at 
$T$=5 MeV and a representative neutron chemical potential $\mu_n$=0 MeV. 
The clusterized system is confined within a volume
$V=2.9 \cdot 10^4$ fm$^3$.

The back-bendings of $\mu_p(\rho_p)|_{(HM)}$ and 
$P(1/\rho_p)|_{(HM)}$ and the tilting shape of $\mu_p(P)|_{(HM)}$ 
indicate that in the considered $\mu_p$ range uniform matter, if it 
would exist alone, would be unstable against phase separation.

If we consider that, at the microscopic scale of the Wigner-Seitz cell,
clusters and free nucleons may be simultaneously present, as expected
from the phenomenon of neutron drip, then the statistical equilibrium 
conditions read:

\begin{equation}
\mu_p^{(HM)}=\mu_p^{(cl)} \; ; \; \mu_n^{(HM)}=\mu_n^{(cl)} \; ; \;
T^{(HM)}=T^{(cl)} \; ; \; p^{(bar)}=p^{(HM)}w_{HM}+p^{(cl)}w_{cl}, 
\label{mixture}
\end{equation}
where $w_{HM}$ and $w_{cl}$ measure the volume fractions associated to the
homogeneous matter and clusters,
\begin{eqnarray}
w_{HM}=\left(V-v_{cl}\right)/V,\nonumber\\
w_{cl}=\left(V-v_0 \left( \rho_n^{(HM)}+\rho_p^{(HM)}\right) 
\left(V-v_{cl} \right)\right)/V, 
\end{eqnarray}
with $v_0$ standing for the intrinsic volume of a nucleon calculated such as
$v_0=1/\rho_0$ and $v_{cl}=v_0\sum_{i=1}^{N_C} A_i$ 
representing the summed-up clusters intrinsic volumes.

Let us denote by  
$\mu_{p(H)}^{(HM)}$ and
$\mu_{p(L)}^{(HM)}$
the proton chemical potentials corresponding to
the limits of the spinodal region of homogeneous matter.
For $\mu_p < \mu_{p(L)}^{(HM)} \approx -46 $ MeV, 
there will be one single low density solution of uniform matter
which can constitute a mixture at thermal equilibrium with clusterized matter
according to Eqs. (\ref{mixture}) and which, 
for sufficiently small values of $\mu_p$, may differ by orders of magnitude
from the density of clusters.
For $ \mu_{p(L)}^{(HM)}  < \mu_p < \mu_{p(H)}^{(HM)}$, 
three different density states of uniform matter can exist in equilibrium
with the single solution of clusterized matter, leading to three different 
mixtures corresponding to the same chemical potentials but different 
total density, chemical composition and pressure.
It is important to stress that these three uniform matter solutions
may all lead to thermodynamically stable mixtures, as the stability 
properties of the entropy has to be examined after adding all the different
matter components. This is at variance with the case of normal nuclear matter,
where only one solution out of the three corresponds to a thermodynamically
stable state, while the other two are respectively metastable and unstable.
For $\mu_p > \mu_{p(H)}^{(HM)} \approx -24 MeV $ the thermodynamic condition 
is in principle the same as for high and negative chemical potentials. 
Since however the sum of the two densities
corresponding to homogeneous and clusterized matter
exceeds $\rho_0$, the situation goes beyond the
applicability domain of the present model and is, thus, devoid of interest.

Given that one of the purposes of our paper is to address the
unhomogeneous (crust) - homogeneous (core)   transition,
it is possible now to anticipate that for these two representative temperatures 
$T$=1.6 and 5 MeV this will take place when, following a constant proton fraction trajectory, 
the uniform matter state will not correspond anymore to the low density,
but rather to the high density solution.

At higher temperatures the situation is slightly different as shown by 
Fig. \ref{fig:phases_t=10}, which shows the same quantities plotted 
in Fig. \ref{fig:phases_t=5}, for a representative neutron chemical potential
at $T=10$ MeV.
In this case, in the relatively narrow chemical potential region where three 
different homogeneous matter macrostates are found, the corresponding cluster states
are always at high density close to saturation. Then, the only mixture solution not 
overcoming $\rho_0$ is the one given by the low density
solution for uniform matter. 
At this temperature then, the transition to
homogeneous matter does not occur as an increasing proportion of homogeneous matter
that becomes abruptly dominant in the mixture as we have just seen at lower temperatures, 
but rather as an increasing size of 
the largest cluster which abruptly starts to fill the whole available volume.
The two transition mechanisms are physically very close to each other:
an infinitely large cluster has the same energetic and entropic properties of 
homogeneous matter at saturation density, provided the parameters of the 
effective interaction are compatible with the cluster energy parameters.

Since the thermodynamic potential of the mixture can be expressed as
the sum of the thermodynamic potentials corresponding to each sub-system:
\begin{equation}
G^{(bar)}(\beta,\beta \mu_n,\beta \mu_p,V)=
G^{(cl)}(\beta,\beta \mu_n,\beta \mu_p,V)+G^{(HM)}(\beta,\beta \mu_n,\beta \mu_p,V),
\end{equation}
the total neutron and proton density of the mixture is obtained by summing up
the corresponding particle numbers, 
\begin{eqnarray}
N_{n,p}&=&\left(
\frac{\partial G^{(cl)}}{\partial\left( \beta \mu_{n,p} \right)} + 
\frac{\partial G^{(HM)}}{\partial\left( \beta \mu_{n,p} \right) } 
\right)|_{\beta,\beta\mu_{p,n},V}
\nonumber
\\
&=&N_{n,p}^{(cl)}+N_{n,p}^{(HM)},
\label{eq:TotalN}
\end{eqnarray}
with the extra condition that nucleons cannot be found in the space occupied 
by the clusters and vice-versa. 

The densities then result:
\begin{eqnarray}
\rho_{n,p}&=&\frac{N_{n,p}}{V}
\nonumber
\\
&=&\frac{N_{n,p}^{(cl)}}{V-v_0 \left( \rho_n^{(HM)}+\rho_p^{(HM)}\right)
\left(V-v_{cl} \right)} 
\cdot
\frac{V-v_0 \left( \rho_n^{(HM)}+\rho_p^{(HM)}\right) \left(V-v_{cl} \right)}{V}
\nonumber
\\
&+&\frac{N_{n,p}^{(HM)}}{V-v_{cl}} \cdot \frac{V-v_{cl}}{V}
\nonumber
\\
&=&\frac{N_{n,p}^{(cl)}}{V-v_0 \left( \rho_n^{(HM)}+\rho_p^{(HM)}\right)
\left(V-v_{cl} \right)} \cdot w_{cl}
+ \frac{N_{n,p}^{(HM)}}{V-v_{cl}} \cdot w_{HM}. 
\label{excl_volume}
\end{eqnarray}

The excluded volume correction plays a sizeable role only in the vicinity of
$\rho_0$, and is in particular responsible for the un-homogeneous -
homogeneous matter transition at high temperature.
More precisely, pure homogeneous nuclear matter phase occurs if 
$\left(\rho_n^{(HM)}+ \rho_p^{(HM)}\right) > \rho_0$ and, respectively,
$\left(V-v_{cl}\right) \searrow 0$.

\begin{figure}
\begin{center}
\includegraphics[angle=0, width=0.5\columnwidth]{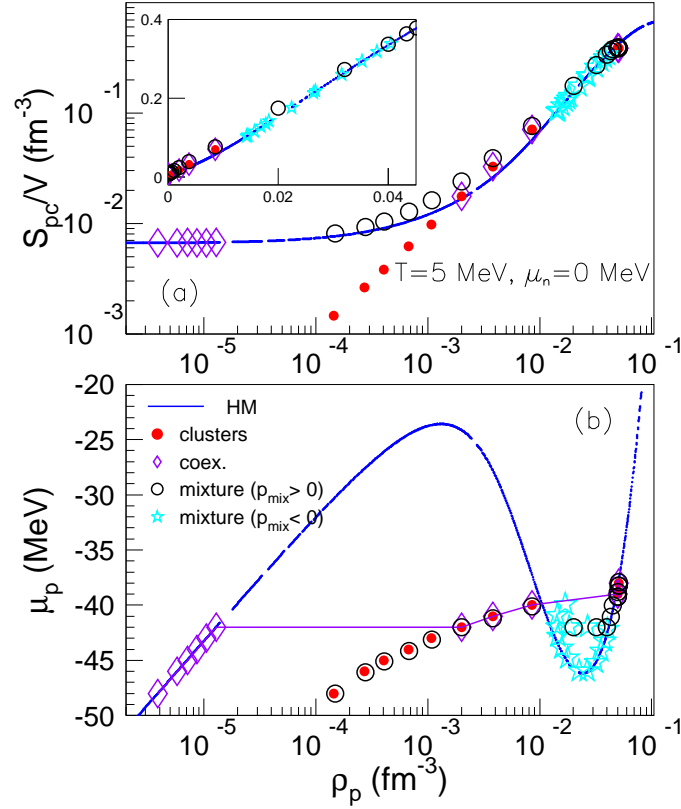}
\end{center}
\caption{(Color online)
$\mu_p$ versus $\rho_p$ (bottom) 
and $ \bar S[\beta,-\beta \mu_n]/V$ versus $\rho_p$ (top)
for the homogeneous-clusterized matter coexistence (open diamonds) and mixture
(open circles if $p_{mix}>0$ 
and stars if $p_{mix}<0$) at $T$=5 MeV along the constant $\mu_n=0$ MeV path. 
The homogeneous matter and 
net-charge neutralized cluster matter curves are plotted with
solid line and, respectively, small solid symbols. 
}
\label{fig:sbar_t=5}
\end{figure}

\begin{figure}
\begin{center}
\includegraphics[angle=0, width=0.5\columnwidth]{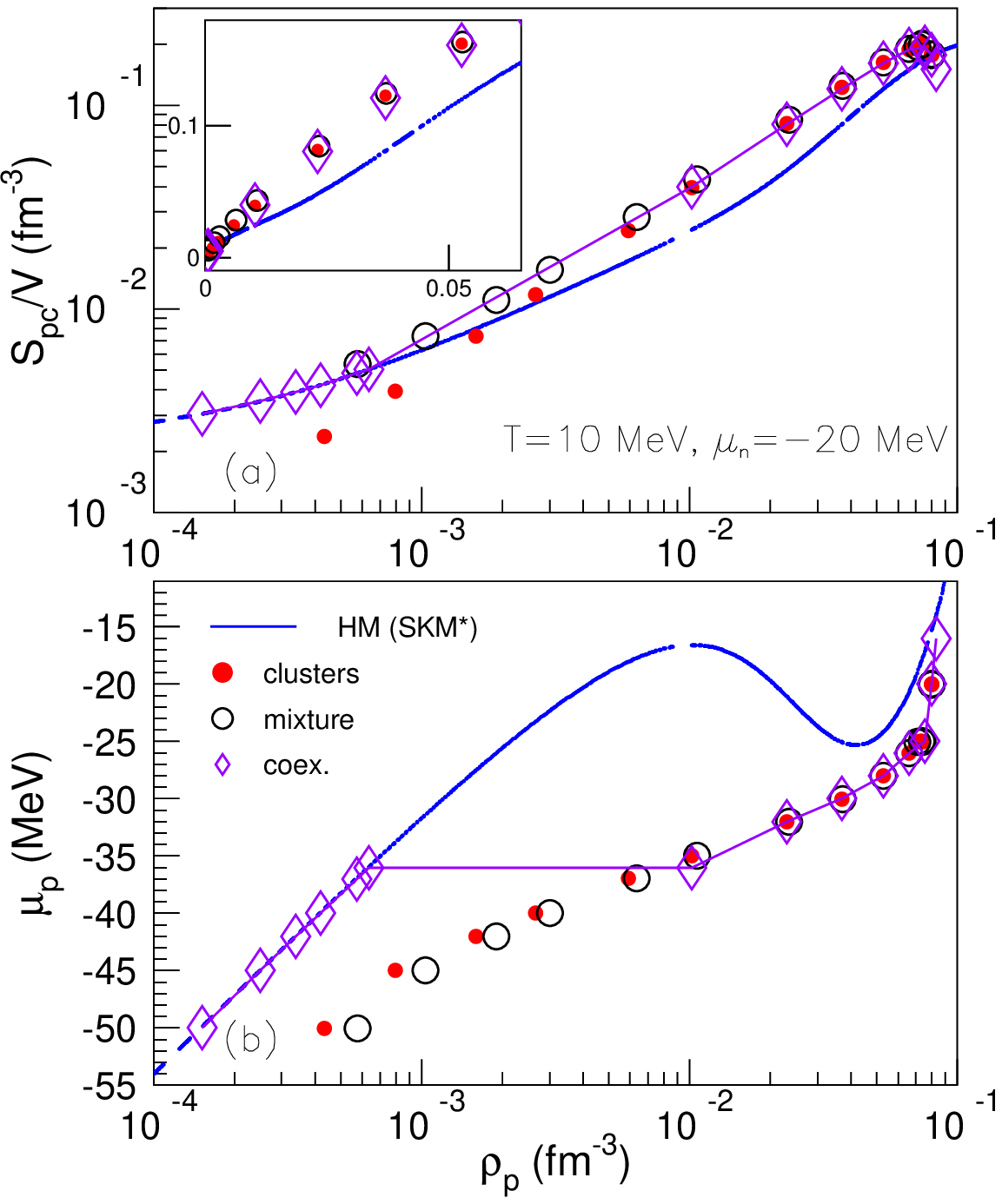}
\end{center}
\caption{(Color online)
The same as in Fig. \ref{fig:sbar_t=5} for $T$=10 MeV and $\mu_n=-20$ MeV.
}
\label{fig:sbar_t=10}
\end{figure}

The equation of state and constrained entropy of the mixture 
for ($T$=5 MeV, $\mu_n=0$ MeV) and ($T$=10 MeV, $\mu_n$=-20 MeV)
are indicated by circles in Figs.
\ref{fig:sbar_t=5} and \ref{fig:sbar_t=10}. 
We can see that the simultaneous presence of the 
two components constitutes always an entropy gain respect to both purely homogeneous
and purely clusterized matter, and corresponds therefore always to the equilibrium 
solution. 

At the lowest temperature the $\mu_p$ versus $\rho_p$ equation of state
is bivalued in a given density domain. This can be understood as follows. 
At temperatures low enough that the cluster component shows a liquid-gas transition,
which can be recognized from the plateau in the  $\mu_p(\rho_p)$ relation at 
constant $\mu_n$, two very different cluster-matter mixtures can be obtained
from Eqs. (\ref{mixture}) and (\ref{excl_volume}) 
at the same total density but different chemical potentials. These two possible
mixtures correspond to a homogeneous matter solution lying inside the spinodal
combined respectively to a low density-low chemical potential cluster solution  
(stars in Fig. \ref{fig:sbar_t=5}), or to a higher density-higher chemical
potential solution.
Since these two possibilities correspond to the same total density, the equilibrium
solution will be the one maximizing the associated entropy. It comes out that the
lowest chemical potential solution always correspond to a negative total pressure for the 
mixture, while the pressure associated to the highest chemical potential solution
is always positive, meaning that the entropy is higher (see the inset in 
Fig. \ref{fig:sbar_t=5}). 

The states denoted by stars
in Fig. \ref{fig:sbar_t=5} are therefore unstable or metastable and will not be further
considered. The presence of such multiple solutions in this density and temperature 
domain may be an indication that some degrees of freedom are missing in the model
and cluster deformations should be added in this region where pasta structures
are expected \cite{horowitz,watanabe_prl,newton,sebille}.

To summarize, we have modelised the simultaneous presence of free nucleons and light
and heavy nuclei as two independent components which are allowed to exchange
particles and energy following the laws of statistical equilibrium, 
and whose proportions are further determined by the 
geometrical constraint given by the excluded volume.

This means that the densities will have different values in each
sub-system and that the partial pressures will be also different, 
following Eqs. (\ref{mixture}) and (\ref{excl_volume}).

The chemical potential equality between the two components 
insures the global maximization of the total entropy of the mixture.
As a result, a smooth decreasing weight  of the cluster component 
is first observed with increasing density, followed by a successive smooth 
but steeper decrease as density approaches $\rho_0$.  
This last result can be intuitively understood from the simple physical fact that 
when density increases, the space accessible to a clusterized system decreases,
and so does the effective number of degrees of freedom. 
As we will discuss in greater detail in the following, this smooth behavior
of the mixture does in no way imply that the crust-core transition is smooth.
Indeed because of the characteristic backbending 
behavior of the homogeneous matter equation of state, which reflects the
well-known instability respect to spontaneous density fluctuations,
the density content of the homogeneous component can discontinuously change
as a function of the chemical potential.  

\subsubsection{4. Mixture versus phase coexistence}

The crust-core transition of a cold neutron star is 
often \cite{Lattimer85,cold_NS,pethick,douchin} depicted
as a first-order transition from a solid phase of nuclei on a Wigner lattice 
to a liquid phase of homogeneous nuclear matter.
By continuity, a similar phenomenology is thought to occur at the finite temperature
and more isospin symmetric conditions involved in supernova matter
\cite{LS91,hempel2010}.

This hypothesis arises from essentially two considerations that we now turn to examine:
\begin{itemize}
\item
The intermediate configuration between the solid crust and the liquid core, namely 
finite nuclei immersed in an homogeneous neutron background, intuitively resembles
to a coexistence between a clustered and a homogeneous baryonic phase.
\item
Normal nuclear matter is known to present an instability respect to cluster formation 
and phase separation. This spinodal instability is associated to a first-order transition. 
By continuity the same is expected in the baryonic part of stellar matter.
\end{itemize}
 
Let us first concentrate on the first statement.
As we have already mentioned, the simultaneous presence of two different states 
characterized by a different average value for an order parameter 
(here: baryonic density) constitutes a phase coexistence if and only if 
the order parameter fluctuations (here: density inhomogeneities) 
occur on a length scale which is macrosocopic respect to the range of the 
interaction (here: much larger than the femtometer scale of typical nuclei
size). This condition
on macroscopic scales is necessary because
the Gibbs construction neglects the interface energy and entropy, which is essential to
describe thermodynamics of finite systems \cite{gross}. 
Since this is clearly not the case in stellar matter, surface
and Coulomb terms are sometimes added \cite{LS91} to approximately 
account for finite-size corrections. 
However we have shown in the last section that the modelization of a microscopic
mixture between clusters and homogeneous matter is also a way to describe the intermediate
configurations between core and crust. Therefore we may ask if the two modelizations
of phase mixture and phase coexistence are equivalent or at least lead to equivalent results 
for the observables of interest.
 
To clarify this point, let us consider homogeneous and clusterized matter as 
two different possible phases
for stellar matter which may coexist at a first-order transition point.
The advantage of working with an hybrid ensemble as the proton-canonical
ensemble defined above, is that the 
two-dimensional tangent construction implied by Gibbs conditions in a 
multi-component first-order phase transition \cite{glendenning,pagliara}
is reduced to a simple one-dimensional Maxwell construction on the unique 
free extensive variable \cite{ducoin_paper1}.
It is important to stress that this procedure is just a technical
simplification and the obtained equilibrium is exactly identical to the usual 
two-dimensional Gibbs construction in the grancanonical ensemble. 
This construction in a Legendre-transformed potential should not be confused
with a Maxwell construction on the $p(\rho)$ equation of state performed on 
trajectories characterized by constant proton fractions
$Y_p=\rho_p/\rho$, which is currently used in the literature \cite{LS91} and 
has no physical justification.
Indeed in a first-order phase transition in a two-component system the
chemical composition of the two coexisting
phases is never the same and the total pressure is a monotonically increasing
function of the total density \cite{glendenning}. 
This means that a Maxwell construction on $p(\rho)$  at constant $Y_p$ does 
not give the correct Gibbs equilibrium.

The inset in the right upper panel of Fig. \ref{fig:phases_t=5} shows that
for the representative $(T,\mu_n)$ choice of the figure the two components
are found at the same value of proton chemical potential and pressure
in a single point located at 
$(P^t,\mu_p^t)\approx(0.005$ MeV/fm$^{3}$ $,-41.8$ MeV$)$. 
Only at this point 
the Gibbs equilibrium rule for coexisting phases is fulfilled,
\begin{equation}
\mu_p^{(HM)}=\mu_p^{(cl)} \; ; \; \mu_n^{(HM)}=\mu_n^{(cl)} \; ; \;
T^{(HM)}=T^{(cl)} \; ; \; p^{(HM)}=p^{(cl)} . 
\label{gibbs}
\end{equation}

As it can be seen in Fig. \ref{fig:phases_t=5}, at this point the 
density of the two components differs of more than two orders of magnitude.
If clusters and homogeneous matter could be viewed as two different phases
for stellar matter, the global system would be composed of homogeneous matter
only for $\mu_p<\mu_p^t$, and of clusters only for $\mu_p>\mu_p^t$, with a 
discontinuous (first-order) transition at the transition point 
$(P^t(T,\mu_n),\mu_p^t(T,\mu_n))$.
The effect of this construction on the equation of state is shown on 
Fig. \ref{fig:sbar_t=5} by the open diamonds.
Within the picture of a first-order phase transition, stellar matter at the chosen 
representative $(T,\mu_n)$ point would be homogeneous at low density up to 
$\rho_p=\rho_p^m$, with $\rho_p^m\approx 1.2\cdot 10^{-5}$ fm$^{-3}$; 
for intermediate densities 
$\rho_p^m \leq \rho_p \leq \rho_p^M$, 
with $\rho_p^M \approx 2\cdot 10^{-3}$ fm$^{-3}$,
it would be a combination of matter and clusters varying in linear proportions; 
and at high density $\rho_p\geq \rho_p^M$ it would be solely composed of clusters, 
which would extend over the whole volume at the highest densities.
As we can see, this equation of state is very different from the physically 
meaningful one corresponding
to a continuous mixture of matter and clusters over the whole density domain.
However, the two models lead to an almost equivalent entropy, as it is shown by
the upper part of Fig. \ref{fig:sbar_t=5}. 

Similar considerations apply at other temperatures and chemical potentials.
As a second example, Fig. \ref{fig:sbar_t=10} compares the properties of the mixture
with the ones of an hypothetical first-order transition for $T=10$ MeV and $\mu_n=-20$ MeV,
the same thermodynamic conditions as in Fig. \ref{fig:phases_t=10}.
Only within the mixture picture free protons in weak proportions are present 
in the matter, as it is physically expected. 
An inspection of the behavior of the constrained entropy at this high temperature 
(higher part of Fig. \ref{fig:sbar_t=10}) clearly shows that the continuous mixture 
of nucleons and clusters not only correctly recovers the homogeneous matter properties 
at high density, but is also very effective in compensating the mean-field instability of
homogeneous matter at lower densities, leading to a concave entropy over the whole
density range (see the inset in the upper panel). 
Thus the usual argument that a phase coexistence would be needed to maximize the entropy 
and cure the convexity anomaly of the entropy cannot be applied: the same entropic gain 
is obtained considering a continuous mixture, without applying an artificial construction
which neglects all finite-size effects.

We note by passing that at high temperature there are two $(p,\mu_p)$ points 
satisfying the Gibbs conditions Eq. (\ref{gibbs}) 
for a given value of $\mu_n$ (see Fig. \ref{fig:phases_t=10}).  
However at the second point at high pressure the two
components correspond to almost the same density. Making a mixture or a phase coexistence 
does not therefore produce any sensitive difference in the equation of state 
(lower part of Fig. \ref{fig:sbar_t=10}) in this density region.
  
\begin{figure}
\begin{center}
\includegraphics[angle=0, width=0.5\columnwidth]{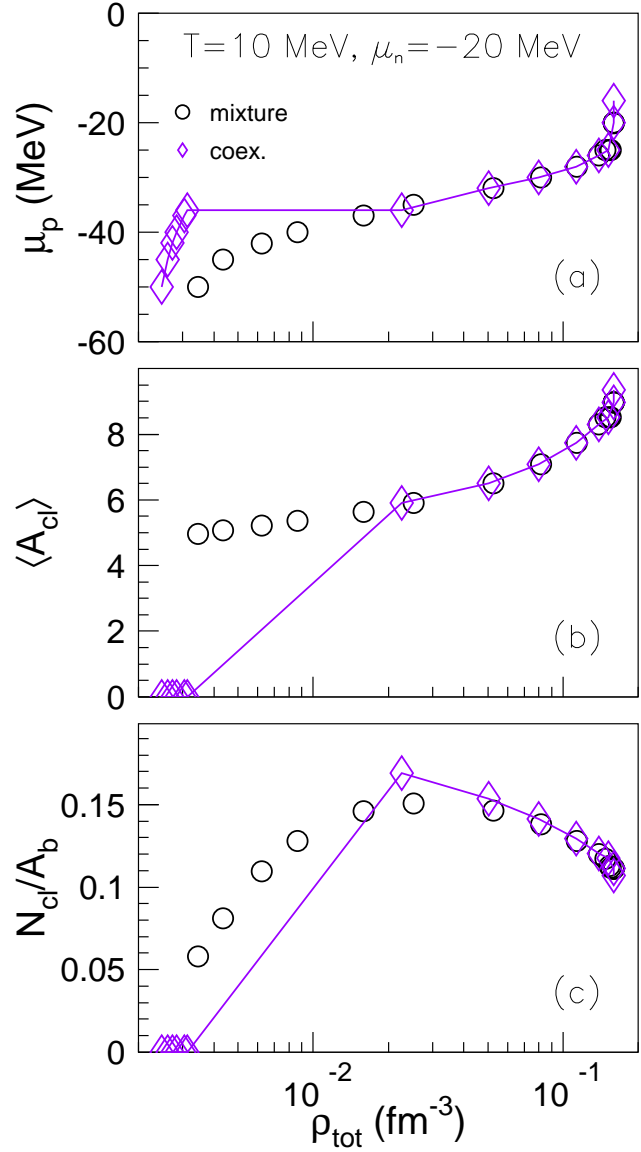}
\end{center}
\caption{(Color online)
Mixture versus coexistence for $T$=10 MeV and  $\mu_n=-20$ MeV.
Up: $\mu_p$ versus $\rho_{tot}$; 
Middle: average mass of a cluster versus  $\rho_{tot}$; 
Bottom: average cluster number per total baryon number versus  $\rho_{tot}$.
}
\label{fig:mix-coex_t=10_mun=-20}
\end{figure}

The matter composition obtained with a continuous mixture is compared to the one
corresponding to a discontinuous transition obtained with the Gibbs
construction in Fig. \ref{fig:mix-coex_t=10_mun=-20}.
We can see, together with the different behavior of the equation
of state, that the predictions for the average cluster size and cluster 
multiplicity show quantitative differences. 
It is difficult to know a priori the consequence of such altogether
slight differences on the macroscopic modelization of supernova matter, and we
cannot exclude that the use of a mixture equation of state would not be 
really distinguishable from the currently used approximations employing first-order
phase transitions. 
Even if this would be the case, the discontinuities in all thermodynamic quantities 
implied by these artificial first-order transitions lead to numerical instabilities
in supernova codes \cite{fantina} which are difficult to handle, and which 
could be avoided using continuous equations of state.

To defend the possibility of a first-order crust-core transition, one may argue that the
behavior of the entropy as a function of the density shown in Figs. \ref{fig:sbar_t=5}
and \ref{fig:sbar_t=10} is certainly model dependent, and that in general residual convexities
could still be present after considering the mixture of nucleons and clusters, especially
at the lowest temperatures. Such convexities would represent residual instabilities
and should then be cured with a Gibbs construction, leading again to a first-order
phase transition from a low-density, cluster dominated  mixture to a high density,
matter dominated mixture as in the usual representation of a neutron star.
Such an instability is known to be present in homogeneous nuclear matter, as shown
by the back-bendings of Figs. \ref{fig:phases_t=5} and \ref{fig:phases_t=10}, 
and it may well survive if clusterization was allowed in a microscopic way 
\cite{typel} going beyond
the mean-field approximation in a microscopically consistent way.

The argument exposed above is the second argument which can be invoked to justify a discontinuous transition,
which was mentioned at the beginning of this section, and that we now come to discuss.

It is certainly correct that our model depicting the simultaneous presence of
nucleons and clusters
as a non-interacting mixture of two different components associated to completely independent
degrees of freedom and self-energies is highly schematic, and improvements are in order.

In particular the way the excluded volume is parametrized
has an influence on the quantitative balance between nucleons and clusters, with a
possible consequence on the precise location of the transition which is not
completely under control.

Fully self-consistent treatments of the quantal many-body problem at finite temperature
as in Refs. \cite{samaddar,schwenk,heckel,typel}, 
including the possibility of self-consistent clustering at low densities
are extremely promising, and will in the long run provide a much better 
description of stellar matter than the phenomenological model proposed here.
Concerning the issue of the order of the transition however, 
we believe that the availability of such 
sophisticated calculations would not modify our conclusions.
Indeed, as we have already stated, the thermodynamic stability has to be checked 
on the global thermodynamic potential Eq. (\ref{independent}), 
after all the different constituents are added up. 
If gammas and neutrinos are completely decoupled from the baryonic sector, this is not true
concerning electrons and positrons because of the neutrality constraint $\rho_p=\rho_e$.
Because of this, the properties of the electron free energy will influence the convexity
properties of the baryonic entropy as a function of  the density.

\begin{figure}
\begin{center}
\includegraphics[angle=0, width=0.48\columnwidth]{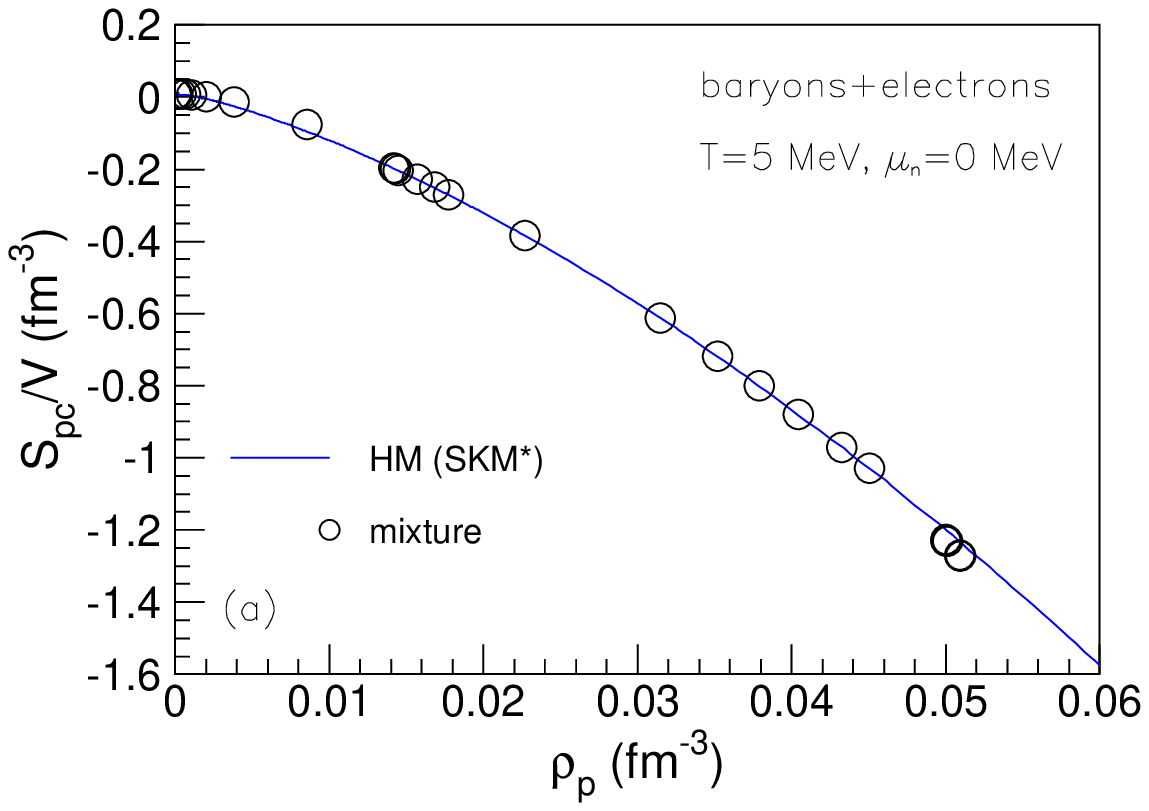}
\includegraphics[angle=0, width=0.48\columnwidth]{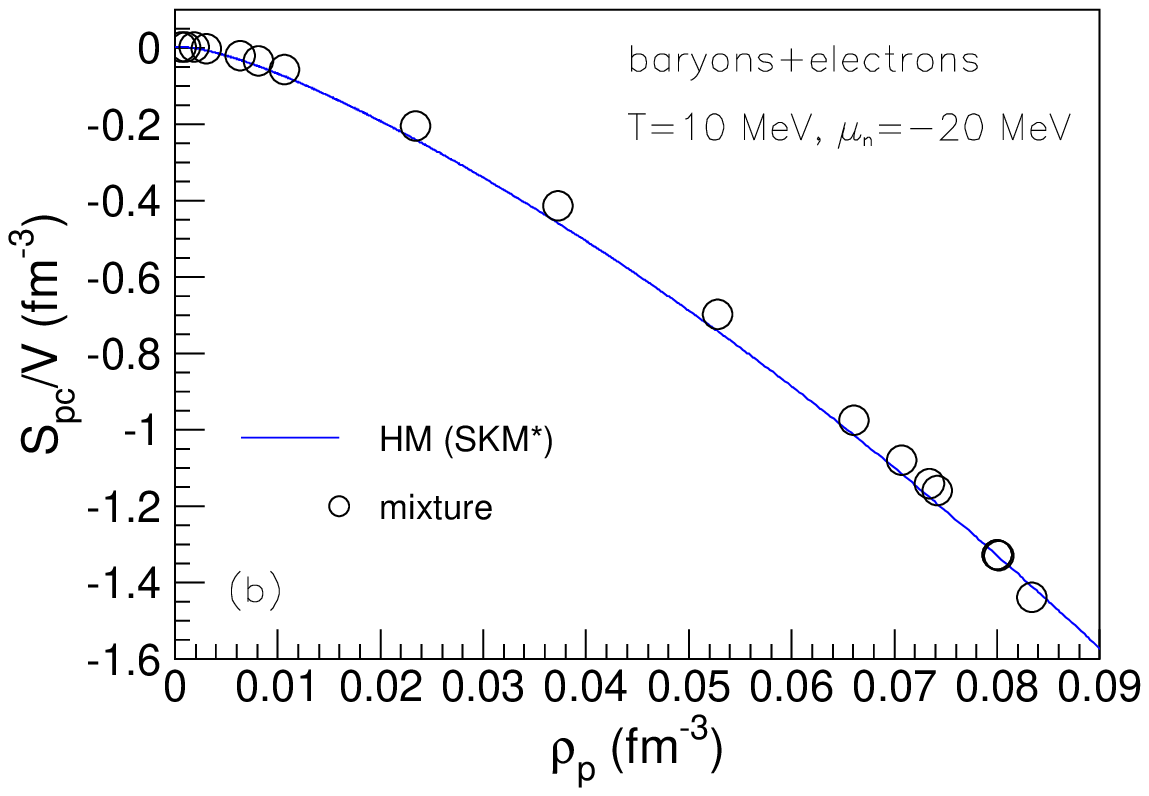}
\end{center}
\caption{(Color online) 
Constrained entropy for the ensemble baryons plus electrons 
at $T$=5 MeV along the constant $\mu_n=0$ MeV path (left panel)
and at $T$=10 MeV and $\mu_n=-20$ MeV (right panels)
for the homogeneous matter (full line)
and the mixture of matter and clusters (open circles). 
 }
\label{fig:sbar_baryel}
\end{figure}

Fig. \ref{fig:sbar_baryel} shows the thermodynamic potential of the
proton-canonical, neutron-grancanonical ensemble in the same thermodynamic
conditions as in Figs. \ref{fig:sbar_t=5} and \ref{fig:sbar_t=10}, 
including the contributions of electrons (detailed in the next section) and explicitly 
including the neutrality constraint

\begin{equation}
s_{pc}^{tot}\left (\beta,\beta\mu_n,\rho_p\right )
=\frac 1V \ln {\cal Z}^{(cl)}_{gc}(\beta,\beta\mu_n,\beta\mu_p) 
+\frac 1V \ln {\cal Z}^{(HM)}_{gc}(\beta,\beta\mu_n,\beta\mu_p) 
+\frac 1V \ln {\cal Z}^{(el)}_{(\beta,\beta\mu_e)} 
- \beta (\mu_p+\mu_e)\rho_p ,
\end{equation}

We can see that taking into account the presence of clusters has virtually no effect
on the convexity of the global entropy, which is largely dominated by the convexity 
properties of the electron thermodynamic potential. Since this latter is highly concave,
whatever the treatment of the baryonic sector, the global entropy will stay concave 
in the whole density domain. The effect of the electrons is thus to compensate the 
instability of nuclear matter and quench the possibility of a first-order phase transition.

This result can be physically understood \cite{ising_star,ducoin_npa2007}: 
because of the low electron mass, the electron gas 
is much more incompressible than the nuclear component. 
A fluctuation in the baryonic density creates a fluctuation in the charge
density, which because of the neutrality constraint implies
a fluctuation in the net electron density. 
Such fluctuations are naturally present in the length scale of the
Wigner-Seitz cell, and are at the origin of the electrostatic interaction energy
which is non-zero even if the system is neutral, see Eq. (\ref{eq:WS}). 
However these fluctuations cannot extend over macroscopic lengths because they would imply 
macroscopic electron inhomogeneities, which are contrasted by the 
high electron incompressibility.

To conclude, phases of different global baryon density cannot exist in stellar matter if
baryonic matter is constituted of heavy particles as nucleons and clusters, 
as it is the case at the densities of interest in the present paper.
Since this statement is valid for all chemical potentials and temperatures, 
including $T=0$, we conclude that the crust-core phase transition cannot 
be of first-order at any temperature.

\subsection{B. Leptons and photons}

In the temperature and density domains relevant for this paper,
$T > 1$ MeV and $\rho>10^6$ g $\cdot$ cm$^{-3}$, 
leptons (electrons and neutrinos) are relativistic, 
in particle-antiparticle pair equilibrium and in
thermal equilibrium with nuclear matter \cite{Lattimer85,LS91}.

Given that the net charge of the electron gas (that is the number of electrons
minus the number of positrons) has to neutralize the positive charge of protons
in uniform and clusterized matter, the electron chemical potential $\mu_e$ 
is determined by the equality among the electron density, 
calculated in the relativistic Fermi gas model,
and the proton density,
\begin{equation}
\rho Y_p=\frac{g_e}{6 \pi^2} \left( \frac{\mu_e}{\hbar c} \right)^3
\left[ 
1+ \frac1{\mu_e^2} \left( \pi^2 T^2 -\frac32 m_e^2 c^4\right)
\right],
\label{eq:mu_e}
\end{equation}
where $\rho$ represents the baryon density,
$g_e=2$ is the spin degeneracy and $m_e$ is the rest mass.

The electron gas pressure, entropy and energy densities are functions of $T$
and $\mu_e$ and their expressions are the following,
\begin{equation}
p^{(el)}=\frac{g_e \mu_e}{24 \pi^2} \left(\frac{\mu_e}{\hbar c} \right)^3
\left[
1+\frac1{\mu_e^2} \left( 2\pi^2 T^2 -3 m_e^2 c^4\right)+ \frac{\pi^2
  T^2}{\mu_e^4}
\left( \frac{7}{15} \pi^2 T^2 -\frac12 m_e^2 c^4\right)
\right],
\label{eq:P_e}
\end{equation}
\begin{equation}
s^{(el)}=\frac{g_e T \mu_e^2}{6 \rho \left(\hbar c \right)^3} 
\left[
1+\frac1{\mu_e^2} \left( \frac{7}{15} \pi^2 T^2 -\frac12 m_e^2 c^4 \right)
\right],
\label{eq:s_e}
\end{equation}
and, respectively,
\begin{equation}
e^{(el)}=\frac{g_e \mu_e}{8 \pi^2 \rho} \left(\frac{\mu_e}{\hbar c} \right)^3
\left[
1+\frac1{\mu_e^2} \left( 2 \pi^2 T^2 -m_e^2 c^4 \right) +
\frac{\pi^2 T^2}{\mu_e^4} \left( \frac{7}{15} \pi^2 T^2 -\frac12 m_e^2 c^4 \right)
\right].
\label{eq:e_e}
\end{equation}

Neutrinos also form a relativistic Fermi gas in pair equilibrium and thermal
equilibrium with nuclear matter. 
This means that Eqs. (\ref{eq:mu_e}, \ref{eq:P_e}, \ref{eq:s_e} and \ref{eq:e_e})
with $m_{\nu}=0$ and $g_{\nu}=1$ instead of 
$m_e$ and $g_e$ describe their
density, pressure, entropy and energy densities.
Under the $\beta$-equilibrium hypothesis, neutrinos chemical potential is
dictated by the chemical potential of neutrons, protons and electrons,
\begin{equation}
\left(\mu_p+m_p c^2 \right) +\left(\mu_e+m_e c^2 \right)=
\left(\mu_n+m_n c^2 \right)+\mu_{\nu}.
\label{eq:betaeq}
\end{equation}

If not explicitely mentioned otherwise (section III.C), 
all over this paper we shall work out of $\beta$-equilibrium. 
Neutrinos contribution to the total energy, entropy and pressure 
will be disregarded, as well. The motivation of this choice
is that equilibration with respect to weak interaction is often not achieved
over the time scales of many astrophysical phenomena, but we will discuss 
this point further in section III.C.

Photons are assumed to be in thermal equilibrium with the other star
ingredients and their pressure, entropy and energy densities are,
\begin{equation}
p^{(\gamma)}=\frac{\pi^2 T^4}{45 \left( \hbar c\right)^3},
\label{eq:P_gamma}
\end{equation}
\begin{equation}
s^{(\gamma)}=\frac{4 p^{(\gamma)}}{\rho T},
\label{eq:s_gamma}
\end{equation}
and, respectively,
\begin{equation}
e^{(\gamma)}=\frac{3 p^{(\gamma)}}{\rho}.
\label{eq:e_gamma}
\end{equation}

\section{III. Star matter relevant results}

\subsection{A. Constant chemical potential paths}

In the evolution of supernovae explosions, nuclear composition plays
an extremely important role.
First of all, isotopic abundances in the light and medium mass region
influence nucleosynthesis via r-, rp- and photodisintegration processes.
Then, nuclei with masses larger than 40-60 determine the rate of electron
capture \cite{Hix2003,Pinedo2004},
the most important weak nuclear interaction
in the dynamics of stellar core collapse and bounce,
responsible for the neutron enrichment of baryonic matter and 
neutrino emission.
On the other hand, nuclei absorb energy via internal excitation making the
increase in temperature and pressure during star collapse 
be less strong than in the case of a
uniform nucleon gas \cite{bethe79}.
Nuclei influence also the maximum density reached during the collapse.
In the absence of nuclei the collapse stops before reaching $\rho_0$, while
the limiting value is larger than  $\rho_0$ if nuclei are present \cite{zeldovich}.
Finally, by  dissociation under the shock wave,
nuclear composition affects the shock propagation 
\cite{pinedo2006}.

Under this perspective, it is obvious that realistic 
star matter equations of state require accurate 
treatment of the nuclear statistical equilibrium.
Working within the single nucleus approximation (SNA),
the first generation models \cite{Lattimer85,LS91} 
do not offer a satisfactory description of nuclear miscellanea 
and may be to some extent responsible for the failure of present
supernovae simulations to produce explosions, even if realistic neutrino transport\cite{kitaura} 
and convective instabilities in multi-dimensional hydrodynamics\cite{marek} 
are probably the key ingredient of a successful explosion of highly massive cores.
Most recent statistical models \cite{souza,mishustin,hempel2010} fix this
inconvenience by 
working in a grancanonical approximation (GCA) which
allows for continuous distributions in the
nuclear cluster mass and charge. 
This means that practically everywhere in the temperature - density domain, the
'cluster gas' may contain a variety of nuclei of very different sizes.
In our model these, additionally, may change from one configuration to the other, meaning that the 
composition fluctuation is also accounted for.
  
\begin{figure}
\begin{center}
\includegraphics[angle=0, width=0.7\columnwidth]{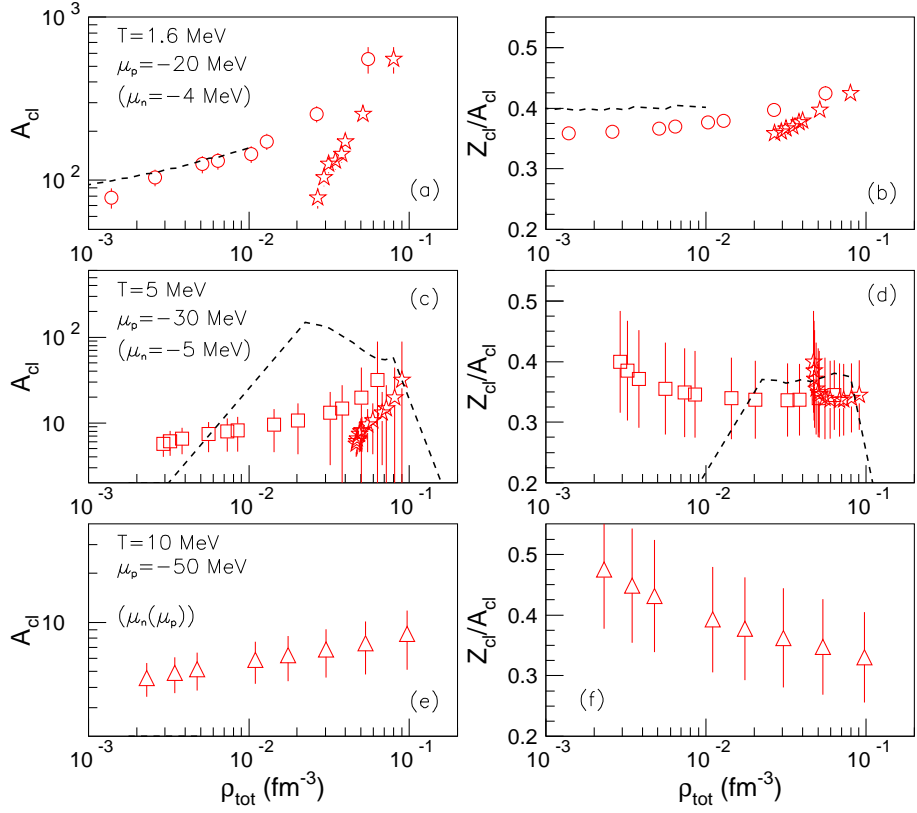}
\end{center}
\caption{(Color online)
Evolution with total baryonic density of the cluster size (left) and 
isospin composition (right) along trajectories of constant $\mu_p$ for
T=1.6, 5 and 10 MeV.
Open circles, squares and triangles: predictions of the present model corresponding
to a positive total pressure of the mixture $p_{mix}>0$;
open stars: the same as above for a negative total pressure of the mixture $p_{mix}<0$;
dashed lines: Lattimer-Swesty data from Ref. \cite{LS-webdatabase}.
}
\label{fig:frag_wignerseitz}
\end{figure}

To understand the correlation between the thermodynamic properties and the 
matter composition obtained with our
model,
Fig. \ref{fig:frag_wignerseitz} presents the evolution with density of the 
average cluster size and isotopic composition at the three temperatures 
$T=1.6,5,10$ MeV analyzed before, and for some 
chosen characteristic values of the chemical potentials, in comparison 
to the Lattimer-Swesty (LS) model employing the single nucleus approximation 
(SNA) \cite{LS-webdatabase}. 
As a technical detail, we mention that at $T$=1.6 and 5 MeV the
  clusterized matter was treated within a
  canonical-constant $Z$ ensemble, while a grand-canonical one was employed at
  $T$=10 MeV. In all considered cases, the cell volume spanned 
  the domain $1.9 \cdot 10^4 - 1.4 \cdot 10^8$ fm$^3$.

The left part of the figure shows the evolution with temperature and density 
of the average cluster size, compared to the single nucleus size of LS,
 corresponding to the same effective nucleon-nucleon potential, SKM*.
The general trend, which is independent of the chosen values of the chemical 
potential, is an increase with density and a decrease with temperature 
of the cluster size. This behavior is smooth, even though from the thermodynamic 
viewpoint at $T=1.6$ MeV and $T=5$ MeV the calculation corresponds to
the coexistence zone for the cluster component 
(stars in the left part of Figs. \ref{fig:phd_t=1.6} and \ref{fig:phd_t=5}),
while $T=10$ MeV is supercritical for clusters. This smoothness is due to 
the presence of the homogeneous matter component as a continuous mixture, 
as we have discussed in Section II.A.3.
The only indication of the underlying cluster phase transition can be seen
from the fluctuations of the size distribution, indicated by the vertical
bars in Fig. \ref{fig:frag_wignerseitz}, which are maximal at $T=5$ MeV in 
the middle of the coexistence region as it is well known from multifragmentation
studies in finite nuclei \cite{cneg,fluctuation,bimodality}.
The open stars in Fig. \ref{fig:frag_wignerseitz} denote the unstable or
metastable mixture solutions discussed in section II.A.3
, obtained when 
the homogeneous matter component is characterized by a density very similar 
to the density of the clusters. As we have already mentioned, the presence
of these multiple solutions with similar entropy content may be an indication 
that allowing for pasta-like clusters may provide a higher entropy solution.

Coming to the comparison to LS, the two models nicely agree at low density
and temperature, but important differences are seen elsewhere. 
 In the LS model the crust-core transition is a discontinuous
(first-order) disappearance of the single heavy cluster which merges at high density
and temperature with homogeneous matter: this is why $A_{cl}$ decreases with 
density at $T=5$ MeV in LS, and clusters abruptly disappear at the
temperature dependent transition points. This is at variance with our calculation,
where both components exist, though in very different proportions, in the whole
temperature and chemical potentials space. 

 Considering alpha particles together with the heavy nucleus when 
discussing the 'cluster component' in the LS case and defining
$A_{cl}=(A_{heavy}+4N_{\alpha})/(N_{\alpha}+1)$
would not change the situation.
The explanation relies on the fact that, for these trajectories, 
when a heavy cluster exists the number
of alpha particles in a cell is negligible.

The right part of Fig. \ref{fig:frag_wignerseitz} gives the average cluster 
chemical composition, while the fluctuations of this quantity are
given by the vertical bars. We can see that at low temperature (below the 
cluster limiting point) the isospin content of clusters is fairly constant 
both in our calculation and in the LS model. This is expected, since by definition 
a trajectory moving inside a coexistence region implies a transformation 
$\mu_p=\mu_p^{trans}=$cte and $\mu_n=\mu_n^{trans}=$cte, that is 
$<Z>/<A>=$cte. Since a huge part of the density plane at low temperature
belongs to the coexistence region if the cluster component is considered
alone (see Figs. \ref{fig:phd_t=1.6} and \ref{fig:phd_t=5}), the chosen trajectory
can span the whole relevant density domain staying inside the coexistence, that is at 
constant chemical composition. The little deviations observed in our calculation
from this constant behavior at high density have to be considered as numerical
accidents due to the non-perfect achievement of the thermodynamic 
limit.

In the case of $T=10$ MeV, the chemical composition is entirely determined 
by the chosen transformation and has no specific physical meaning. It is interesting
to remark that, except at the lowest temperature, this quantity presents very huge
fluctuations that cannot be addressed within the single nucleus approximation SNA.

Changing the chosen chemical potential value obviously modifies the behavior 
of the chemical composition with density, but does not change our 
qualitative conclusions.

\begin{figure}
\begin{center}
\includegraphics[angle=0, width=0.95\columnwidth]{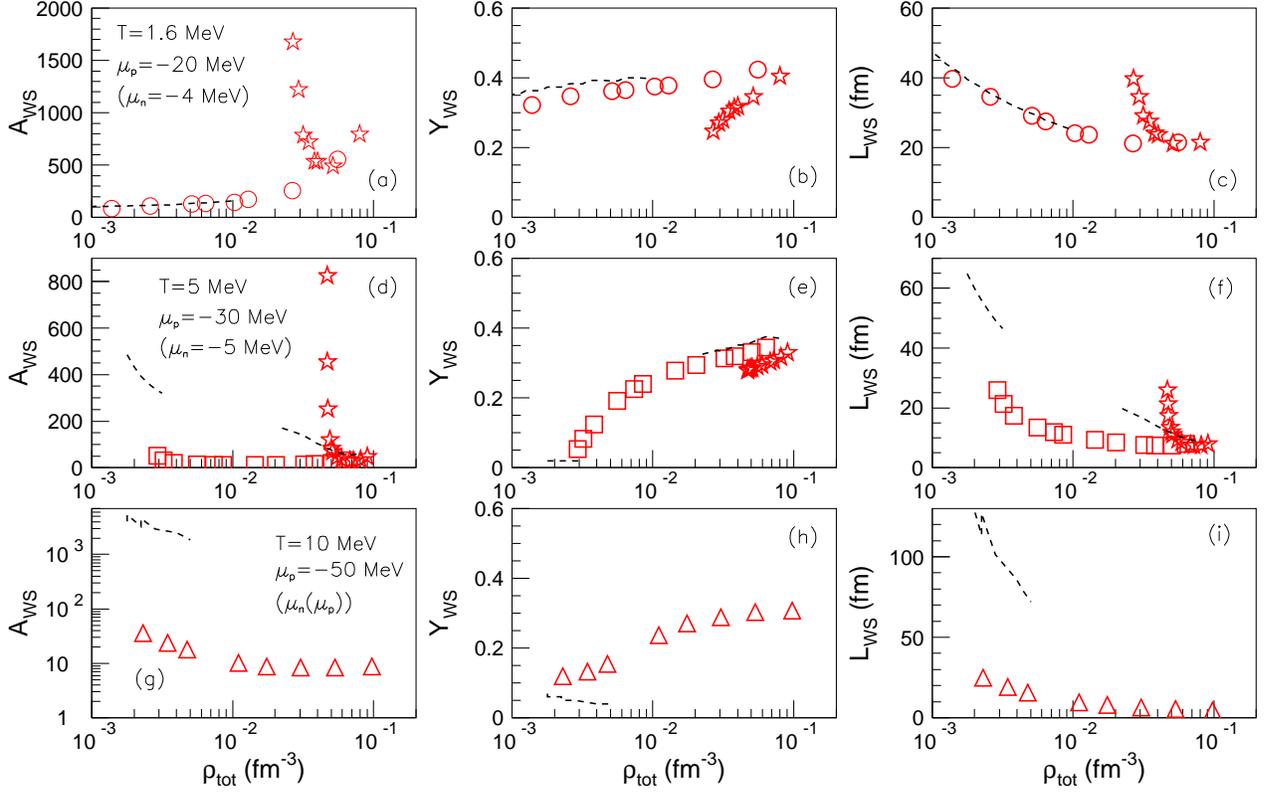}
\end{center}
\caption{(Color online)
Evolution with total baryonic density of the total baryonic number (left), 
proton fraction (middle) and 
linear dimension (right) of the Wigner-Seitz cell for $T$=1.6, 5 and 10 MeV
along trajectories of constant $\mu_p$.
Open symbols: predictions of the present model;
dashed lines: Lattimer-Swesty data from Ref. \cite{LS-webdatabase}.
}
\label{fig:cell_wignerseitz}
\end{figure}

Fig. \ref{fig:cell_wignerseitz} displays the characteristics of the Wigner-Seitz cell in terms
of baryonic number, proton fraction and linear size, in the same thermodynamic conditions of 
Fig. \ref{fig:frag_wignerseitz}. 
The Wigner-Seitz size is defined as the total baryonic mass
associated to each fragment,
\begin{equation}
 A_{WS}=\frac {\langle \sum_{i} A_i\rangle}{N_{cl}},
\end{equation} 
where the sum   runs over clusters and free nucleons, 
and $N_{cl}=<N_C>$ is the average cluster number obtained
in the simulation. 
It is important to stress that the concept of a Wigner-Seitz cell is very
different in our approach and in SNA.
While in SNA the Wigner-Seitz cell is a primitive cell, 
{\em i.e.} the smallest volume element which contains complete information 
on the infinite system and from which the last one can be built,
in GCA we work with an ensemble of tens or hundreds of
elementary cells. 
None of these cells is primitive, as their cluster partitions may differ.
Only within fluctuations  they may be seen as replica of an 
{\em average elementary cell}, the counterpart of the Wigner-Seitz cell in SNA. 

The decrease of $A_{WS}$ with increasing density, as observed both in LS
and in our calculations at high temperature, can be understood as a trivial compression effect,
which implies that reducing the volume the matter composition does not drastically change.
A different behavior is seen at low temperature, due to the rapid increase of the heaviest 
cluster size towards the transition to the core. 

As before, open stars denote unstable mixture solutions and we cannot exclude that the crust-core
transition would be more abrupt if additional degrees of freedom were added 
in our model at high density.
Indeed at the peak a small cluster coexists with a huge quantity of slightly
diluted matter. 
This situation is energetically similar to a very big bubble in a dense 
$\rho=\rho_0$ medium, but the absence of deformation
degrees of freedom may suppress the entropy of such exotic configurations.

Concerning the comparison with the LS model, the two calculations fairly agree
only at low temperature.  
As a general trend, taking into account the whole equilibrium distribution of clusters (GCA) 
produces smaller fragments (see Fig. \ref{fig:frag_wignerseitz}, left panel) 
with higher multiplicity respect to the single nucleus SNA approximation.
This is in qualitative agreement with other authors findings 
\cite{mishustin,souza,hempel2010}.
The absence of points at intermediate density at $T=5$ MeV is not an effect 
of the discontinuous transitions in LS, but is simply due to the fact that 
LS calculations are performed in discrete $Y_p$ steps, 
and therefore some points are missing when looking for $\mu=$cte trajectories.
The much higher $A_{WS}$ seen in LS calculations at high temperature is due to the fact
that no fragments other than alphas are present there in the SNA approximation 
(see Fig. \ref{fig:frag_wignerseitz}).
  In this situation 
($\rho_{tot}<3 \cdot 10^{-3}$ fm$^{-3}$ for $T$=5 MeV and the whole considered 
total density domain at $T$=10 MeV), 
the cell is built around an alpha particle such that the total baryonic mass of a
cell becomes,
\begin{eqnarray} 
A_{WS}&=&(N_n+N_p+4N_{\alpha})/N_{\alpha} \nonumber \\
&=&12+4(\omega_n+\omega_p)/\omega_{\alpha},
\end{eqnarray}
where $N_n$, $N_p$, $N_{\alpha}$ and $\omega_n$, $\omega_p$, $\omega_{\alpha}$ 
denote respectively the number of free neutrons, free protons 
and alpha particles per unit volume and the corresponding mass fractions.
This makes the Wigner-Seitz mass increase because of our definition 
$A_{tot}/V=A_{WS}N_{cl}/V$.

The global proton fraction of the Wigner-Seitz cell is shown in the middle
panels of Fig. \ref{fig:cell_wignerseitz} as a function of temperature and
total density. 
At the lowest temperature fragments dominate in the global composition, 
which implies that the behavior of the cell closely follows the behavior of
the proton fraction of the clusters shown in Fig. \ref{fig:frag_wignerseitz}. 
At higher temperature, clusters start to be immersed in a nuclear gas which is neutron rich. 
As density increases, proton drip becomes allowed slightly increasing the 
proton fraction of the free nucleons. Since matter dominates over fragments at high
density and temperature, $Y_{WS}$ reflects this behavior.

Finally the right part of Fig. \ref{fig:cell_wignerseitz} shows the average 
distance among fragments, which can
be physically interpreted as the linear size of the Wigner-Seitz cell:

\begin{equation}
L_{WS}=\left ( \frac {V}{N_{cl}}\right )^{1/3}.
\end{equation} 

In the case of SNA, at temperatures where the heavy nucleus is not present any
more, this quantity is replaced with the average distance among alphas.
$L_{WS}$ decreases with density and temperature as expected. 
Again, only in the thermodynamic conditions
where a single massive cluster dominates in the equilibrium distribution 
(that is: at the lowest temperature)
the results agree with the SNA approximation which assumes a single cluster in 
the whole phase diagram.

\subsection{B. Results at constant proton fraction}

The construction of the phase diagram and mixture properties requires to work in 
a grancanonical formulation, that is fix the intensive
observables $T$, $\mu_n$ and $\mu_p$, or alternatively in a canonical formulation 
working in terms of $T$, $\rho_n$ and $\rho_p$, where
$\rho_n$ and $\rho_p$ stand for the total neutron and proton densities, 
{\em i.e.} the summed up contributions of uniform and clusterized matter. 
For the purpose of studying the stability properties of the different phases, 
hybrid ensembles where only 
one charge is let free to fluctuate have also been employed.
 
These variables are however not convenient to use for 
expressing the equation of state of
stellar matter, as the chemical potentials are not measurable
and the total baryonic density is of outmost importance for astrophysical applications.
  
The mostly spread choice is then to work in terms of ($T$, $\rho$ and $Y_p$), where
$\rho=\rho_n+\rho_p$ is the total baryonic density and
$Y_p=\rho_p/\rho$ is the proton fraction.
The net charge neutrality requires proton and electron fractions to be equal, 
$Y_p=Y_e$.

\begin{figure}
\begin{center}
\includegraphics[angle=0, width=0.48\columnwidth]{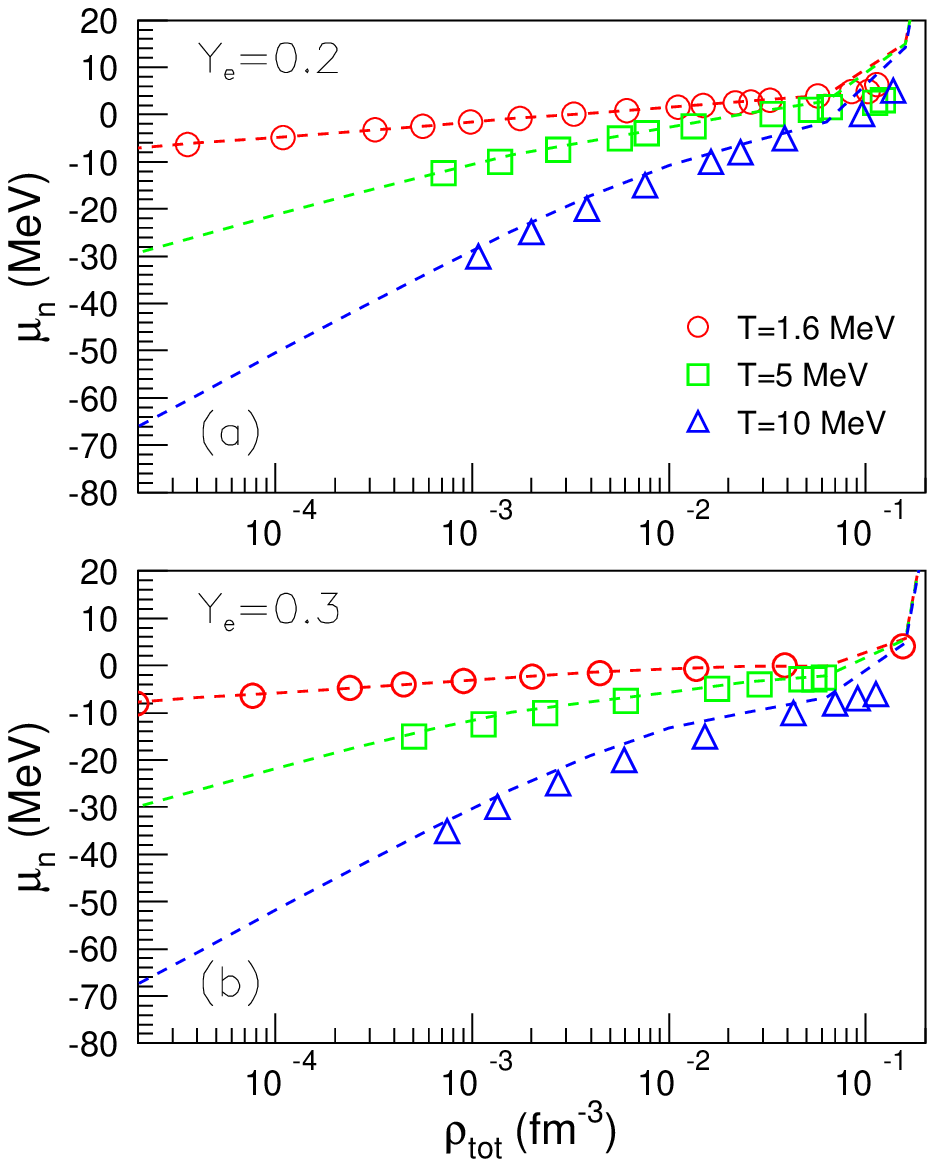}
\includegraphics[angle=0, width=0.48\columnwidth]{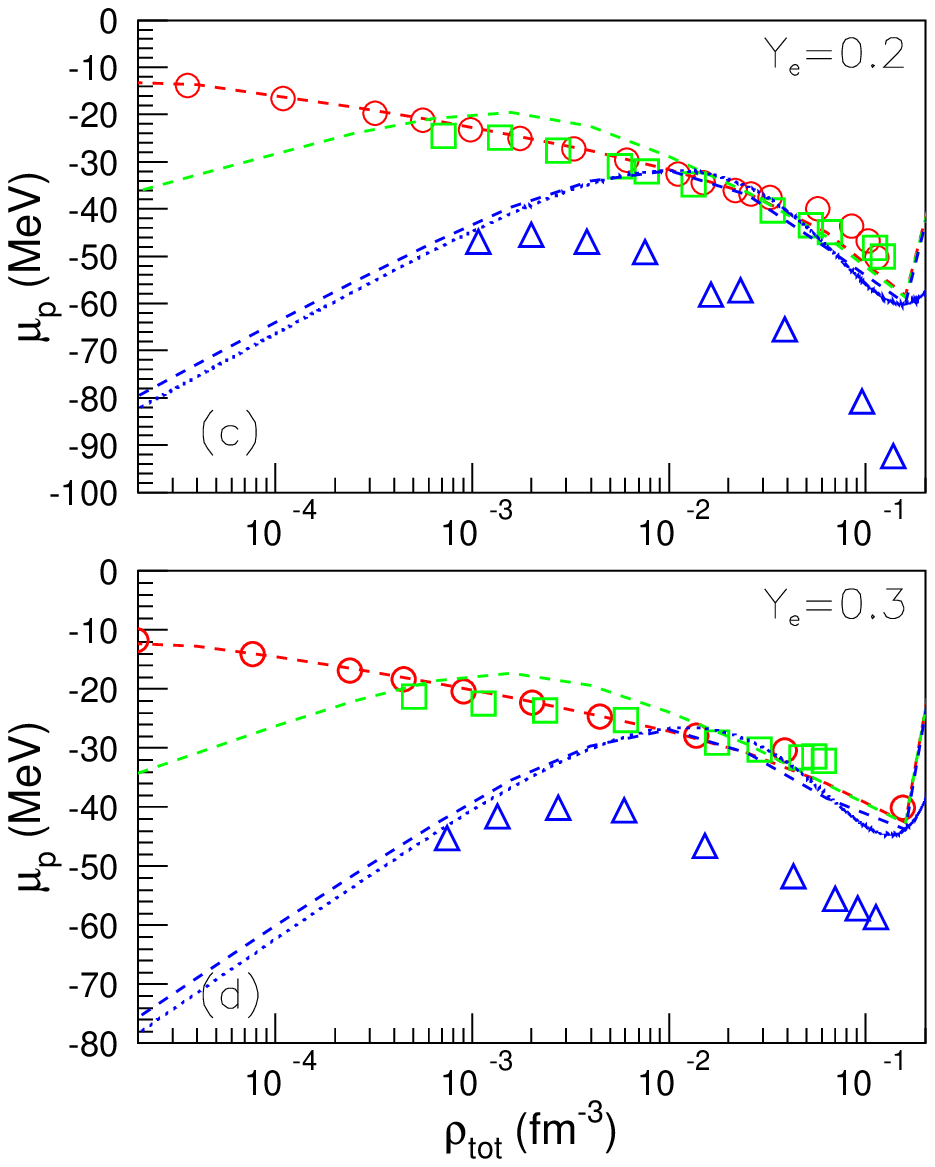}
\end{center}
\caption{(Color online)
  $\mu_n$ (left panels) and $\mu_p$ (right panels) versus total
  baryonic density for different values of the constant proton fraction 
  $Y_e$=0.2, 0.3
  and temperature $T$=1.6, 5 and 10 MeV. 
  Present model predictions are illustrated with open circles ($T$=1.6 MeV),
  squares ($T$=5 MeV) and triangles ($T$=10 MeV);
  dashed lines correspond to Lattimer-Swesty results \cite{LS-webdatabase}.
  In both cases the uniform matter component is described by SKM*.
  The thin dotted line corresponds to the case in which,
  and $T$=10 MeV,
  nuclear matter would exclusively consist out of free nucleons. 
}
\label{fig:mu-rhotot}
\end{figure}

Let us adopt these coordinates and investigate which are the domains of
$\mu_n$ and $\mu_p$ spanned by the total system with two arbitrarily chosen
values of $Y_p$=0.2 and 0.3 if the total density ranges from $\rho_0$ to 
$\rho_0/1000$. 
The results are plotted in Fig. \ref{fig:mu-rhotot} for the same values of
temperature considered before, $T$=1.6, 5 and 10 MeV. 
Present model predictions (open circles, squares and triangles) 
are systematically compared to LS predictions \cite{LS-webdatabase} 
(dashed lines).
While for $\mu_n(\rho)$ our predictions agree with LS, 
for $\mu_p(\rho)$ this is true only in the limit of low and medium
temperatures. 
At the highest considered temperature ($T$=10 MeV) and $\rho>\rho_0/100$
present model predictions differ from the LS ones by few tens of MeV.

This difference between the models can be understood as an effect of 
the presence of clusters at high temperature implied by our approach.
The thin dotted line corresponds to a calculation where 
the cluster contribution in the mixture is artificially switched off,
and the nuclear matter is made out exclusively of free nucleons. 
Quite remarkably, in this case our calculation agrees with LS predictions, 
which can be understood knowing that the amount of mass contained in clusters 
in LS at high temperature is negligible, as we have seen in the previous section.
We want to stress that the total entropy of the mixture is always 
higher than the total entropy of homogeneous matter (see Fig. \ref{fig:sbar_t=10}), 
meaning that our mixture solution is preferred at equilibrium.

\begin{figure}
\begin{center}
\includegraphics[angle=0, width=0.45\columnwidth]{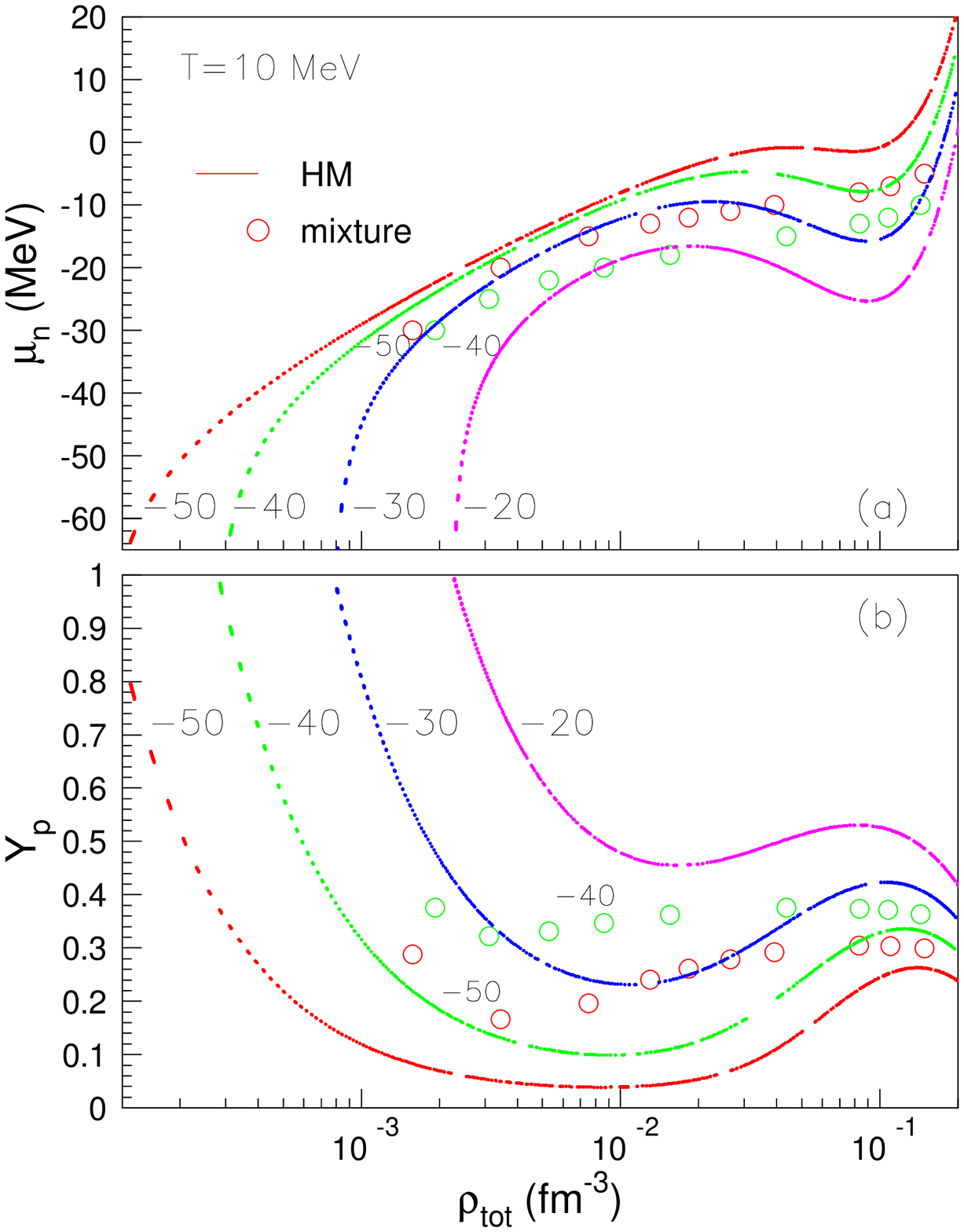}
\includegraphics[angle=0, width=0.45\columnwidth]{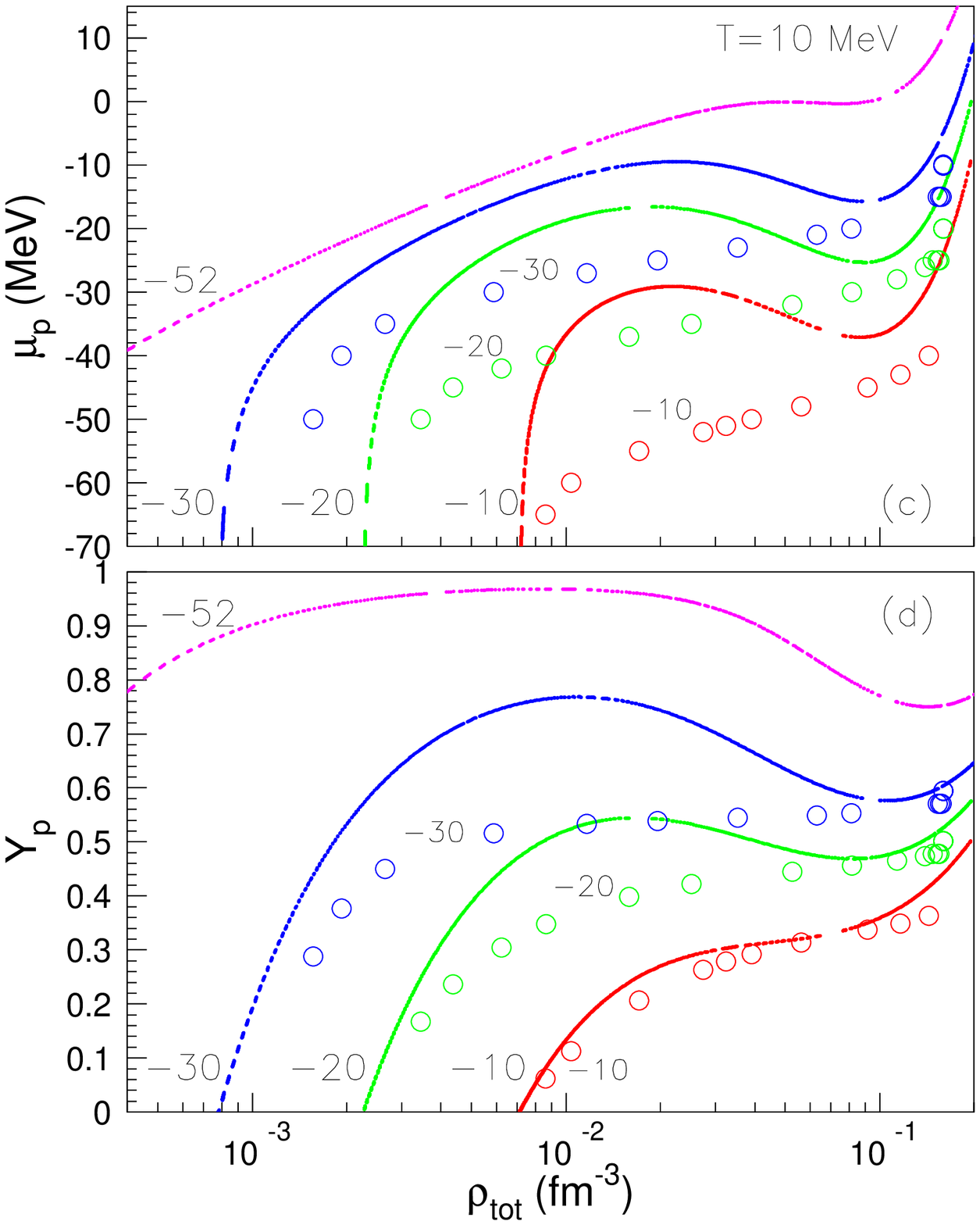}
\end{center}
\caption{(Color online)
$\mu_{n,p} (\rho)|_{\mu_p,\mu_n}$  (top) and 
$Y_p (\rho)|_{\mu_p,\mu_n}$ (bottom) curves
at T=10 MeV.
The values of the constant chemical potentials are expressed in MeV.
Solid lines correspond to the homogeneous matter; 
open circles correspond to the mixture between homogeneous and clusterized
matter.
}
\label{fig:yp_rhotot}
\end{figure}

Surprisingly, this difference between the models is not seen for the 
neutron chemical potential (left side of Fig. \ref{fig:mu-rhotot}). A very similar 
behavior was already observed in Ref. \cite{hempel2010}. 
This can be understood from inspection of Fig. \ref{fig:yp_rhotot},
which displays the relation between $\mu_n$, $\mu_p$, $Y_p$ and $\rho$
at $T=10$ MeV for the homogeneous matter and clusterized component.

Let us look at the homogeneous matter first.
Because of the symmetry properties of the effective interaction, 
the behavior of $\mu_n(\rho)$ at constant $\mu_p$
is the same as the behavior of $\mu_p(\rho)$ at constant $\mu_n$, and shows the characteristic
back-bending implied by the homogeneous matter instability. Because of that,
the behavior of the proton fraction $Y_p(\rho)$ at constant $\mu_n$ is also back-bending,
and the behavior of $Y_p(\rho)$ at constant $\mu_p$ is inverted respect to
it. A constant $Y_p=0.3$
path then explores, as a function of the density, the back-bending region of the iso-$\mu_p$,
while an analogous behavior for the iso-$\mu_n$ would be seen in the (unphysical) symmetric 
trajectory $Y_p=0.7$. As a consequence, $\mu_p$ shows a maximum as a function of the density 
for this low value of the proton fraction while $\mu_n$ is a monotonically increasing function,
as observed in Fig. \ref{fig:mu-rhotot}. 
Turning to the clusterized component, this latter does not show any
instability meaning that the chemical
potentials are monotonically increasing functions of the density. 
Then the behavior of $Y_p$ as a function of the density at constant 
(neutron or proton) chemical potential is essentially flat,
and the net effect is to translate the maximum of $\mu_p$ in the constant 
proton fraction trajectories,
while no effect is seen in  $\mu_n$ because of its monotonic behavior at this low $Y_p$.
This explains the different behavior of the two chemical potentials at constant proton fraction
seen in Fig.\ref{fig:mu-rhotot}.

We now turn to analyze some quantitative predictions of our model concerning 
both matter composition and equations of state, in thermodynamic conditions 
which are relevant for supernova matter.
 
\subsubsection{1. Matter composition}

\begin{figure}
\begin{center}
\includegraphics[angle=0, width=0.95\columnwidth]{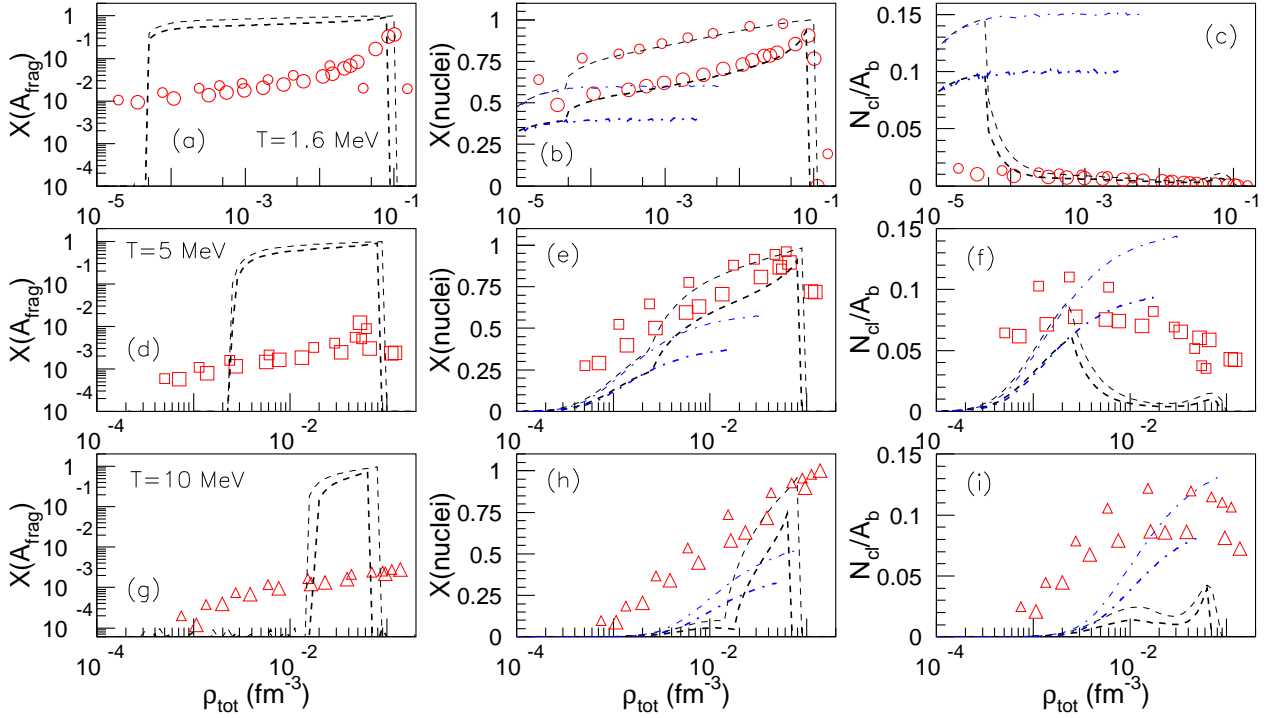}
\end{center}
\caption{(Color online)
  Evolution with total baryonic density of the average fragment mass fraction
  (left panels), cumulated cluster mass fraction (middle panels) and
  cluster number per baryonic number (right panels)
  for $T$=1.6, 5 and 10 MeV.
  Large and small open symbols stand for present model predictions corresponding
  to $Y_e$=0.2 and, respectively, 0.3;
  think ($Y_e$=0.2) and thin ($Y_e$=0.3) dashed lines 
  illustrate Lattimer-Swesty data from Ref. \cite{LS-webdatabase};
  think ($Y_e$=0.2) and thin ($Y_e$=0.3)  
  dot-dashed lines correspond to the Horowitz-Schwenk results based on the
  virial expansion of the equation of state \cite{HS}.
}
\label{fig:massfractions}
\end{figure}

Fig. \ref{fig:massfractions} illustrates the total baryonic density dependence
of the  average fragment mass fraction $<A_{cl}>/A_{WS}$ (left panels),
cumulated fragment mass fraction $\sum_{i=1}^{N_C}A_i/A_{WS}$
(middle panels), 
and average number of clusters per baryonic number (right panels) 
for the three values of temperature already considered, $T$=1.6, 5 and 10 MeV, 
densities ranging from $10^{-5}$ to $10^{-1}$ fm$^{-3}$ 
and $Y_p$=0.2 and 0.3.
LS predictions (dashed lines) correspond to the unique heavy nucleus of SNA 
whose mass fraction is immediately calculated out of the mass fractions of free 
neutrons, free protons and alphas, 
$\omega_{heavy}=1-(\omega_n+\omega_p+\omega_{\alpha})$ (left panels)
and, respectively,
summed-up contributions of the heavy nucleus and alpha particles 
$(\omega_{heavy}+\omega_{\alpha})$ (middle panels). 
The fact that alpha particles are considered for LS also when computing
$N_{cl}/A_{WS}$ is responsible for the double-humped structure of the
corresponding distributions (right panels).

The huge difference between SNA and GCA in what regards
$A_{cl}/A_{tot}$ vs. $\rho$ is due to the different acceptations of the
cell concept in the two frameworks.
As we have already mentioned, contrary to SNA 
in GCA no primitive cell exists: because of the fluctuations implied by finite temperature,
the system cannot be viewed as an infinite number of replicas of a single elementary 
volume element containing one single cluster. 
 
Quite remarkably, the discrepancies originating from cell fluctuations  
are to a large extent washed-out when alpha particles are included (middle panels) 
and SNA and GCA agree much better. 
However, at $T$=5 MeV and, to a larger extent, at $T$=10 MeV 
GCA accounts for more mass bound in massive and light
nuclei than SNA. This result may be considered a step forward in making
numerical simulations of stellar evolution able to reproduce the 
supernovae explosions. 
The same qualitative information on cell population
is revealed by the cluster number per baryon number
depicted in the right panels, for which the maximum deviation is observed at
the highest temperature.
The bimodal character of the SNA distributions is an artefact of the fact that
only two types of nuclei, $\alpha$s and the heavy nucleus, are allowed to
exist, each of them being preferentially produced in another total density domain.
A fully microscopic description of nuclear matter 
composed of neutrons, protons and alpha particles has been
recently proposed by Horowitz and Schwenk \cite{HS} and relies on the
virial expansion of the equation of state. 
The results corresponding to the density dependence of the cumulated cluster
mass fraction $4 n_{\alpha}/(n_n+n_p+4 n_{\alpha})$ and
average cluster number per baryonic number $n_{\alpha}/(n_n+n_p+4 n_{\alpha})$
are plotted with dot-dashed lines in the middle and, respectively, 
right panels.
The density domains over which this calculation is possible was fixed by the
condition of having sub-unitary nucleon fugacities. From a qualitative point of view, 
our model agrees with HS in the sense that both of them show a smooth
behavior of $X_{nuclei}$ and $N_{cl}/A_b$ as a function of $\rho$. 
From a quantitative point of view,
the absence of nuclei heavier than alpha particles makes
HS systematically underestimate, in the density domains where such heavy
cluster may exist,
the percentage of mass bound in clusters
with respect to both present model and LS.
On the other hand, in the temperature-density regions
where the largest cluster allowed by LS is the alpha particle, there is no 
clear relation among the predictions of the two models.
Thus, at $T$=5 MeV and $\rho \leq 3 \cdot 10^{-3}$ fm$^{-3}$
HS and LS predictions agree perfectly while
for $T$=10 MeV and $1 \cdot 10^{-3} \leq \rho \leq 1.8 \cdot 10^{-2}$ fm$^{-3}$
HS results exceed those corresponding to LS.
 
\subsubsection{2. Homogeneous-inhomogeneous transition}

\begin{figure}
\begin{center}
\includegraphics[angle=0, width=0.85\columnwidth]{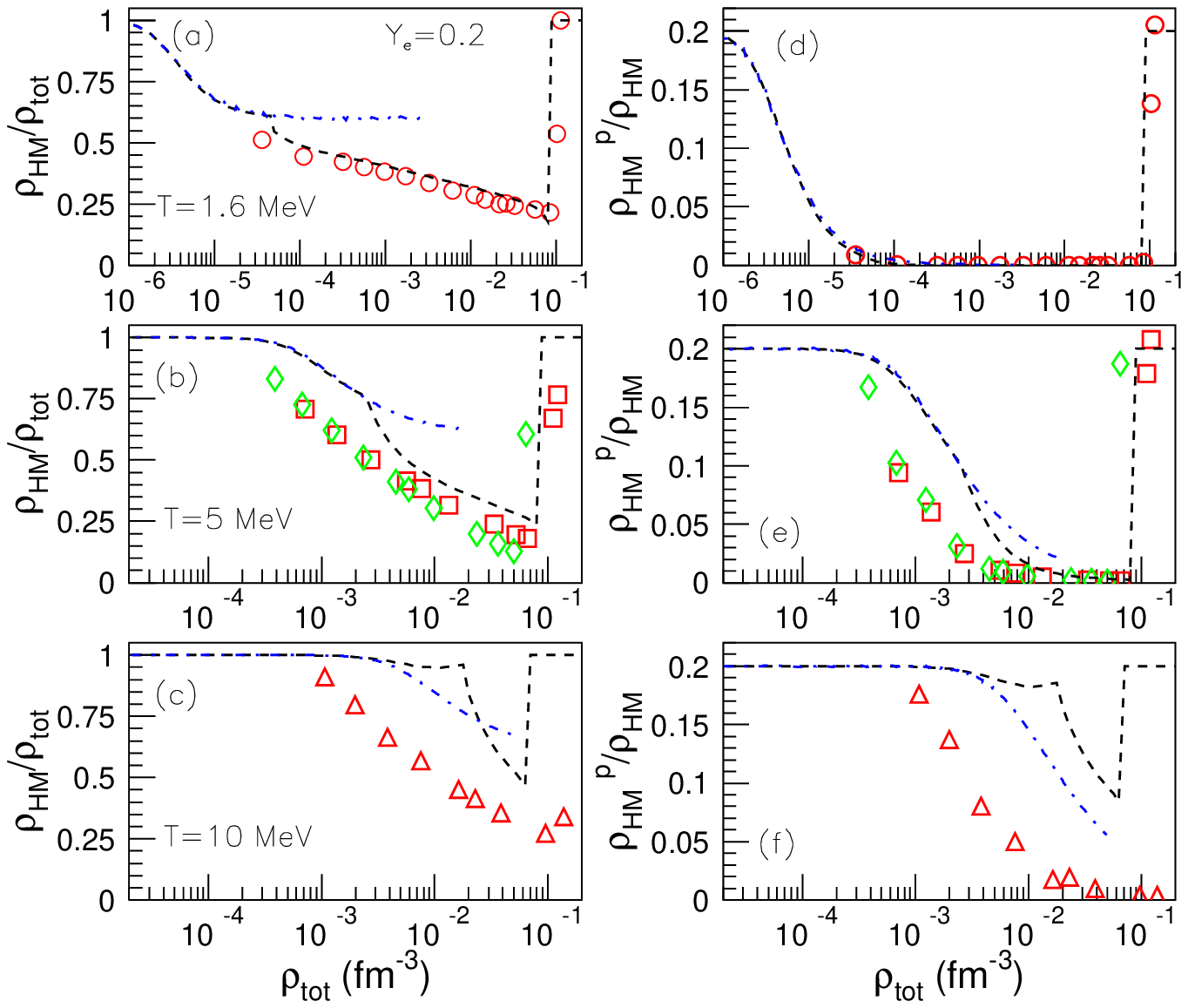}
\end{center}
\caption{(Color online)
Evolution with total baryonic density of 
relative baryonic density in uniform matter (left panels) and 
free proton fraction (right panels) 
for $T$=1.6, 5 and 10 MeV and $Y_e$=0.2.
Open circles, squares and triangles correspond to the predictions of the
present model when uniform matter is calculated according to SKM*;
dashed lines stand for Lattimer-Swesty data from Ref. \cite{LS-webdatabase};
dot-dashed lines correspond to Horowitz-Schwenk results based on the virial
expansion of the equation of state \cite{HS};
open diamonds at $T$=5 MeV correspond to present model predictions in
the case in which the uniform matter
is calculated according to Sly230a \cite{sly230a}.
}
\label{fig:fragvsmatter_ye=02}
\end{figure}

\begin{figure}
\begin{center}
\includegraphics[angle=0, width=0.85\columnwidth]{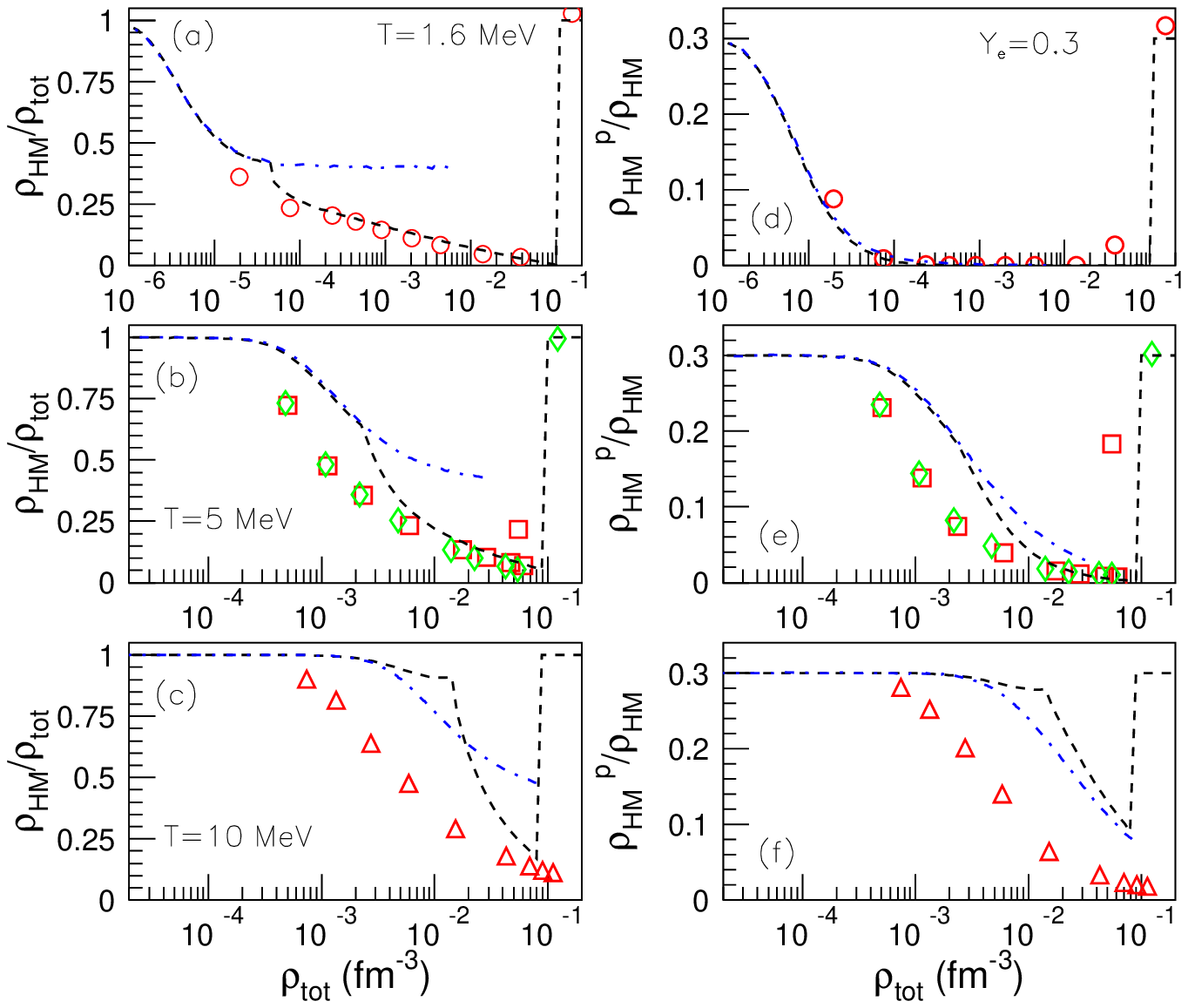}
\end{center}
\caption{(Color online)
The same as in Fig. \ref{fig:fragvsmatter_ye=02} for $Y_e=0.3$.
}
\label{fig:fragvsmatter_ye=03}
\end{figure}

We have already reported that,
in the vicinity of $\rho_0$, the strong interplay between clusters and 
uniform matter which are not allowed to occupy the same volume
is responsible for a homogeneous-unhomogeneous matter transition. 
Its fingerprints have been already visible in 
Figs. \ref{fig:frag_wignerseitz}, \ref{fig:cell_wignerseitz}, 
and \ref{fig:massfractions}.

A more suited representation is given in the left panels of 
Figs. \ref{fig:fragvsmatter_ye=02} and  \ref{fig:fragvsmatter_ye=03}
where the relative baryonic density of uniform matter
$\rho^{(HM)}/(w^{(HM)}\rho^{(HM)}+w^{(cl)}\rho^{(cl)})$
is expressed as a function of total baryonic density. 
The considered temperatures and proton fractions are 
$T$=1.6, 5 and 10 MeV and, respectively, $Y_p$=0.2 (Fig.\ref{fig:fragvsmatter_ye=02})
and 0.3 (Fig.\ref{fig:fragvsmatter_ye=03}).
The common pattern of these curves is 
a sudden jump towards $\rho^{(HM)}/\rho=1$ 
taking place at $\rho \lesssim \rho_0$
and a much smoother increase towards 1 at much smaller  
values of total density. 
This means that close to normal nuclear density and at low total densities,
 uniform nuclear matter dominates.
While for the low density domain the SNA  statement is true in the strict sense, 
in GCA it should be rather understood as the dominance of free nucleons over clusters.
As we have already seen, in GCA clusters survive, by construction, in the
whole density-temperature domain, but at low densities they represent a
negligible percentage of matter.

With the exception of $T$=1.6 MeV where Monte-Carlo GCA convergence problems 
prevent accurate determination of the transition point, 
SNA and GCA predictions agree well.
Both models show that the transition density depends on the temperature, 
but, at least in the presently investigated range, only slightly on $Y_p$.
 
The interesting point here is that this transition is spontaneously obtained 
in our model without performing an external artificial Maxwell construction
at high density, as it is necessary in standard NSE models
\cite{mishustin,blinnikov,souza}. 
 
Refs. \cite{horowitz01,vidana} show that the homogeneous-unhomogeneous
transition density, as well as the neutron skin thickness, depends upon the
density dependence of the symmetry energy and, more precisely, upon 
 $L=3 \rho_0 \left( \partial S_2 \left(\rho \right)/
\left(\partial \rho \right) \right)|_{\rho=\rho_0}$ 
where $S_2 \left(\rho \right)=\left( \partial^2 \left( E/A \right)
/2\partial \beta^2 \right)|_{\beta=0}$ with $\beta=(\rho_n-\rho_p)/\rho$.
To check to what extend this cold matter result holds at finite temperature,
we have calculated the transition region for $T$=5 MeV also in the case in
which homogeneous matter is described by Sly230a \cite{sly230a} instead of
SKM*. The result plotted with open diamonds in
Figs. \ref{fig:fragvsmatter_ye=02} and \ref{fig:fragvsmatter_ye=03} 
shows no significative modification with respect to the previous ones.
Given that $L$(SKM*)=46 MeV and $L$(Sly230a)=43.9 MeV, we partially confirm
the conclusions of Refs.  \cite{horowitz01,vidana}. 
In addition to this, within our model the transition density 
as well as the proton enrichment of the uniform matter could, at least in
principle, depend also on how the clusterized matter component is implemented.  

The isospin composition of uniform matter as a function of $\rho$
is illustrated in the right panels of Figs. \ref{fig:fragvsmatter_ye=02} 
and \ref{fig:fragvsmatter_ye=03} 
for the same values of temperature and $Y_p$ as above.
Outside the transition region, where homogeneous matter dominates, 
$\rho^{(HM)}_p/\rho^{(HM)}$
approaches the total system proton fraction, $Y_e$.
Inside the transition region, clusters proton enrichment 
(right panels of Fig. \ref{fig:frag_wignerseitz}) 
makes uniform matter be depleted in protons. 
As the clusterized sub-system, free proton fractions keep the memory of 
total system proton fraction. 
In what regards the sensitivity on how homogeneous matter
is implemented, the open diamonds show that SKM* and Sly230a give comparable
results.

In what concerns the virial expansion of the equation of state as
  performed in Ref. \cite{HS}, it is clear that the condition of sub-unitary
  nucleon and alpha fugacities prevents the exploration of
  $\rho_{HM}/\rho_{tot}$ and $\rho_{HM}^p/\rho_{HM}$
  vs. $\rho$ up to normal nuclear density.
  Nevertheless, over the accessible density domain the behavior of the
  two quantities clearly
  reproduces the already discussed patterns, that is (1) within a
  restrained density domain, which shrinks with increasing temperature,
  clusterized matter (alpha matter, in this case) 
  dominates over homogeneous matter and 
  (2) homogeneous matter is depleted in protons.
  As expected, the evolution with density of the two quantities 
  is smooth and the lack of nuclei heavier than alpha leads to 
  less mass bound in clusters with respect to both LS and the 
  presently proposed model. 
  The relative amount of alpha matter and its isospin symmetry make
  homogeneous matter have a proton fraction similar ($T$=1.6 MeV),
  higher ($T$=5 MeV) or 
  in between ($T$=10 MeV) the values predicted by LS
  and the present model.

\subsubsection{3. Equations of state}

In astrophysical simulations, the equations of state, {\em i.e.} the
functional dependences between different state observables, are extremely
important ingredients.

\begin{figure}
\begin{center}
\includegraphics[angle=0, width=0.9\columnwidth]{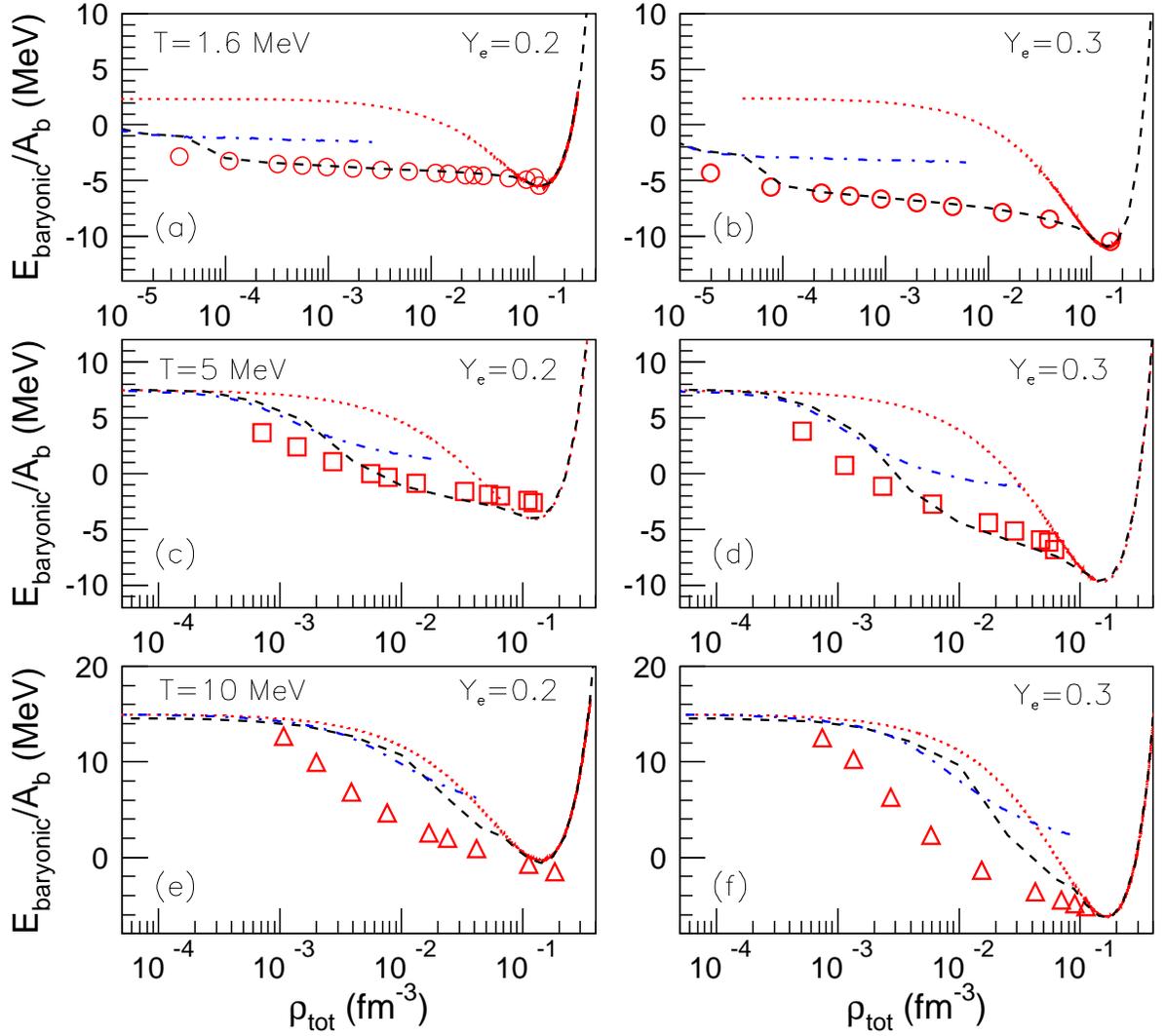}
\end{center}
\caption{Evolution with total baryonic density
 of the baryonic energy per baryon 
 at different temperatures $T$=1.6, 5 and 10 MeV for $Y_e$=0.2
 (left panels) and $Y_e$=0.3 (right panels).
 Open symbols correspond to present model predictions;
 dashed lines stand for Lattimer-Swesty data \cite{LS-webdatabase};
 dot-dashed lines illustrate Horowitz-Schwenk results \cite{HS};
 dotted lines correspond to the case in which
 only homogeneous nuclear matter would exist. 
}
\label{fig:eb-rhotot}
\end{figure}

\begin{figure}
\begin{center}
\includegraphics[angle=0, width=0.48\columnwidth]{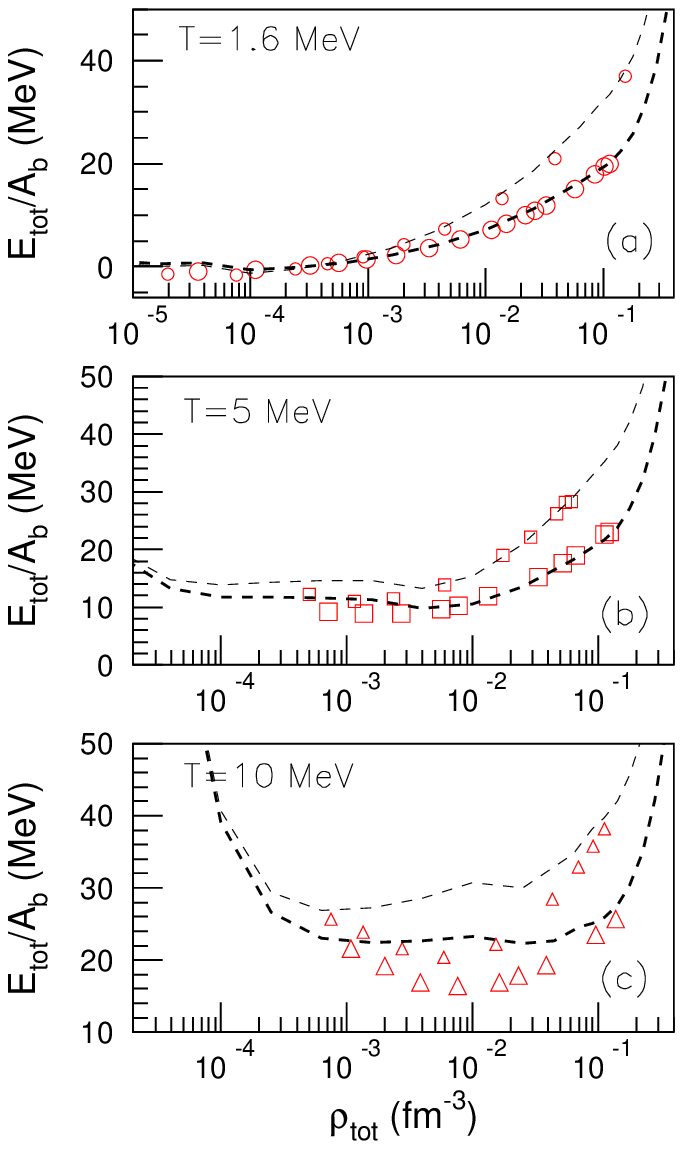}
\end{center}
\caption{Evolution with total baryonic density
 of the total energy per baryon 
 at different temperatures $T$=1.6, 5 and 10 MeV and $Y_e$=0.2, 0.3.
 Open symbols illustrate present model predictions while 
 dashed lines stand for Lattimer-Swesty data
 \cite{LS-webdatabase}. 
 Large (small) symbols and thick (thin) lines correspond to $Y_e$=0.2 (0.3).
 }
\label{fig:etot-rhotot}
\end{figure}

Figs. \ref{fig:eb-rhotot} and \ref{fig:etot-rhotot} 
present the baryonic energy per baryon, 
$E^{(bar)}/A_b=\left( E^{(cl)}+E^{(HM)} \right)/A_b $, 
and, respectively, total energy per baryon, 
$E_{(tot)}/A_b=\left( E^{(bar)} +E^{(el)} +E^{({\gamma})}\right)/A_b$,
as a function of total baryonic density for 
$T$=1.6, 5 and 10 MeV and $Y_p$=0.2 and 0.3.

As one may notice, at all considered temperatures 
the $E^{(bar)}/A_b (\rho)$ curves behave similarly:
at low densities, they are flat as dominated by the ideal gas
character of diluted homogeneous matter, while 
in the transition density, they manifest a sharp fall.
LS results additionally show that for $\rho > \rho_0$ $E^{(bar)}/A_b$
rises steeply with $\rho$.
The dependence with isospin asymmetry and temperature is trivial:
more symmetric systems are more bound and 
average baryon energy increases with the temperature.
Concerning the comparison between our model and LS, one can say that at
$T$=1.6 MeV they agree perfectly.
At higher temperatures the two models results deviate in the total baryonic
density range corresponding to the homogeneous-unhomogeneous matter
transition.
Moreover, the discrepancy augments with the temperature.
Taking into account that uniform matter is described by the same equation of
state in LS and present model, the only possible explanation of the observed
behavior concerns the clusterized component.
As we have already discussed that the present model accounts for
a systematically larger mass fraction bound in clusters with respect with LS,
the clusters binding energy is expected to act in the sense of diminishing 
$E^{(cl)}$.       
This is confirmed by inspection of the thin lines  in Fig. \ref{fig:eb-rhotot}, 
which represent the baryonic energy
per baryon number, obtained by artificially switching off the cluster component. 
For $T$=10 MeV, the thin line perfectly agrees with LS results, showing that
in SNA clusters practically do not contribute to the EOS at high temperatures.

We have already seen
(middle panels of Fig. \ref{fig:massfractions})
that in the high density sub-region of the transition density region
HS accounts for a smaller fraction of mass bound in clusters 
with respect LS, while in the low density sub-region the opposite holds.
This relation is replicated in the density dependence of the baryonic energy
per baryon: in the high density sub-region $(E_b/A_b)|_{HS}>(E_b/A_b)|_{LS}$
while in the low density region  $(E_b/A_b)|_{HS}\leq (E_b/A_b)|_{LS}$. 
Concerning the comparison of HS versus the present model, a similar conclusion may
be drawn: as far as our model accounts for a higher percentage of bound mass,
it predicts smaller values of $E_b/A_b$. In addition to this, HS shows less
$Y_e$-sensitivity than the other two models.

As the electron and photon contributions are trivial, 
it is obvious that all features of the baryonic energy density will be the same
in the representation of the total energy per baryon.  
Indeed, Fig. \ref{fig:etot-rhotot} shows that 
at low temperatures and whatever densities, the 
present model results agree with the LS ones. The same is true for 
high temperatures and densities corresponding to homogeneous matter.
Otherwise, GCA results deviate from
the SNA ones.

\begin{figure}
\begin{center}
\includegraphics[angle=0, width=0.5\columnwidth]{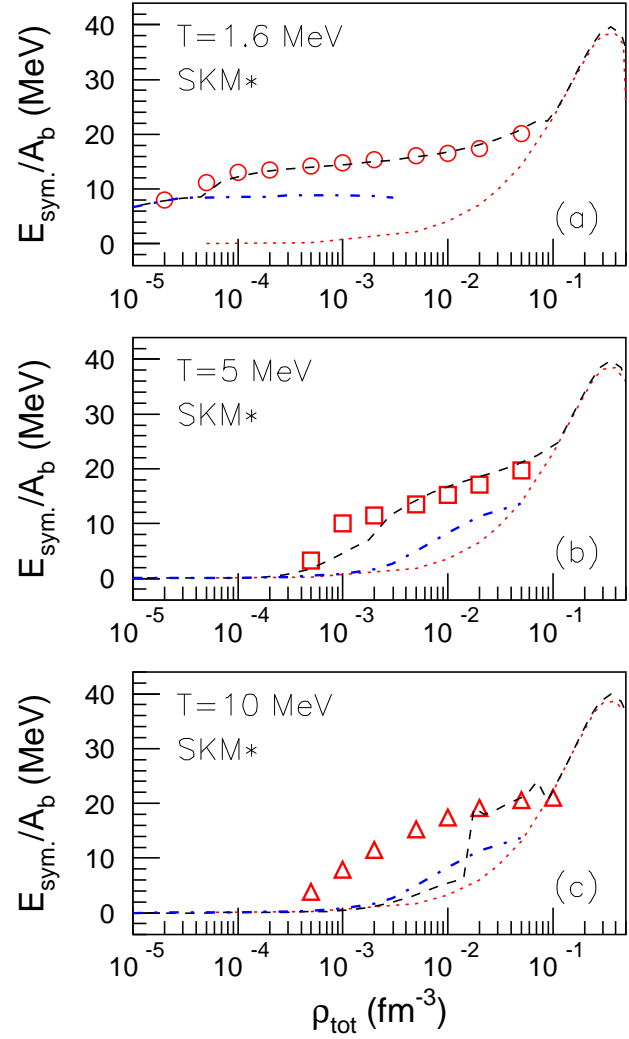}
\end{center}
\caption{Evolution with total baryonic density and temperature
 of the symmetry energy per baryon calculated according to Eq. (\ref{eq:symen_clust}).
 See Fig. \ref{fig:eb-rhotot} for the line and symbol code.
}
\label{fig:symen}
\end{figure}

Information on the density dependence of the baryonic energy for different
system asymmetry allows one to infer the density dependence of the symmetry
energy, an extremely important quantity which is so far largely unknown.
Indeed, recasting the first term in the expansion 
of the internal energy per baryon in powers of asymmetry
$\delta=(N-Z)/A$,
\begin{equation}
E_{sym}(\rho,T)/A= \frac12 \frac{\partial^2 (E/A)}{\partial \delta^2}|_{\delta=0},
\label{eq:esym}
\end{equation} 
in a finite difference formula, one gets,
\begin{equation}
E_{sym}(\rho,T)/A=\frac{E(\rho,T,Y_p^{1})/A-E(\rho,T,Y_p^{(2)})/A}{(1-2Y_p^{1})^2-(1-2Y_p^2)^2}.
\label{eq:symen_clust}
\end{equation}

Eq. (\ref{eq:symen_clust}) is exact because of the quadratic dependence of the 
baryonic density with the isospin asymmetry.
The model predictions on the baryonic density dependence of the
symmetry energy per baryon  (open symbols) are illustrated in
Fig. \ref{fig:symen} along
with the Lattimer-Swesty (dashed lines), Horowitz-Schwenk (dot-dashed lines) 
and homogeneous matter (dotted lines) results. 
The common feature of the three models is that, when a significant amount of
matter is bound in clusters, $E_{sym}/A$ deviates from expectations based on 
homogeneous matter behavior. The deviation is visible in diluted matter and is
washed out while approaching $\rho_0$, as nuclei and homogeneous matter
are described by similar EOS. 
In addition to this remarkable symmetry energy increase, 
clusterization induces also a temperature dependence of $E_{sym}/A(\rho)$, 
not present in homogeneous matter. 
At the highest temperature, where our model
accounts for much more clusters with respect to LS and HS, the agreement is mainly
qualitative. At $T$=1.6 and 5 MeV the results of the present model
perfectly agree with those of LS and both exceed the HS predictions.  
The same conclusions 
have been recently pointed out in Ref. \cite{typel}.
There it is shown that clusterization also reduces the sensitivity of the EOS 
to the parameters of the effective interaction.

\begin{figure}
\begin{center}
\includegraphics[angle=0, width=0.9\columnwidth]{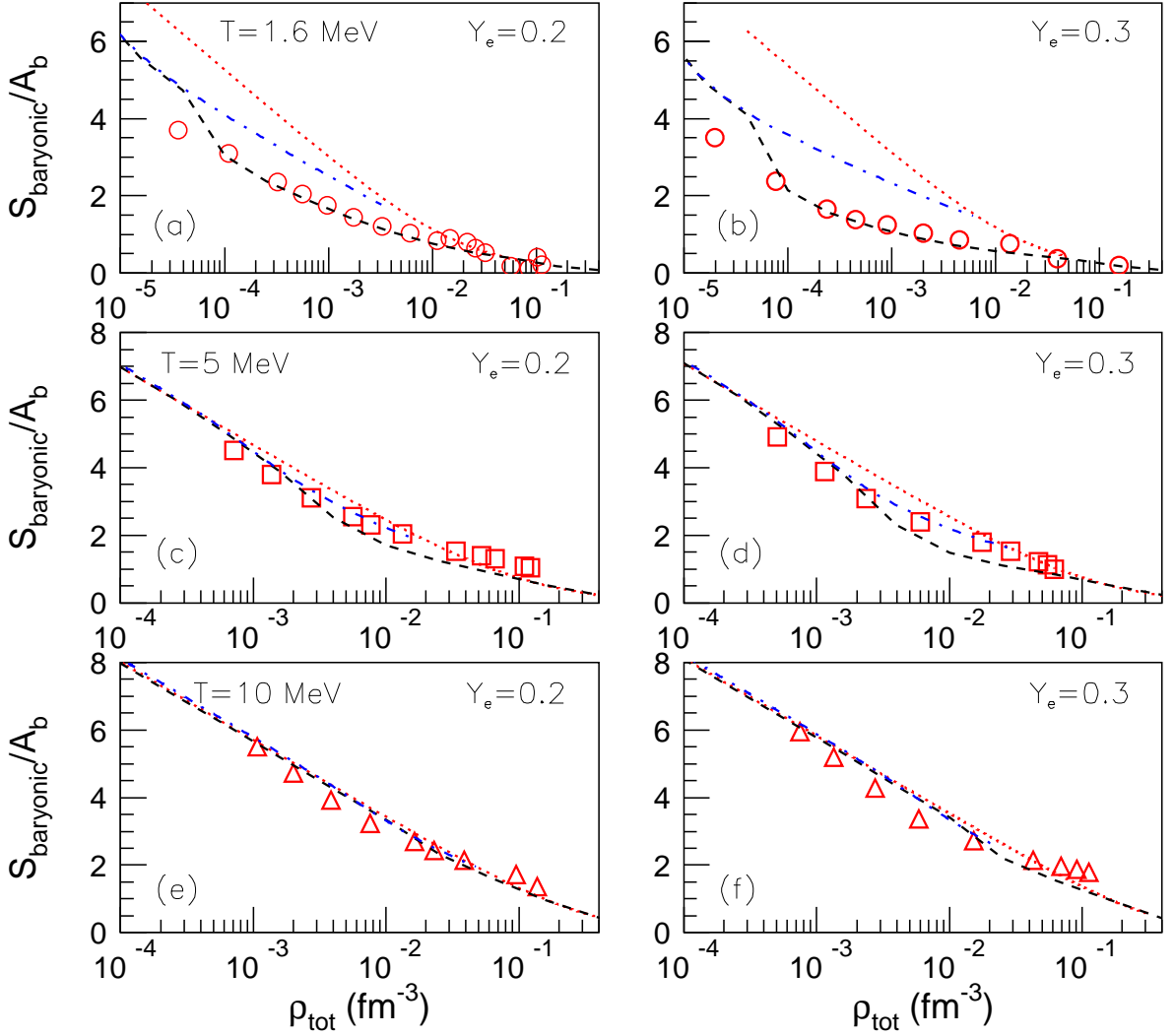}
\end{center}
\caption{The same as in Fig. \ref{fig:eb-rhotot} but for the baryonic entropy.
  See Fig. \ref{fig:eb-rhotot} for the line and symbol code.
}
\label{fig:sb-rhotot}
\end{figure}

\begin{figure}
\begin{center}
\includegraphics[angle=0, width=0.48\columnwidth]{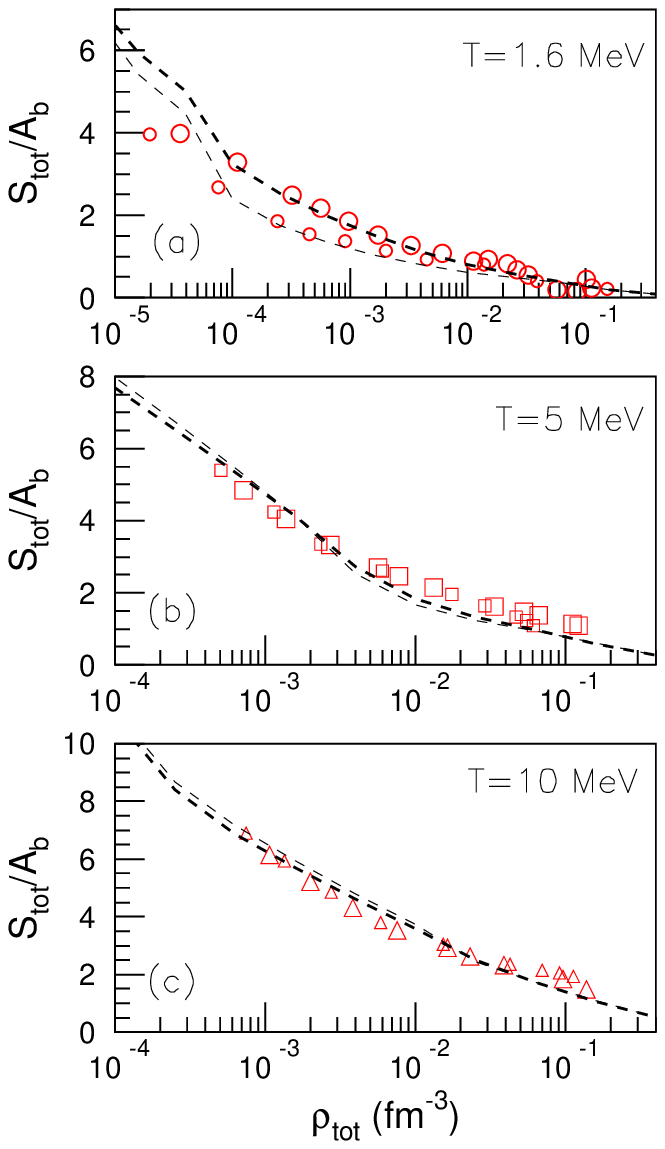}
\end{center}
\caption{The same as in Fig. \ref{fig:etot-rhotot} but for the total entropy.
  See Fig. \ref{fig:etot-rhotot} for the line and symbol code.
}
\label{fig:stot-rhotot}
\end{figure}

Figs. \ref{fig:sb-rhotot} and \ref{fig:stot-rhotot}
illustrate the total baryonic density dependence of
the baryonic entropy per baryon,
$S^{(bar)}/A_b=(S^{(cl)}+S^{(HM)})/A_b$
and total entropy per baryon
$S_{tot}/A_b=\left(S^{(bar)}+S^{(el)}+S^{(\gamma)}\right)/A_b$ 
for $T$=1.6, 5 and 10 MeV and $Y_e$=0.2, 0.3.
Both curve families present a monotonic decrease with $\rho$, a monotonic
increase with $T$ and an isospin asymmetry dependence which vanishes with  
increasing $T$.
Again, at high temperatures and densities corresponding to the transition region,
the predictions of our model are typically different than LS results, while
the two calculations agree well outside the transition region and at low temperatures.
The fact that at $T$=10 MeV the thin line, 
corresponding to the case in which baryonic matter would
exclusively consist of uniform matter, sits perfectly on top of LS,
reflects the different roles played
by the clusterized component in the two models.
The fact that at low temperatures HS results exceed the LS ones may be
  explained by the much increased cluster number within the first
  model (see the right panels of Fig. \ref{fig:massfractions}). 
  This discrepancy diminishes with increasing temperature such at
  $T$=10 MeV, HS and LS results are practically identical. 
As in the case of energy, the total entropy inherits the characteristics of
the baryonic term. Baryonic and total entropies show less sensitivity to 
clusters respect to the energy.

\begin{figure}
\begin{center}
\includegraphics[angle=0, width=0.9\columnwidth]{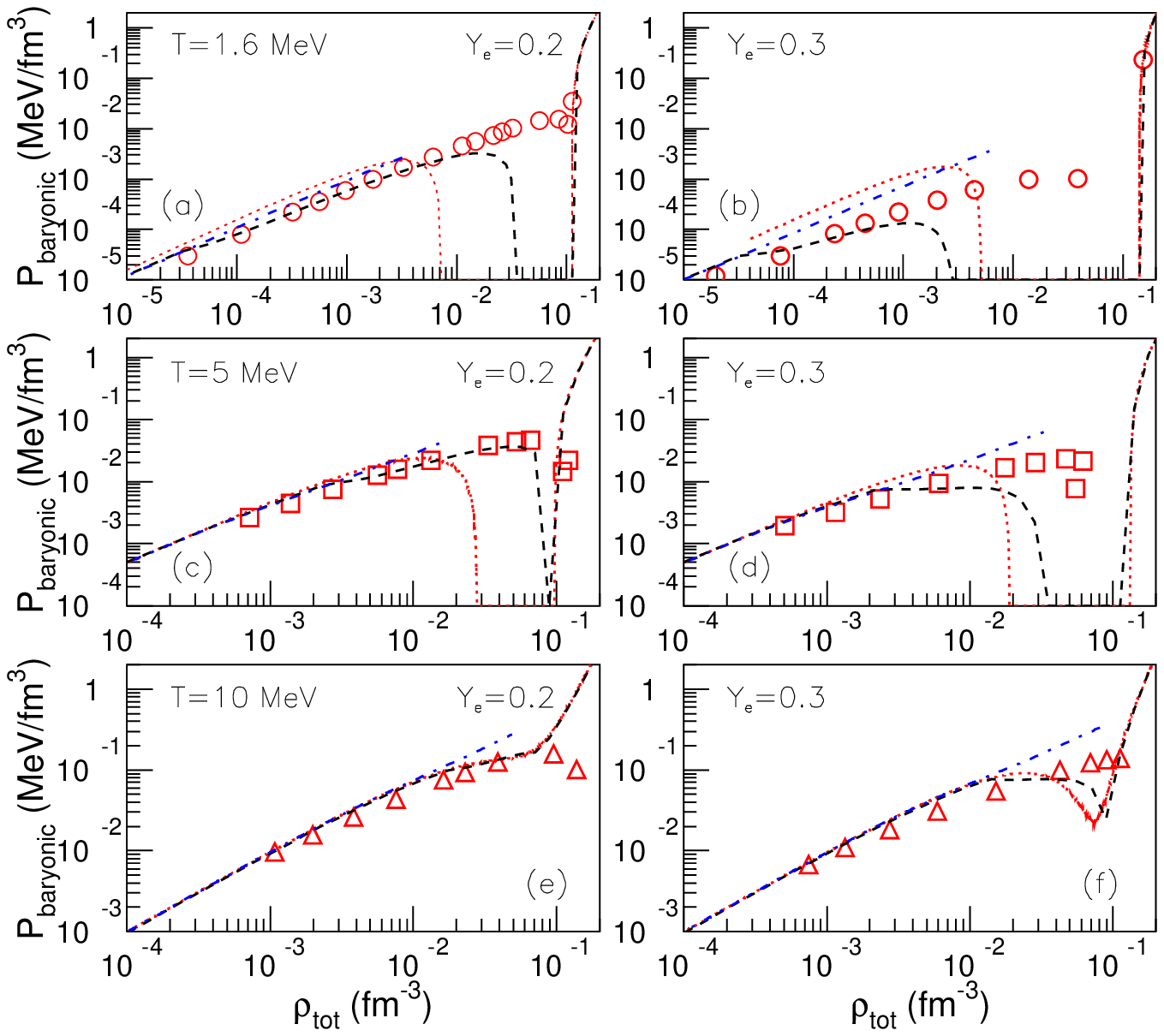}
\end{center}
\caption{The same as in Fig. \ref{fig:eb-rhotot} but for the baryonic pressure,
$p_b=w_{cl} p^{(cl)}+w_{HM}p^{(HM)}$.  
See Fig. \ref{fig:eb-rhotot} for the line and symbol code.
}
\label{fig:pb-rhotot}
\end{figure}

\begin{figure}
\begin{center}
\includegraphics[angle=0, width=0.48\columnwidth]{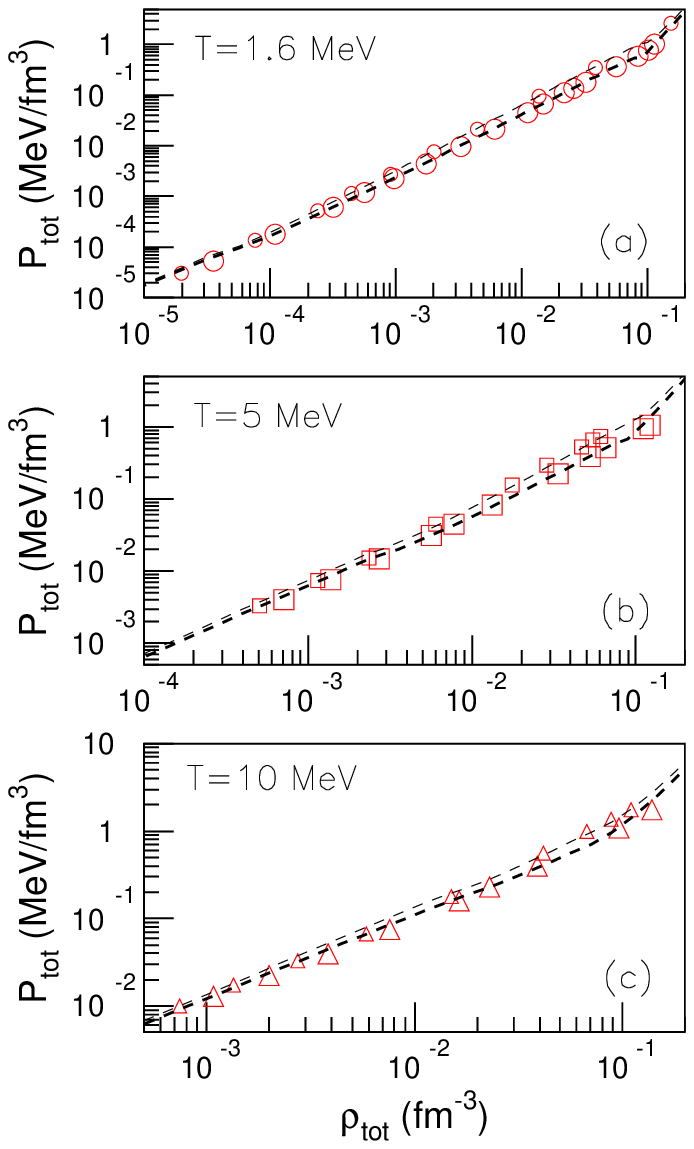}
\end{center}
\caption{The same as in Fig. \ref{fig:etot-rhotot} but for the total pressure,
$p_{tot}=w_{cl}p^{(cl)}+w_{HM}p^{(HM)}+p^{(lattice)}+p^{(el)}+p^{(\gamma)}$.
It is remarkable to notice that by summing up the electron and gamma
contribution, the total EOS shows almost no sensitivity to the baryonic EOS. 
See Fig. \ref{fig:etot-rhotot} for the line and symbol code.
}
\label{fig:ptot-rhotot}
\end{figure}

Figs. \ref{fig:pb-rhotot} and \ref{fig:ptot-rhotot} present
the dependence on the total baryonic density
of the baryonic pressure 
$p^{(bar)}= w_{cl} p^{(cl)}+w_{HM} p^{(HM)}$ and, respectively,
total pressure $p=p^{(bar)}+p^{(lattice)}+p^{(el)}+p^{(\gamma)}$ 
for the same temperatures and proton fractions considered before.
For dilute matter $p^{(bar)}(\rho)$ is a linear function, as the system
recovers the ideal gas limit.
A spectacular feature of the LS results is that, 
up to a certain temperature, 
at a certain value of $\rho$ which depends on the temperature
$p^{(bar)}(\rho)$ manifests a sudden fall. 
 
This behavior is due to those solutions of the uniform matter component 
which are unstable and, thus, characterized by
negative values of $p^{(HM)}$.
If baryonic matter would be exclusively made out of homogeneous matter, 
$p^{(bar)}(\rho_{tot})|_{Y_e}$ would correspond to the dotted lines in 
Fig. \ref{fig:pb-rhotot}. Their shape is qualitatively similar to LS but,
except for $T$=10 MeV, quantitative discrepancies exist over the whole density
range, as one would have actually expected given the cluster contribution.
 
Quite interesting, the falling pattern of HM and LS pressure curves
practically does not exist within our model.
The explanation relies on the mixture stability criterion we have adopted 
in order to maximize the (constrained) entropy (see Section IIA1).
A reminiscence of the fall may, nevertheless, be noticed in the short plateaus
or back-bendings. 
HS results plotted with dot-dashed lines in the left panels
do not manifest signatures of negative pressure, either, meaning that the
nuclear matter obtained by performing a virial expansion of the equation of
state is stable.

The disappearance of the sudden fall at high temperatures 
(in our case $T$=10 MeV) may be understood looking again at
Fig. \ref{fig:phd_t=10}. This figure shows that if the $\mu_n-\mu_p$ trajectory
through the phase diagram fixed by the clusterized system is systematically
situated under the high density spinodal limit, then only one
solution for the uniform nuclear matter will exist, namely the low density stable solution. 

Fig. \ref{fig:ptot-rhotot} correspond to $p_{tot}$
vs. $\rho$, and show that after considering the electron and photon
contributions the discrepancies originating from the baryonic component of the
EOS are to a large extent washed out. 
The same conclusion was reached by Refs. \cite{mishustin,hempel2010}.

\subsection{C. Neutrino opacity}\label{subsec:neutrino_opacity}

\begin{figure}
\begin{center}
\includegraphics[angle=0, width=0.4\columnwidth]{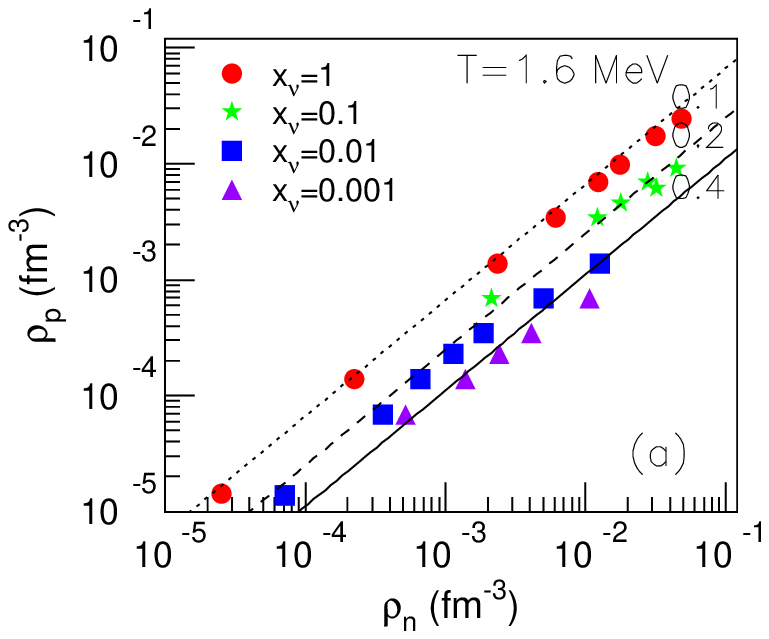}
\includegraphics[angle=0, width=0.4\columnwidth]{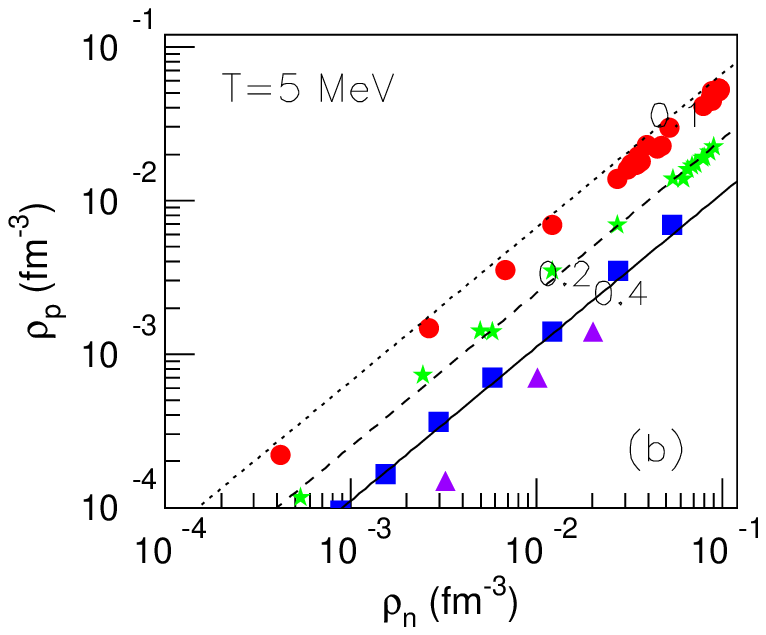}
\includegraphics[angle=0, width=0.4\columnwidth]{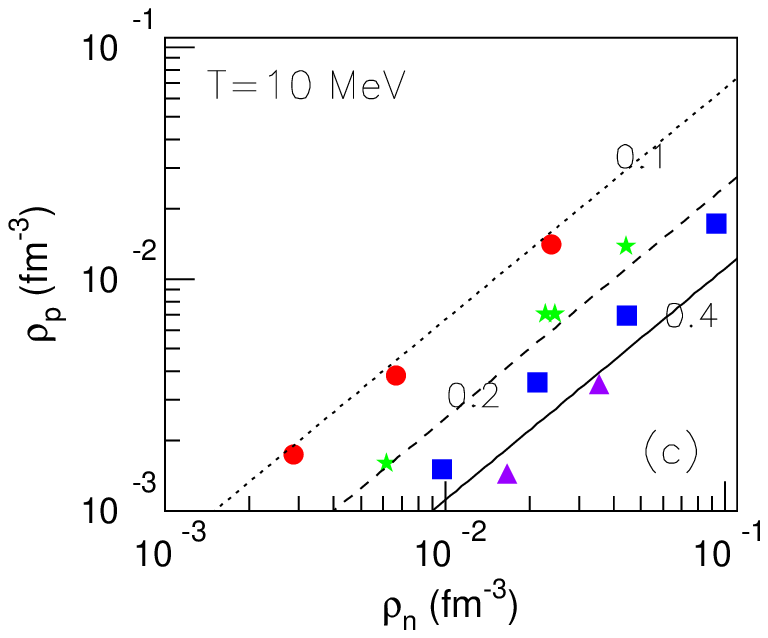}
\end{center}
\caption{(Color online)
Constant neutrino opacity ($x_{\nu}=1, 10^{-1}, 10^{-2}, 10^{-3}$) paths in
the total density plane for $T$=1.6, 5 and 10 MeV under the assumption of 
$\beta$-equilibrium. Neutrino opacity is calculated according to 
Eq. (\ref{eq:neutrinoopacity}).
The lines correspond to paths of constant $Z/A$, whose values are indicated on
the figure. 
}
\label{fig:neutrinos}
\end{figure}

\begin{figure}
\begin{center}
\includegraphics[angle=0, width=0.75\columnwidth]{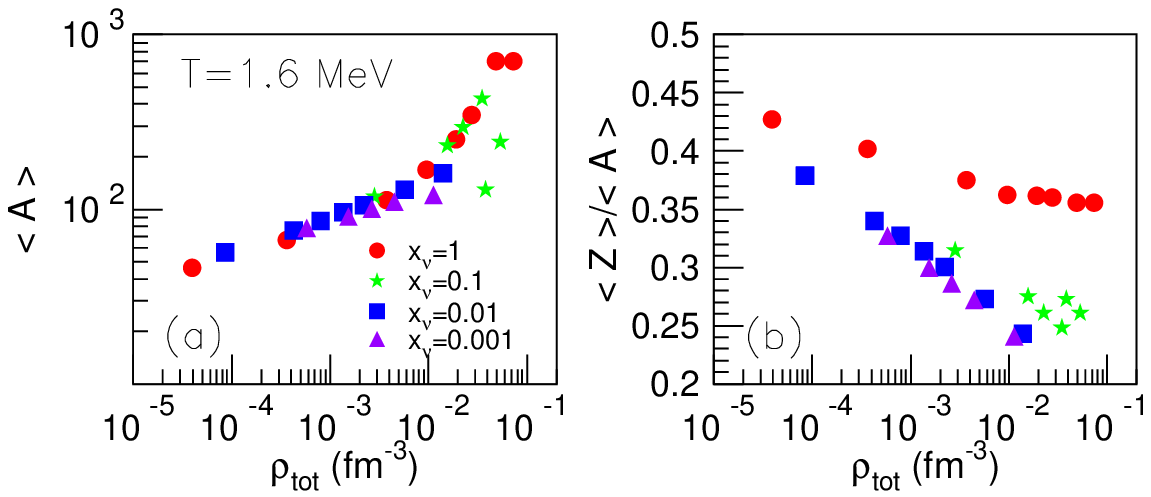}
\includegraphics[angle=0, width=0.75\columnwidth]{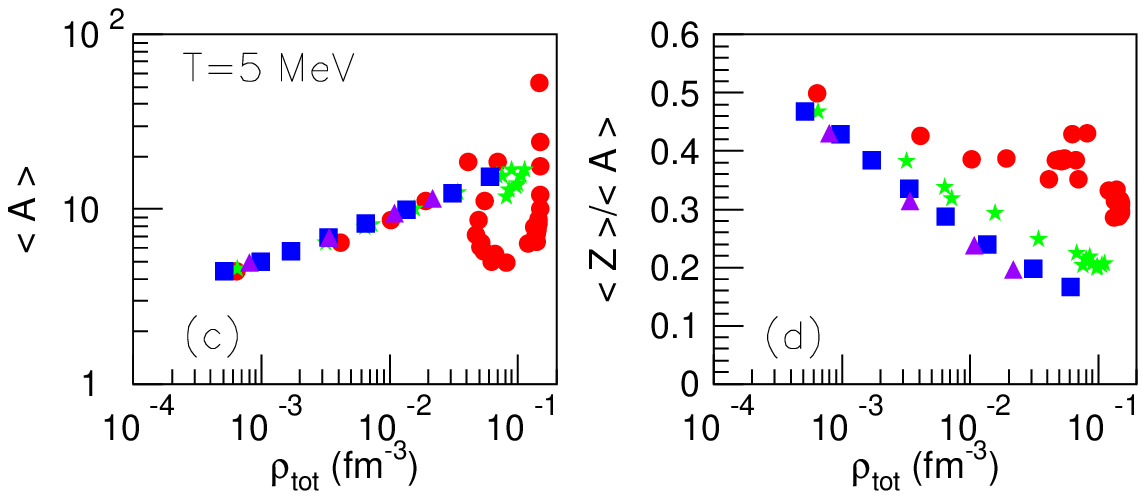}
\includegraphics[angle=0, width=0.75\columnwidth]{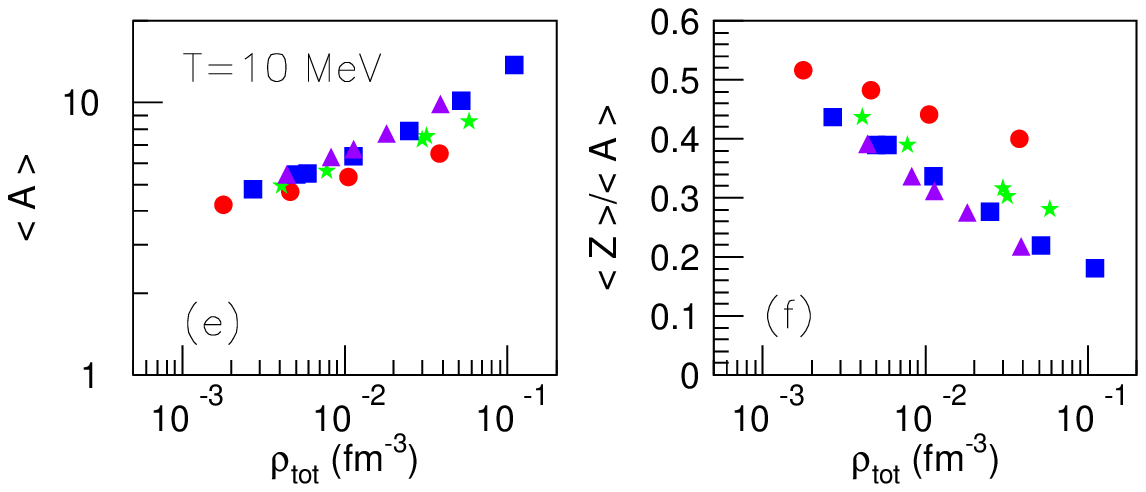}
\end{center}
\caption{(Color online)
Average cluster size (left panels)
and average cluster isospin (right panels)
as a function of total baryonic density for $T$=1.6, 5
and 10 MeV along constant neutrino opacity ($x_{\nu}=1, 10^{-1}, 10^{-2},
10^{-3}$) paths under the assumption of 
$\beta$-equilibrium.
No selection is performed with respect to the mixture stability.
}
\label{fig:amed-opacity}
\end{figure}

Abundant neutrino fluxes are emitted during the collapse of 
a supernova core and in the first $10^5$ years of life of the subsequently born
neutron star, being thus the main responsible for the star cooling. 
The mean free path of the neutrinos depends on the EOS and baryonic composition.
During the collapse phase, it decreases as the star radius shrinks from 
~100 km to ~10 km, such that it eventually becomes smaller than the star radius.
This final regime corresponds to the so-called full neutrino trapping, 
while the initial stage corresponds to neutrino transparency and is
characterized by a vanishing neutrino chemical potential.
Full neutrino trapping is expected to occur also in post-bounce supernovae,
when the matter is still dense. 
The transition between neutrino trapping and neutrino free streaming 
should in principle be a continuous process, obtained by solving the Boltzmann equation
for neutrino transport coupled with the hydrodynamics of the supernova evolution\cite{lieben}.
In many actual simulations, schematic flux-limiting schemes are adopted in the diffusion approximation\cite{bowers}
allowing regimes of partial neutrino trapping.
In these modelizations the neutrino diffusion coefficient depends on the neutrino mean free path, 
which in turn obviously depends on the matter composition through 
the different elastic and inelastic scattering processes of neutrinos on protons and nuclei\cite{lieben,sonoda}.

In Ref. \cite{ducoin_npa2007} it was additionally suggested that 
the matter composition itself may depend in turn on the percentage of trapped neutrinos, thus
introducing a self-consistency problem.

To explore this issue, we study in this section the composition of the baryonic matter in our model 
as a function of the trapped neutrino density.
 
The total number of  neutrinos produced per unit volume $\rho^{prod}_{\nu}$ 
can be calculated from the (local) proton fraction
considering hot star matter as produced from the deleptonization of an initial iron core
according to \cite{ducoin_npa2007}

\begin{equation}
\rho^{prod}_{\nu}=\rho \left[ 
\left( \frac{Z}{A} \right)_0 -\left( \frac{Z}{A} \right)
\right], \label{eq:neutrinoopacity}
\end{equation}

where $\left( Z/A \right)_0$ stands for the $Z/A$ in $^{56}$Fe.

The percentage of trapped neutrinos $x_\nu$ depends on time and on 
the local neutrino mean free path.
Determining this quantity consistently would require a full hydrodynamic calculation
with a complete treatment of the weak processes implied, which is completely out of scope of the 
present paper. 
To study the interplay between trapping and matter composition we simply take the 
neutrino density $\rho_\nu$ at the given thermodynamic condition as a free parameter,
linked to the opacity $x_\nu$ by   
$x_\nu\equiv \rho_\nu/\rho^{prod}_{\nu}$.
The scope of this analysis will then be to check whether the matter composition
depends on the neutrino density, which would imply an extra self-consistent coupling
between the baryon and the lepton sector. 
Indeed if the cluster properties strongly vary during the transition between neutrino 
trapping and neutrino streaming, it would be necessary to couple the neutrino propagation 
with the calculation of the EOS, which would be a very heavy numerical task.

Eq. (\ref{eq:mu_e}) in Section IID shows that for a relativistic Fermi gas 
the particle density is univocally linked to its chemical potential.
In the case of neutrinos, one can safely use the simpler
$T$=0 expression of an 
ultra-relativistic degenerate Fermi gas:
\begin{equation}
\mu_\nu=\hbar c \left ( 6 \pi^2 \rho_\nu \right ) ^{1/3}.
\label{munu}
\end{equation}
 
The beta equilibrium equation (\ref{eq:betaeq}) with a neutrino chemical  
potential $\mu_\nu(\rho_\nu)$ fixed from eq.(\ref{munu}) 
such as to produce the chosen neutrino density $\rho_\nu=x_\nu \rho^{prod}_{\nu}$,
defines a trajectory in the $(\mu_n,\mu_p)$, or equivalently $(\rho_n,\rho_p)$ plane.

These trajectories are shown in Fig. \ref{fig:neutrinos}  for
different values of  $x_{\nu}$ , covering
the whole domain between full trapping $x_\nu=1$ and zero trapping $x_\nu=0$.

As one may remark, the constant $x_{\nu}$ paths roughly correspond to constant
$Z/A$ paths (lines on the figure).
Moreover, 
the global isospin asymmetry of the matter increases monotonically 
with the percentage of trapped neutrinos and no significant dependence on temperature 
is observed.
For instance, over the whole considered temperature domain 
the full neutrino trapping path lies along the $Z/A=0.1$ path, 
while $x_{\nu}=10^{-3}$ corresponds to $Z/A$
slightly higher than 0.4. 
These results are in good agreement with Ref. \cite{ducoin_npa2007}, 
where only homogeneous
nuclear matter was considered and no clusters were included.

The extra information beared by the present work is that we can correlate
the percentage of trapped neutrinos with the characteristics of the clusters.

Fig. \ref{fig:amed-opacity} displays the average cluster size (left panels) 
and isospin composition (right panels)
as a function of total baryonic density  
at different temperatures along constant neutrino opacity paths. 
For the sake of completeness, no selection is now made with respect to the
stability of the mixture.
We can see that the size strongly depends on the thermodynamic conditions 
(temperature and density) but,
in the cases in which homogeneous matter is stable,
 it is virtually independent of the opacity to neutrinos. 
This result can be understood from the fact that changing the percentage 
of trapped neutrinos at beta-equilibrium essentially
amponts to changing the matter isospin asymmetry $Y_p$ for a given total
density $\rho$ (see Fig. \ref{fig:neutrinos}). 
In turn, this affects the chemical composition of clusters 
but not the cluster size, as we have already observed in discussing 
Fig. \ref{fig:frag_wignerseitz}.
Since the interaction probability of neutrinos with matter essentially depends on isoscalar
quantities as the average cluster density and size \cite{horowitz_nu,sonoda}, 
our result indicates that the
problem of neutrino-baryon interaction can be effectively decoupled from the 
problem of the matter composition.

\section{IV. Conclusions}

In this paper we have presented a phenomenological model for stellar matter 
at finite temperature and sub-saturation densities which describes 
matter inhomogeneities as a continuous mixture of 
a distribution of loosely interacting clusters in statistical equilibrium, 
with free nucleons treated within the non-relativistic mean-field
approximation. 
The two components interact through the electrostatic energy in the 
Wigner-Seitz approximation, and a configuration-dependent excluded volume term 
in the spirit of the Van-der-Waals gas. 
 
Such mixture is demonstrated to minimize (maximize) the associated energetic
(entropic) thermodynamic potential in the whole 
considered range of densities and temperature such that a stable solution can
always be found. 
As a consequence, the transition from the homogeneous core to the 
inhomogeneous crust of a finite temperature neutron star is naturally
obtained in the model, at variance with other approaches where discontinuous 
transitions have to be invoked.  

In qualitative agreement with previous works based on 
nuclear statistical equilibrium (NSE), 
we find that the inclusion of a statistical distribution
of clusters in low-density stellar matter modifies in an important way the
average matter composition respect to standard treatments as the LS equation
of state. 
For densities close to the transition density which are not accessible to NSE
models in a thermodynamically consistent way, we show that clustering 
has also sizeable consequences of thermodynamic quantities and equations of state. 
In particular both the symmetry energy and the baryonic energy are 
considerably altered by the 
presence of clusters, in agreement with the microscopic results of Ref. \cite{typel}.

These results open different perspectives for future work which will be
pursued in the next future.

Our formalism where the cluster partition sum is sampled within an exact,
though numerical, Monte-Carlo procedure, allows to account for cluster
correlations. 
We have already shown in a previous paper \cite{clustermatter}
that going beyond the non-interacting NSE approximation for cluster
equilibrium has an important
effect on the cluster phase diagram, and in the present work the
configuration-dependent excluded volume correction, 
which phenomenologically accounts for the nuclear interaction, was shown to be responsible
of the crust-core transition. 
The Coulomb part of the interaction, which is presently treated in the usual
one-body Wigner-Seitz approximation, could also be calculated exactly
accounting for charge fluctuations inside the Wigner-Seitz cell. 
From the existing literature \cite{ising_star,watanabe_screening} we expect 
that properly accounting for Coulomb correlations induced by the charge
fluctuations at the microscopic level
may have an important influence on matter composition and thermodynamic
features, particularly at high density
and temperature where the configuration fluctuations are maximal.

In the regime of baryonic density close to saturation, 
we observe metastable and unstable solutions corresponding
to bubble-like structures where small spherical clusters coexist with 
slightly diluted homogeneous matter.
The possible persistence at high temperature of pasta-like structures, 
well documented at low temperature \cite{watanabe_prl,newton,sebille}, 
is still an open question and it will be interesting to see
if such exotic structures appear if deformation degrees of freedom are 
allowed in the configuration geometry and cluster energy functional.

Concerning the cluster energy functional, in this exploratory work we have
employed a very simple and  simplistic liquid-drop based expression. 
It will however be very easy to implement a composite functional
giving the experimental masses at zero temperature and vanishing density, 
and evolving towards a parametric
formulation at high density and temperature and for unknown neutron-rich nuclear species.

Concerning the homogeneous matter functional, we plan to include pairing 
correlations within the finite temperature HFB approach as a natural 
extension of the present formalism, which will allow having a trustable
equation of state at the lowest temperatures.

Finally, once the different physical inputs will be well settled, 
a longer range project consists 
in constructing a complete EOS table for direct implementation in supernova
and neutron stars codes, in the framework of the COMPSTAR project \cite{compstar}.  

\section{acknowledgements}
Ad. R. R acknowledges partial support from the Romanian National Authority for 
Scientific Research under grant {\it IDEI nr. 267/2007} and PN 09370105/2009
and kind hospitality from LPC-Caen within IFIN-IN2P3 agreement nr. 07-44.
F.G. acknowledges partial support from ANR under Project NExEN.

\end{document}